\RequirePackage{fix-cm}

\documentclass[a4paper,11pt]{article}
\usepackage{jheppub} 

\usepackage[normalem]{ulem}
\usepackage{xcolor}
\usepackage{colortbl}
\usepackage{multirow}
\usepackage{slashed}
\usepackage{dsfont}
\usepackage{caption}
\usepackage{subcaption}
\usepackage{mathtools}
\usepackage{bm}
\usepackage{comment}
\usepackage[capitalise]{cleveref}

\usepackage{tikz-feynman}
\tikzfeynmanset{compat=1.1.0}
\usepackage{subcaption}

\usepackage[page,toc,titletoc,title]{appendix}

\definecolor{colorRTD}{rgb}{.2,.2,.7}

\definecolor{colorRTD2}{rgb}{.2,.3,.7}

\newcommand{\be}{\begin{eqnarray}}
\newcommand{\ee}{\end{eqnarray}}
\newcommand{\nn}{\nonumber}
\newcommand{\ID}{NNID }

\newcommand{\inv}{\mspace{-1mu}{}^{-1}}

\newcommand{\diag}{\text{diag}\mspace{-1mu}}

\newcommand{\ba}{\begin{eqnarray}}
\newcommand{\ea}{\end{eqnarray}}

\numberwithin{equation}{section}
\renewcommand*{\theequation}{%
	\ifnum\value{section}=0 %
	\thechapter
	\else
	\thesection
	\fi
	.\arabic{equation}%
}

\preprint{}

\date{\today}

\title{The Intrinsic Dimension of Collider Events and Model-Independent Searches in 100 Dimensions}
\author[1,2]{Raffaele~Tito~D'Agnolo,}
\affiliation[1]{Universit\'e Paris-Saclay, CNRS, CEA, Institut de Physique Th\'eorique, F-91191 Gif-sur-Yvette, France}
\affiliation[2]{Laboratoire de Physique de l’École Normale Supérieure, ENS, Université PSL, CNRS, Sorbonne
Université, Université Paris Cité, F-75005 Paris, France}
\author[3]{Alfredo Glioti,}
\affiliation[3]{INFN Sezione di Roma, Piazzale Aldo Moro 2, I-00185 Rome, Italy}
\author[1]{Gabriele Rigo,}
\author[4]{Alessandro Valenti}
\affiliation[4]{Department of Physics, University of Basel, Klingelbergstrasse 82, CH-4056 Basel, Switzerland}

\abstract{%
The phase space of hadron collider events spans hundreds of dimensions, generating an intricate geometry that we are just starting to explore. The number of possible new physics signals is exponential in the number of dimensions and detecting all of them is currently impossible for any human or artificial intelligence. In this work we introduce a method to search for new physics model-independently in this high-dimensional space.
It is based on the measurement of the most basic property of the manifold of collider events, its dimensionality.
Our proposed technique does not suffer from a look-elsewhere effect that grows exponentially with the number of dimensions of the dataset, and by construction is insensitive to energy scale uncertainties. 
We illustrate its potential by finding new physics in simulated events with hundreds of phase space dimensions, taking as input single particles rather than jets. This study sets the stage for new model-independent search strategies based on global properties of collider data manifolds. 
}

\begin{document}

\setcounter{page}{1}
\setcounter{footnote}{0}
\setcounter{equation}{0}

\maketitle

\section{Introduction}
\label{sec:intro}
Proton–proton collisions at the Large Hadron Collider (LHC) fill the detector with hundreds of particles, populating a highly complex manifold\footnote{Without any infrared regulator this is not a real manifold. Its dimension is unbounded due to the emission of infinitely many soft photons. However the detector itself regulates the divergence. We thank J. Thaler for pointing this out to us.} in their phase space. Identifying all potential new physics signals within this manifold is extremely challenging and goes beyond what traditional analysis strategies or novel machine-learning approaches can reliably achieve.
In this work we propose a methodology that allows to take a first step in the search for new physics with hundreds or even thousands of observables, when the signal model is unknown. 

Our approach is based on the measurement of the number of intrinsic dimensions of phase space. To derive this quantity from data we need to define a metric on this space. In this work we use the energy mover’s distance (EMD) proposed in \cite{Komiske:2019fks}, which connects to several known collider observables \cite{Komiske:2020qhg} and allows a transparent interpretation of our results in terms of measurable energy scales. 
We compute the intrinsic dimension of the data manifold using a nearest-neighbors estimator \cite{Levina2004MaximumLE, ceruti2012danco, facco2017estimating, denti2022generalized} described in Section~\ref{sec:ID}, and we refer to the resulting observable as \ID (Nearest Neighbor Intrinsic Dimension) to distinguish it from other existing ways to measure the intrinsic dimension of a dataset (see \cite{campadelli2015intrinsic, CAMASTRA201626} for a review).

Our proposal has been developed in the context of a growing interest in the geometry of collider events. New observables were presented in~\cite{Komiske:2020qhg,Cesarotti:2020hwb,Ba:2023hix}, a series of optimal transport distances between energy flow distributions were introduced in~\cite{Cai:2020vzx, Cai:2021hnn, doi:10.1137/21M1400080} and a number of applications have followed. In particular, a different notion of intrinsic dimension, compared to the one used in this paper, was applied to jets in the CMS open data in~\cite{Komiske:2019jim,Komiske:2022vxg} and other strategies for new physics searches were discussed in~\cite{Mullin:2019mmh,CrispimRomao:2020ejk,Davis:2023lxq}. We comment in detail in Section~\ref{sec:ID} on the differences between our notion of intrinsic dimension, the correlation dimension used in~\cite{Komiske:2019fks, Komiske:2019jim,Komiske:2022vxg} and other estimators for the dimension of a data manifold.  

Additionally, the EMD has been used to speed-up collider computations~\cite{Cai:2020vzx,Cai:2021hnn, Tsan:2021brw,Kitouni:2022qyr} and classify jets~\cite{Alipour-Fard:2023prj}. A covariant description of $N$-body phase space has been introduced in~\cite{Larkoski:2020thc} and has led to the definition of a Riemaniann metric on this manifold in~\cite{Cai:2024xnt, Cai:2025fyw}. 

From this discussion it is clear that we are not the first ones to take a ``global look'' at collider events, but we still want to focus on the most basic property of the phase space manifold, its dimensionality. There are multiple reasons to make this choice. First of all, to the best of our knowledge, we are the first to use the intrinsic dimension of the data to search for new physics. This approach has several qualitative advantages, that we detail below, over existing model-independent searches in high-dimensional spaces.

Besides, the \ID that we discuss in this paper, which is well-known in computer science~
\cite{pettis1979intrinsic, Levina2004MaximumLE, Maaten2008DimensionalityRA, Rozza2012NovelHI, campadelli2015intrinsic, facco2017estimating, ansuini2019intrinsic, pope2021intrinsic,Valeriani2023TheGO, Gliozzo2024IntrinsicDimensionAF} and statistical physics
~\cite{Grassberger:1983zz, Hentschel:1983zhc}, can be computed from a maximum-likelihood procedure founded on solid statistical principles, with well-understood convergence properties~\cite{NIPS2004_74934548}.  

Finally, in its simplest form the \ID is dominated by the shortest distances in the dataset as opposed to the correlation dimension used in~\cite{Komiske:2019fks,Komiske:2019jim,Komiske:2022vxg} which is dominated by the largest distances.
Furthermore, one can generalize its simplest definition to explore the kinematics of the phase space manifold at different energy scales.\footnote{This can be done also with correlation dimensions~\cite{Komiske:2019fks,Komiske:2019jim,Komiske:2022vxg}, but we find in toy models that the \ID better captures the short-distance physics.}

After introducing the intrinsic dimension and its properties in Section~\ref{sec:ID}, in Section~\ref{sec:NP} we build a test for new physics based on it, which includes also information on the average distance between neighboring points in phase space. In Section~\ref{sec:applications} we find that our test is sensitive to a wide variety of signals, to the point that it can be considered a model-independent search for new physics. Any signal that we have tested has a characteristic shape for its intrinsic dimension as a function of energy which differs from the Standard Model (SM), even when the kinematics of the events are similar. 

A large number of model-independent new physics search strategies exists in the literature~\cite{CMS:2020zjg, Aguilar-Saavedra:2017rzt, Collins:2018epr, DAgnolo:2018cun, DeSimone:2018efk, Hajer:2018kqm, Farina:2018fyg, Heimel:2018mkt, 1809.02977, Cerri:2018anq, Collins:2019jip, Roy:2019jae, Dillon:2019cqt, Blance:2019ibf, Romao:2019dvs, Mullin:2019mmh, DAgnolo:2019vbw, Nachman:2020lpy, Andreassen:2020nkr, Amram:2020ykb, Romao:2020ojy, Knapp:2020dde, ATLAS:2020iwa, Dillon:2020quc, CrispimRomao:2020ucc, Cheng:2020dal, Khosa:2020qrz, Thaprasop:2020mzp, Aguilar-Saavedra:2020uhm, Alexander:2020mbx, Benkendorfer:2020gek, Pol:2020weg, Mikuni:2020qds, vanBeekveld:2020txa, Park:2020pak, Faroughy:2020gas, Stein:2020rou, Kasieczka:2021xcg, Chakravarti:2021svb, Batson:2021agz, Blance:2021gcs, Bortolato:2021zic, Collins:2021nxn, Dillon:2021nxw, Finke:2021sdf, Shih:2021kbt, Atkinson:2021nlt, Kahn:2021drv, Aarrestad:2021oeb, Dorigo:2021iyy, Caron:2021wmq, Govorkova:2021hqu, Kasieczka:2021tew, Volkovich:2021txe, Govorkova:2021utb, Hallin:2021wme, Ostdiek:2021bem, Fraser:2021lxm, Jawahar:2021vyu, Herrero-Garcia:2021goa, Lester:2021aks, Aguilar-Saavedra:2021utu, Tombs:2021wae, Mikuni:2021nwn, Chekanov:2021pus, dAgnolo:2021aun, Canelli:2021aps, Ngairangbam:2021yma, Aguilar-Saavedra:2022ejy, Buss:2022lxw, Bradshaw:2022qev, Birman:2022xzu, Raine:2022hht, Letizia:2022xbe, Fanelli:2022xwl, Finke:2022lsu, Verheyen:2022tov, Dillon:2022tmm, Alvi:2022fkk, Dillon:2022mkq, Caron:2022wrw, Park:2022zov, Kasieczka:2022naq, Kamenik:2022qxs, Hallin:2022eoq, Araz:2022zxk, Mastandrea:2022vas, Schuhmacher:2023pro, Golling:2023juz, Roche:2023int, Sengupta:2023xqy, Aguilar-Saavedra:2023pde,Vaslin:2023lig, ATLAS:2023azi, Mikuni:2023tok, Golling:2023yjq, Chekanov:2023uot, CMSECAL:2023fvz, Bickendorf:2023nej, Finke:2023ltw, Buhmann:2023acn, Freytsis:2023cjr, Metodiev:2023izu, Banda:2025nrv, Puljak:2025gtj,CMS:2025lmn, CMS:2024nsz, BrightThonney2025,grosso2025sparseselforganizingensembleslocal}. Most of them have been developed in the last decade in response to the tension between our theory priors and LHC results. Theoretical considerations (as the electroweak hierarchy problem) led us to expect new physics at the LHC, but hundreds of experimental searches have measured only the Standard Model (SM). 

There are several qualitative differences between our approach and other existing ideas. To better understand the main difference, it is useful to divide existing techniques into two broad categories. The first one, developed for instance in~\cite{Collins:2018epr, Collins:2019jip, Benkendorfer:2020gek}, consists in choosing a fixed analysis strategy which is independent of the dimensionality of the dataset. A concrete example could be: ``Take the two leading muons in the event and search for a bump in their invariant mass". This example greatly oversimplifies the ideas in~\cite{Collins:2018epr, Collins:2019jip, Benkendorfer:2020gek}, but captures their two basic features: they do not suffer from a look-elsewhere effect that increases exponentially with the number of dimensions, but they are insensitive to large classes of signals. The second category, which includes~\cite{DAgnolo:2018cun, DAgnolo:2019vbw, dAgnolo:2021aun}, lives at the opposite end of the spectrum. It is sensitive to essentially any signal in the dataset, but suffers from a correspondingly large look-elsewhere effect. Techniques based on autoencoders, such as~\cite{Heimel:2018mkt, Farina:2018fyg, Finke:2021sdf, CMS:2025lmn} or a pre-training dimensionality reduction step~\cite{BrightThonney2025,grosso2025sparseselforganizingensembleslocal, Oord2018, Wang2021, Bronstein2021} can live in either one of the two categories, depending on their implementation. If the number of features extracted in the dimensionality reduction step is increased with the input dimension\footnote{This is done to reliably describe the null hypothesis i.e. the SM.}, then we have a strategy in the second category, with an exponential increase of the look-elsewhere effect. If instead this number is fixed, the analysis strategy falls in the first category, i.e. the look-elsewhere effect is kept under control, but at the price of reducing the number of signals that one is sensitive to.
We explain this point in more detail in Section~\ref{sec:summary}.

Our methodology is conceptually similar to the first category of techniques, since it is a fixed analysis strategy, independent of the dimensionality of the dataset, and does not suffer from a large look-elsewhere effect. The strategy is essentially the comparison of the NNID, and of distances in phase space, between different datasets as a function of energy. We are choosing {\it ab initio} a handful of fixed observables and using them to distinguish signal from background, so by construction we are not sensitive to any possible signal in the dataset. However, this methodology ends up being much more flexible than most existing techniques, because the \ID captures any subtle difference in kinematics between the SM and new physics, well beyond ``bumps'', ``tails'' or other features that one typically looks for, as discussed in Section~\ref{sec:applications}. The approaches in~\cite{BrightThonney2025,grosso2025sparseselforganizingensembleslocal} are quite similar in their final goal, but quite different in terms of implementation. The observables used there are selected using contrastive learning from a higher-dimensional set of input features defined on a event-by-event basis, while we rely on global observables that are only defined on the whole dataset.

A second unique feature of our new physics test is its robustness against an important set of systematic errors. As we discuss in Section~\ref{sec:ID} the \ID is measured from a set of ratios of dimensionful quantities. This makes it completely insensitive to energy scale uncertainties. Additionally, it involves an average over the dataset and this strongly reduces the impact of energy and angular resolution effects. We demonstrate these properties in Section~\ref{sec:applications}.

It is important to point out that our search strategy is not designed for precision searches in the sense of high energy physics, since we need at least one dimension of the dataset to be observably distorted compared to the SM prediction. 
We are not looking for a part in $10^5$ effect on top of a SM distribution that has already been studied in great detail, but rather for signals with large significances. Nonetheless our technique is designed to solve the small signal-over-background problems in high-dimensional spaces that are relevant to high energy physics, where the signal can be rare, weakly expressed and embedded in the data manifold. 

We want a framework to analyze datasets with hundreds or thousands of phase space dimensions, which can go beyond clustering jets and, if possible, even reconstructing tracks. Large signals (in terms of statistical significance) could still be hiding in these spaces, since at the LHC we essentially reconstruct what we know, i.e. well-behaved tracks\footnote{For an alternative approach to the problem which seeks to reconstruct non-helical tracks see~\cite{Condren:2025czc}.} of particles with charge $\sim \pm1$ without long-range correlations that are then clustered into well-behaved anti-$k_T$ jets. Any light new physics creating correlations between tracks, jets or constituents of different jets, or deforming them could still be hiding in the data\footnote{Although a number of efforts in this direction has been made in the context of specific models~\cite{Forsyth:2025wks, Farina:2017cts, Sha:2024hzq, Knapen:2017kly, Li:2019wce, Li:2020aoq, Knapen:2016hky}.}. We consider the intrinsic dimension the natural first step to detect these signals. In this work, as a first test and to develop some analytical intuition, we study the performance of this technique on more traditional signals, but its real goal is to find new physics that we might not even have imagined yet.

\section{Intrinsic Dimension from Nearest Neighbors}
\label{sec:ID}

Consider a sample of $D$-dimensional vectors $\{\bm{x}_l\}_{l=1}^N$, where each vector is drawn from a probability distribution having support on a submanifold of dimension $d \leq D$. Our goal is to estimate $d$ from the data without prior knowledge of their probability distribution. Here, we focus on nearest-neighbor based estimation methods \cite{Levina2004MaximumLE, ceruti2012danco,facco2017estimating, denti2022generalized}, although other approaches are possible (see \cite{campadelli2015intrinsic,CAMASTRA201626} for a review). We choose nearest-neighbor estimators because they are grounded on a firm statistical foundation, allowing for exact asymptotic results and a simple interpretation.

We assume that \emph{locally} the probability density function (pdf) of our dataset on the support of dimension $d$ is constant. In practice we are taking a small neighborhood of a point in the dataset and keeping only the constant term in a Taylor expansion of the pdf. We then imagine to have enough points in the dataset that this constant value does not fluctuate appreciably when we choose different starting points. Notice that this assumption is needed if we want to measure $d$, but it is not needed for our new physics search strategy in Section~\ref{sec:NP}. In that case we want to identify an observable that distinguishes signal from background, but does not necessarily have to be equal to $d$, even in expectation value. In practice, when searching for new physics, we compute a property of the manifold that converges to $d$ when $N\to \infty$, but in some cases we purposefully stay away from this asymptotic limit.
The $N\to \infty$ limit sees any contribution to $d$ including spurious dimensions introduced by the finite momentum and angular resolution of the detector. In a realistic setting, with finite detector resolution and systematic uncertainties, the ``true'' dimension $d$ of the manifold also includes these effects. As a simple illustration, consider $2 \to 2$ events with on-shell final-state particles. When the initial energies are fixed, as at a lepton collider, the kinematics span only two dimensions in phase space so $d = 2$. At a hadron collider, by contrast, the initial parton energies are not fixed, and the same final state spans four dimensions so\footnote{At a lepton collider one may parametrize the configuration by the two angles $\theta$ and $\phi$ of one of the final-state particles, while at a hadron collider one must additionally specify the total energy and rapidity of the two-particle center of mass.} $d = 4$.
However a finite momentum resolution of our detector makes $d \to 6$. 
If the particles have momenta much larger than the detector resolution, these noise directions are significantly smaller than the physical ones, and one may avoid the asymptotic small-scale limit of the estimator to reduce detector effects on the measurement of $d$.
More details on this point are given in Section~\ref{sec:NNID}.

For now we set aside this discussion and focus on the measurement of $d$ in the limit where the pdf $\rho$ is constant. In this limit the number of points falling within a sphere of radius $r$ with center $\bm{x}_k$ can be thought of as a realization of a homogeneous Poisson process with constant intensity function $\rho$, i.e. the probability of finding $m$ points within the sphere of volume $V(r)$ is
\be
P(m; r, \rho) &=& (\rho V(r))^m \frac{\exp(-\rho V(r))}{m!}\, , \nn \\
V(r) &=& \frac{\pi ^{d/2} r^m}{\Gamma(d/2+1)}\, .
\ee
The number of points in the hyperspherical shells between any consecutive neighbors are independent and identically distributed exponential random variables, with constant parameter $\rho$~\cite{kingman1993poisson, facco2017estimating}.

To remove any dependence on the unknown dimensionful variable $\rho$, one can consider the dimensionless ratio between the distance of point $\bm{x}_k$ to its $j$-th and $i$-th nearest neighbors ($j>i$),
\begin{align}
    \mu_{k,ij}  \equiv \frac{r_{k,j}}{r_{k,i}}.
\end{align}
Expressing $\mu_{k,ij}$ in terms of the volume of the hyperspherical shells, it is then straightfoward to obtain its probability distribution~\cite{denti2021distributional},
\begin{align}
    f(\mu_{k,ij}; d) =\frac{ d  \mu_{k,ij}^{- i d - 1} (1-\mu_{k,ij}^{-d})^{j-i-1}}{B(j-i,i)}\, ,
    \label{eq:muDistr}
\end{align}
where $B$ denotes the Euler Beta function.
As promised, Eq.~\eqref{eq:muDistr} does not depend on the unknown variable $\rho$, but it has an explicit dependence on the parameter $d$ that we want to estimate. It is therefore natural to construct the negative log-likelihood,
\begin{align}
    L = - \sum_{k=1}^N \log f(\mu_{k,ij};d) = - N \log d +(1+id) \sum_{k=1}^N \log \mu_{k,ij} - (j-i-1) \sum_{k=1}^N \log\left(1- \mu_{k,ij}^{-d} \right)\, ,
    \label{eq:logL}
\end{align}
where the sum extends over all points in the dataset.
The likelihood in~\cref{eq:logL} can be minimized to obtain the Maximum Likelihood Estimator (MLE) $\hat d_{i,j}$ for the intrinsic dimension $d$. In principle any choice of $i,j$ is measuring the same value of $d$ and should give the same result, but this is true only asymptotically ($N\to \infty$), when $\rho$ is truly constant across the dataset. 
For the specific case $j=i+1$ the minimization can be carried out analytically, resulting in 
\be
\hat d_{i,i+1} &=& \frac{N/i}{\sum_k \log \mu_{k,i(i+1)}}\, , \nn \\
E[\hat d_{i,i+1} ] &=& d \frac{N}{N-1}\, .
\label{eq:IDiplus1}
\ee
In general \cref{eq:logL} should be minimized numerically, but the particular choice of $j$ in Eq.~\eqref{eq:IDiplus1} allows to compute $\hat d_{i,i+1}$ from first principles in QCD as was done in~\cite{Komiske:2022vxg} for a different notion of intrinsic dimension. For $i=1$, $\hat d_{i,i+1}$ in Eq.~\eqref{eq:IDiplus1} is the same as the two nearest neighbors estimator originally introduced in~\cite{facco2017estimating}, although here we obtain it from a ML principle rather than a fit to the empirical CDF of $\mu$. 

To better understand our estimator for $d$ for collider events, it is important to recall the definition and the properties of the metric that we have chosen. The EMD introduced in~\cite{Komiske:2019fks} has dimensions of energy and measures the distance between two events $\mathcal{E}_{A,B}$. Here by events we mean collections of three momenta of massless particles, and in practice we use for each particle its $p_T$, $y$ and $\phi$ coordinates, since we are mainly considering applications to hadron collider physics.
The EMD is the minimum amount of ``work" needed to move all the particles in event $\mathcal{E}_A$ so that the event is rearranged into event $\mathcal{E}_B$. Formally it can be obtained via the minimization process
\be
{\rm EMD}(\mathcal{E}_A, \mathcal{E}_B) = \min_{\{\xi_{k m}\}} \sum_{k,m} \xi_{km} \frac{\theta_{km}}{R} + \left| \sum_{k\in\mathcal{E}_A} E_k - \sum_{m\in\mathcal{E}_B} E_m \right|\, ,
\label{eq:EMD}
\ee
subject to the constraints
\be
\xi_{km}\geq 0\, , \quad \sum_{k\in\mathcal{E}_A} \xi_{km} \leq E_m\, ,\quad \sum_{m\in\mathcal{E}_B} \xi_{km} \leq E_k, \quad \sum_{k,m}\xi_{km} \leq E_{\rm min}\, .
\ee
In the above equations we have introduced a dimensionless parameter $R$ which sets the relative importance of the two terms in the EMD. We have also used the angular distance between two particles
\be
\theta_{km} \equiv \sqrt{(y_k-y_m)^2+(\phi_k - \phi_m)^2}
\ee
and the minimal energy $E_{\rm min} = \min(E_{\mathcal{E}_A}, E_{\mathcal{E}_B})$, where $E_{\mathcal{E}_{A,B}}$ are the total energies in the two events.

One can show~\cite{Komiske:2019fks} that the EMD is a true metric that satisfies the triangle inequality as long as $\theta_{km}$ is a metric and $R\geq \theta_{\rm max}/2$, where $\theta_{\rm max}$ is the largest angular distance in the dataset. Additionally the EMD is an infrared and collinear (IRC) safe observable~\cite{Komiske:2019fks}.

Note that the EMD is not the only possible definition of distance between collider events and more recent proposals, such as~\cite{Larkoski:2023qnv, Gambhir:2024ndc}, reduce the minimization in Eq.~\eqref{eq:EMD} which is a two-dimensional optimal transport problem, to a one-dimensional problem that can be solved analytically. However in this work we use Eq.~\eqref{eq:EMD} since it allows a simple physical interpretation of our results which is needed to construct the new physics search strategy in Sections~\ref{sec:NP} and~\ref{sec:applications}. We leave to future work the exploration of different definitions of distance for the estimation of the NNID.

For us the most relevant feature of Eq.~\eqref{eq:EMD} is that small values of the EMD correspond to small energy differences and small spatial separations on the collider ``celestial sphere''. Similarly, large EMDs correspond to large energy differences and spatial separations. 

We are now ready to develop some intuition on our intrinsic dimension estimator $\hat d_{i,j}$. From now on we focus on the estimator evaluated for $j=2i$, that we call $\hat d_i$,
\be
\hat d_i \equiv \hat d_{i, 2i}\, . \label{eq:2i}
\ee
This choice gives a small statistical error (i.e.\ a stable estimate for $d$) over the largest range in $i$ (and hence energy scales), as verified empirically in~\cite{denti2022generalized} and confirmed by our studies of the models in Section~\ref{sec:applications}.

\subsection{Properties of \ID Estimators at finite $N$}\label{sec:NNID}

Consider the estimator $\hat d_i$ in~\eqref{eq:2i}.
As we increase $i$ at fixed $N$ we are looking at the phase space manifold from further and further away, progressively becoming insensitive to small energy differences and spatial separations, so there is some optimal value of $i$ that cuts out all detector effects. For example, a finite resolution in energy can introduce a fictitious small dimension that we might want to get rid of.

If we take $i$ large enough the manifold will appear to us as a point-like object. Therefore, even if $\hat d_{i,j}$ is independent of $i,j$ asymptotically (that is for $N\to \infty$),  at fixed $N$ we expect $\hat d_{i,j} \to 0$ as $i$ grows. This can be useful to filter out noise, but also to explore the kinematical features of our events at different energy scales. At the qualitative level, we expect $\hat d_{i,j} \to d$, independent of $i,j$, when $N\to \infty$, because in that limit more and more points are added at short distances and any finite choice of $i,j$ is probing infinitesimal distances in the dataset. Therefore in the asymptotic limit $\hat d_{i,j}$ converges to the result that a ``perfect short-distance observer'' would measure.

If we want to retain the dependence of $\hat d_i$ on the energy scales in the dataset also in the asymptotic limit, we can  compute it at fixed $i/N$ rather than $i$.
We can justify quantitatively the choice of varying $i/N$ rather than $i$ as follows. The pdf of the  distance $r_l$ between a point and its $l$-th neighbor is~\cite{denti2021distributional}
\be
f(r_l) = \frac{d}{r_l (l-1)!} (\rho \omega_d r_l^d)^l  \exp(-\rho \omega_d r_l^d) \, , \label{eq:rlpdf}
\ee
with $\omega_d$ the volume of the $d$-dimensional sphere of unit radius. From Eq.~\eqref{eq:rlpdf} one has $E[r_l] \simeq (l/\rho \omega_d)^{1/d}$, so that on average $r_l/r_N \approx (l/N)^{1/d}$. If we imagine that $r_N$ is bounded by the kinematics and varies slowly with $N$, eventually saturating in the asymptotic limit, by fixing the \emph{proportions} $i/N, j/N$ we are probing fixed distances in the dataset. This should lead to curves which are compatible between datasets with different number of points and retains the dependence of $\hat d_i$ on the energy scales in the dataset also in the $N\to \infty$ limit. We verify this intuition on the models studied in Section~\ref{sec:applications}. 

The last property of $\hat d_i$ that we want to highlight and we are going to use in Section~\ref{sec:NP} is its distribution in the asymptotic limit. In the $N\rightarrow \infty$ limit, the estimators $\hat{d}_{i,j}$ minimizing \cref{eq:logL} are expected to follow a multivariate Gaussian distribution under standard regularity conditions. For the case $j=i+1$, this is easily seen and follows directly from the Central Limit Theorem and Delta method applied to \eqref{eq:IDiplus1} \cite{casella1990statistical}. 
For the vector of estimators $\bm{\hat d} = (\hat{d}_{i_1,j_1}, \hat{d}_{i_2,j_2}, \ldots)^T$ across different $(i,j)$ pairs, joint asymptotic normality follows from multivariate M-estimator theory \cite{huber1964robust, huber1973robust}. Asymptotically $\sqrt{N}(\bm{\hat d} - \bm{d}_0) \rightarrow N(\bm{0}, \bm{\Sigma})$, where the covariance matrix $\bm{\Sigma}$ accounts for correlations arising from shared data points and nearest-neighbor relationships across different $(i,j)$ pairs. This is true under ``physically sensible'' regularity conditions, including finite second moments of $\log \mu_{k,ij}$, existence of the Jacobian and non-singularity of the Hessian matrix of \eqref{eq:logL} at the true solution $\bm{d}_0$,

In all the cases discussed later, we empirically verify the normality of our estimators before using this property in any new physics search.

To summarize, when performing the new physics search described in Section~\ref{sec:NP} we are going to compute our estimators $\hat d_i$ as a function of $i/N$ for two reasons: 1) to make them comparable across different datasets, 2) to retain the dependence on $i$ (and thus on energy scale) also in the $N\to \infty$ limit. Additionally we are going to verify that the entries of the vector $\left\{\hat d_i\right\}_{i=1,.., N}$ follow a joint normal distribution that we then use to compute significances.

\subsection{Comparison with Existing Work}\label{sec:exist}

We can compare the \ID to the correlation dimension computed in~\cite{Komiske:2019fks, Komiske:2019jim,Komiske:2022vxg} and other strategies commonly used to measure the dimension of a dataset. First of all we recall the broader context of dimensionality reduction. A number of methodologies~\cite{Jolliffe, Cox2008, Nonlinear, COIFMAN20065, Global, Laplacian} define a loss function which is essentially a projection error and then try to minimize it. The dimensionality of the sample is inferred from discontinuities in the eigenvalues of the loss function. These ideas face serious difficulties when trying to define the notion of discontinuity for a function (the spectrum of eigenvalues) defined over discrete values.

We take the alternative point of view that the data are sampled from a high-dimensional ``intrinsic manifold'' and we estimate their dimensionality from that of its tangent space at each point. Fractal methods, as the one used in~\cite{Komiske:2019fks, Komiske:2019jim,Komiske:2022vxg} to measure the \emph{correlation dimension} of jets, are another application of the same logic. 
Fractal methods rely on the intuition that the number of points in a sphere of radius $r$ should grow as $r^d$ for constant density. Their main drawback is that an accurate estimate of the dimension requires a number of data points that grows exponentially with $d$, as discussed in~\cite{Facco_2017, ID2}. 

In practice, correlation dimensions have a hard time reproducing the dimension that a ``UV observer'' with access to the shortest distances in the dataset would measure. This would be the number that we are interested in if we wanted to know how many neurons we need in the latent space of an autoencoder, or how many additional degrees of freedom one needs to describe a jet compared to the 3 components of the momentum of a massless parton. The correlation dimension in~\cite{Komiske:2019fks, Komiske:2019jim,Komiske:2022vxg} captures a different and physically interesting property of a dataset. It was shown to measure the logarithmic increase in the number of jet constituents as one decreases the jet energy scale~\cite{Komiske:2019jim}. In the regime considered in~\cite{Komiske:2019fks, Komiske:2019jim,Komiske:2022vxg} it is essentially measuring the entropy in the dataset. The \ID is instead insensitive to the density of points, and thus the entropy of the dataset, as long as the density is constant on the scale defined by the typical distance to the $j$-th neighbor.

\begin{figure}[!t]
    \centering
    \includegraphics[width=0.5\linewidth]{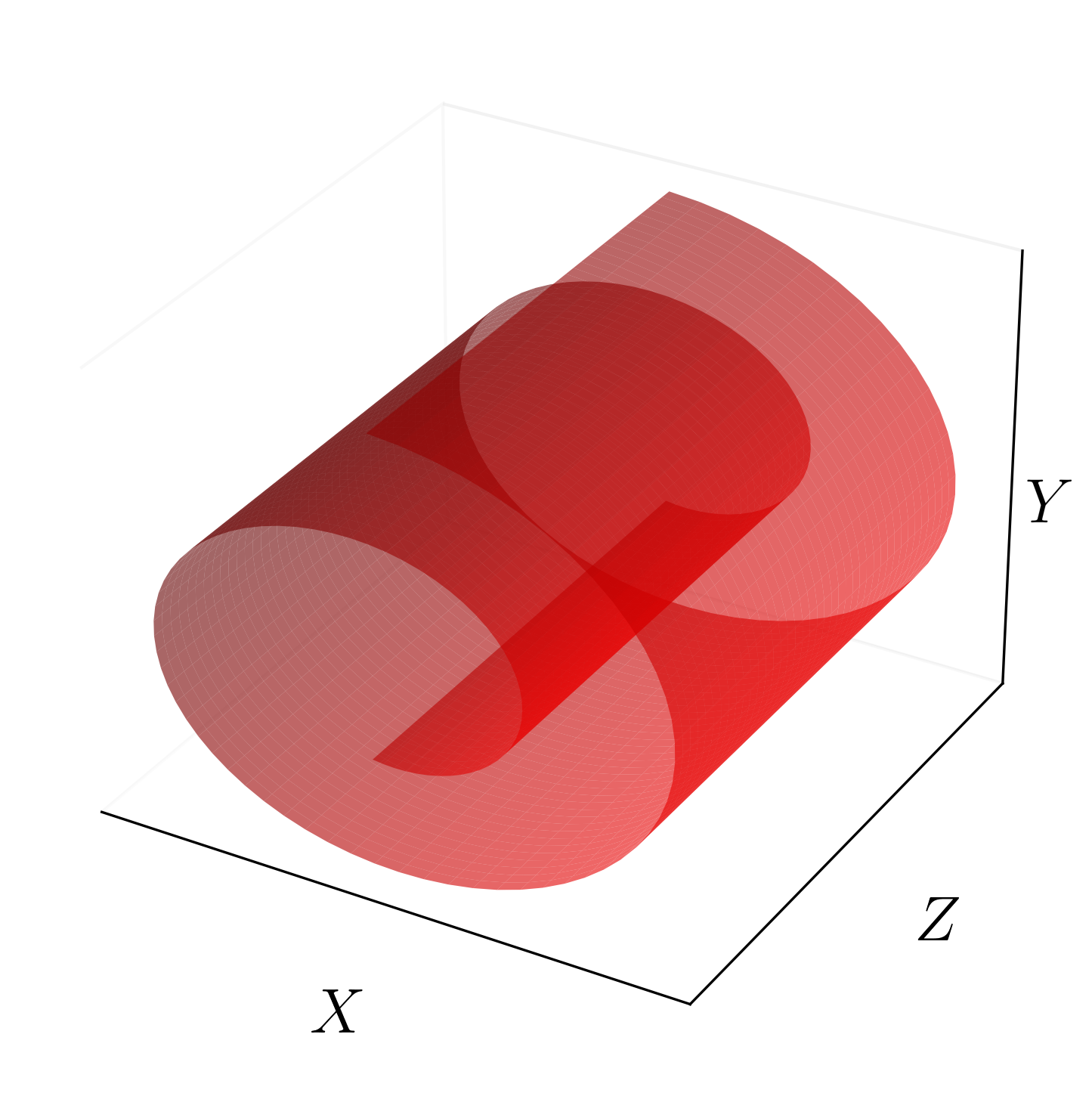}
    \caption{Sketch of a Swiss-Roll manifold.}
    \label{fig:swissrollsketch}
\end{figure}

A second comparison worth exploring is about the dependence of the dimension on the typical energy scales in the problem. The correlation dimension of~\cite{Komiske:2019fks, Komiske:2019jim,Komiske:2022vxg} was defined as
\be
d_C (r_{\rm max}) = \frac{d}{d\log r_{\rm max}}\sum_{1 \leq A \leq B \leq N} \Theta\left[{\rm EMD}(\mathcal{E}_A, \mathcal{E}_B) < r_{\rm max} \right]\, ,
\ee
where $N$ is the total number of events and $r_{\rm max}$ an energy scale. The value of $d_C$ strongly depends on the choice of $r_{\rm max}$, i.e. the largest distances (energy scales) that we include in its calculation. Instead, the \ID estimator based on the two nearest neighbors ($j=i+1=2$) is insensitive to $r_{\rm max}$. As discussed before, we can introduce a dependence on the energy scale in the \ID by considering different values of $i,j$, i.e. using farther neighbors of each point to estimate the dimension of the dataset. If we do this, we should recover the intuition discussed above, i.e. at large $i, j$ the dimension goes to 0, as if we were looking at the manifold from further and further away, while for the closest neighbors ($i,j=1,2$) we should recover the result that we would expect based on first principles, when one has a perfect knowledge of the dataset. We illustrate these points in the toy example below.

\subsection{The Swiss-Roll as a Toy Example}\label{sec:toy}

We consider here a toy example consisting of a \emph{Swiss-Roll} manifold to make all our statements more concrete and easy to visualize. The Swiss-Roll is a two-dimensional manifold embedded in a three-dimensional space.  It is parameterized by
\begin{equation}
    \vec x_\text{SR} = (t\cos t, t\sin t, z)\,,\quad t \in \left[\frac{3}{2}\pi, \frac{9}{2}\pi \right], \quad z \in \left[0, 21\right]\,.
    \label{eq:swisswrollparam}
\end{equation}
We show a sketch in \cref{fig:swissrollsketch}. The (Euclidean) distance between consecutive spiral windings in the $x-y$ plane is $2\pi$, that together with the extension along $z$ sets the typical length scales of the manifold.

We add some noise to this manifold in the form of a random vector extracted from a Normal distribution with mean zero and unit variance, rescaled by a factor $\delta$,
\be
     \vec x &=&  \vec x_\text{SR} + \delta \cdot \vec x_G  \, , \nn \\
     \vec x_G &\sim & \mathcal{N}(\vec 0,\diag(1,1,1))\,.
\ee
In this way, when $\delta = 0$ the manifold is genuinely 2-dimensional, while $\delta\ne 0$ ``fills'' the spiral and makes it 3-dimensional. Later we will choose $\delta = 0.5,1$,  corresponding to a $\approx 10\%, 20 \%$ noise with respect to the typical length scales of the spiral.

\begin{figure}[!t]
    \centering
    \includegraphics[width=0.75\linewidth]{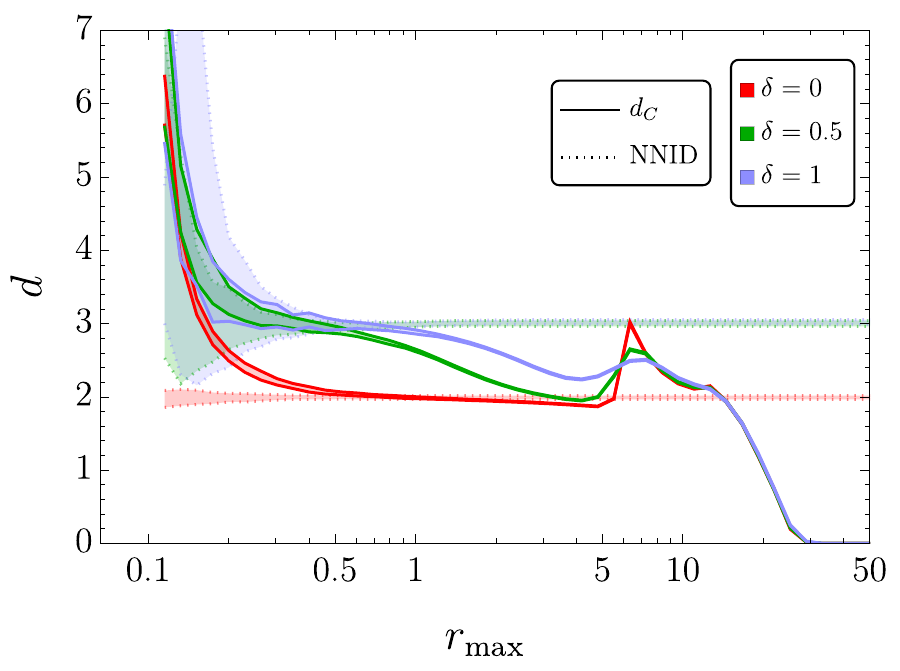}
    \caption{A comparison of the correlation dimension $d_C$ (solid lines) and the \ID (dotted lines) for a Swiss-Roll. The \ID is computed at $i = 1$ and $j=2$ and by definition is insensitive to cutting the largest distances in the dataset. The parameter $\delta$, defined in the text, quantifies the thickness of the Swiss Roll. The shaded areas give the $95\%$ C.L. interval for the two estimators.}
    \label{fig:SwissRollIDvsCorrD}
\end{figure}

We first compare the \ID for the closest nearest neighbors $\hat d_1$ to the correlation dimension by computing both under the constraint on the distances $r \leq r_\text{max}$. We employ the Euclidean distance, and to obtain confidence bands we use 100 datasets each comprised of $10^4$ points. The result, reported in Fig.~\ref{fig:SwissRollIDvsCorrD}, shows that the \ID (dotted bands) is independent of the largest distances in the dataset and always sees either a 2-dimensional manifold (when $\delta=0$) or a 3-dimensional manifold (when $\delta > 0$). The correlator dimension $d_C$ (solid bands) at large $r_{\rm max}$ sees a point-like object, then as $r_{\rm max}$ decreases it starts resolving more features of the Swiss-Roll and therefore increases. It peaks around $r_{\rm max} \simeq 6-7$, the length scale of the spiral, where it can not yet resolve the substructure of the manifold and sees it as a filled cylinder. Then, as we further lower $r_{\rm max}$, it starts resolving the spiral structure and slowly converges to the true result.

\begin{figure}[!t]
    \centering
    \includegraphics[width=0.75\linewidth]{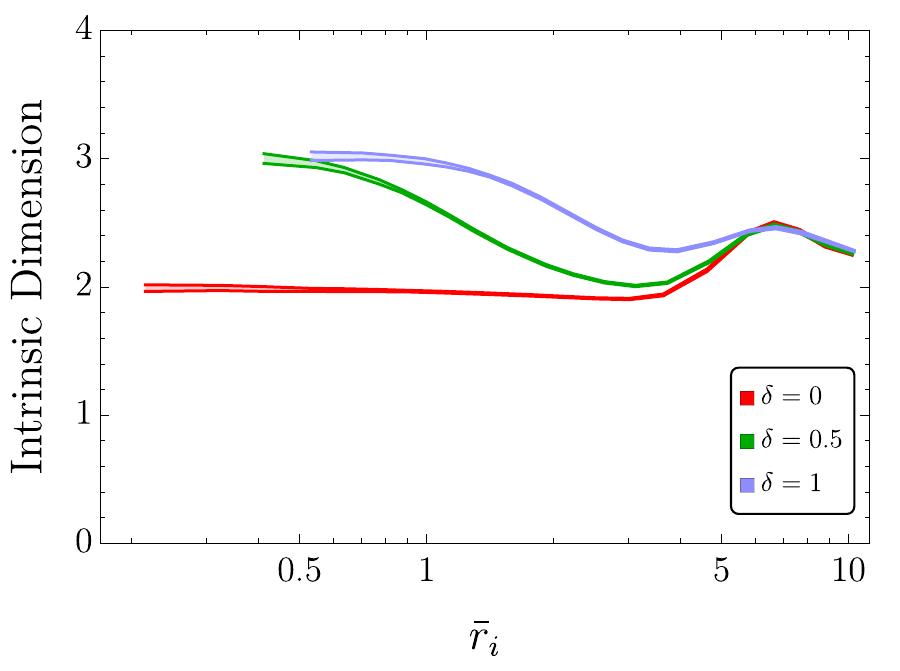}
    \caption{The intrinsic dimension of a Swiss-Roll for different choices of the noise $\delta$, i.e. its thickness. On the $x$-axis we show the average distance of the $i$-th neighbor, on the y-axis the \ID calculated at $j=2i$. The shaded areas give the $95\%$ C.L. interval for the estimator. We truncate our lines at the value of $\overline{r}_i$ below which there are no more neighbors.}
    \label{fig:IDSwissRoll}
\end{figure}

At very small values of $r_{\rm max}$ both measurements of the dimension break down, because we do not have enough points in the dataset for a reliable estimate. For example, in the case of the \ID we are starting to cut out of the dataset the nearest neighbors used for the estimate. It is interesting to notice that the \ID knows that it is failing (i.e. the error bands get larger and larger as we decrease $r_{\rm max}$), while the same is not true for the correlation dimension $d_C$.

To illustrate the dependence of $\hat d_i$ on $i$ that we introduced earlier, we plot in Fig.~\ref{fig:IDSwissRoll} the \ID estimator against the average distance between a point and the nearest neighbor used in the estimate ($\overline{r}_i$), in turn averaged over our 100 datasets. In practice, when varying $\overline{r}_i$ on the $x$ axis of the plot, we are changing the nearest-neighbor number $i$ in the \ID estimation (with fixed $j=2i$). We truncate our lines at the value of $\overline{r}_i$ below which there are no more neighbors.

We observe a dependence of the \ID on distance that resembles the behavior of $d_C$ as a function of $r_\text{max}$ in Fig.~\ref{fig:SwissRollIDvsCorrD}, and the qualitative description of the curve is the same. Interestingly, however, our estimator remains stable even when probing the smallest distances in the dataset, unlike $d_C$ which diverges in that regime. In fact, our estimator correctly converges and stops at $d_{i,2i} =2$ in the case of no noise and at $d_{i,2i} =3$ when noise is present.

\section{A Model-Independent New Physics Test}
\label{sec:NP}

The \ID of events with a fixed number of particles is well defined and can in principle be computed on a piece of paper, simply from kinematics. 
Consider, for example, a generic $2\rightarrow n$ scattering process with no on-shell mediators and in which the masses of all the particles are known, as well as the 3-momenta of the incoming ones. If we can measure the momenta of $m\leq n$ of the outgoing particles, simple kinematic considerations set $d = \min(3n-4, 3m)$. 

Suppose, however, that some of the final momenta are correlated; for example, because two of the $m$ particles in the final state come from a new intermediate on-shell particle $X$ that decays promptly. This reduces the dimension of the support manifold by one, since the momenta of the decay products satisfy $(p_1+p_2)^2 = m_X^2$. Hence, the measured \ID in the data will be lower than the naïve SM expectation, potentially revealing the presence of new physics.

In reality, things are much more complicated. Even if we just consider the simple example above, without systematics or detector effects, our new particle $X$ has some finite width $\Gamma_X$, and the kinematical constraint $(p_1+p_2)^2 \simeq m_X^2$ does not eliminate one dimension, it just shrinks it to a size of order $\Gamma_X$. If we know $\Gamma_X$ in advance we can choose the appropriate $i,j$ in the estimator $d_{i,j}$ so that we are insensitive to energies smaller than $\Gamma_X$. However, if we want to design a model-independent search strategy, we won't know $\Gamma_X$ in advance, or even the nature of the kinematical constraints in the Beyond the Standard Model theory which is hiding in the data. 

Additionally, every measurement has an associated experimental error. This source of noise introduces fictitious dimensions that make the empirical estimate of the \ID deviate from the kinematical one and introduce new energy scales in the problem. Furthermore, hadron collider events coming from the same high-energy scattering process do not have a fixed number of particles due to pile-up, particles from the underlying event, ISR, FSR, and the stochastic nature of the quantum mechanical processes in jet showers. Even if we could isolate two quarks with exactly the same momentum in two different events, the two reconstructed jets would not contain the same number of particles.

In practice the best strategy to find new physics does not seem to analytically predict the \ID of a given class of events or its dependence on $i,j$ and compare it with the data. We should rather leverage the differences in \ID between different processes as a function of energy scale (or in practice of $i/N,j/N$) to detect new physics.

To do it we first construct a reference sample of events that describe our null hypothesis (i.e. the SM). This can be done in several ways, depending on our final goal:
\begin{enumerate}
    \item We take a data sample and split it into subsamples that were collected by the detector at different times. One is the reference and by testing its compatibility with the others we can perform data quality monitoring tasks.
    \item We exploit the approximate symmetries of the SM, as proposed in~\cite{Bressler:2024wzc}. For example we take a data sample that contains at least one high-quality electron and compare it with another sample containing at least one high-quality muon, after correcting for detector effects. 
    \item We generate the reference sample by simulating the SM using Monte Carlo (MC) methods.
\end{enumerate}
The first strategy is straightforward to implement, but its applicability to new physics searches is limited. The second strategy is purely data-driven and can be used to compare complex datasets without clustering jets. It does not allow to probe every possible new physics signal, but it is still quite flexible, as one can repeat the exercise with all sorts of different datasets (two electrons vs two muons, three electrons vs three muons, and so on). Additionally, we remind the reader that we are not doing a precision search in the sense of high energy physics, but rather looking for signals with small $S/B$, but potentially large significances, hiding in the large dimensionality of the space, so the systematic error associated with replacing a muon with an electron is a small effect for us. 

The last strategy is the most flexible, i.e. the one which is sensitive to the largest number of new physics signals, but requires a reliable MC simulation. We do not view this as a real drawback, for two reasons. On the one hand MC simulation has progressed a lot since the early days of the LHC, to the point that it is routinely used in precision measurements~\cite{CMS:2024lrd, ATLAS:2024erm}. Secondly, we do not need to simulate the SM with high accuracy, as the dimension of the dataset is a robust quantity, as we demonstrate in Section~\ref{sec:applications}. Its sensitivity to new physics degrades only slightly going from parton-level to detector-level events, it is not affected by momentum scale uncertainties and only mildly by momentum resolution effects, even if at detector level we use hundreds of particles to compute the \ID as opposed to the few partons coming out of the hard collision. 

\subsection{Statistical Test} 
\label{sec:statTest}

Equipped with the information contained in the \ID estimators $\hat d_i$ and in the neighbor distances $\overline{r}_i$, one can perform model-independent searches for new physics. Here we consider the simplest possible implementation of such a test, which has the virtue of being easy to interpret and computationally very efficient in the calculation of significances, since it exploits analytic results in the asymptotic limit. In the previous Sections we have described at length the \ID and its properties, but in the test we include also information on the average distance between neighbors $\overline{r}_i$. Its properties have already been studied in~\cite{Komiske:2019fks}. For our purposes it is sufficient to recall that since $\overline{r}_i$ is an average we also expect it to follow a Gaussian distribution asymptotically.

Our test compares \ID and $\overline{r}_i$, as a function of $i/N$, between collider data and a reference hypothesis (the SM). We consider a single collider dataset, while the reference ensemble is split into smaller subsamples and used to estimate the mean and covariance under the null hypothesis of the asymptotic Gaussian distribution of \ID and $\overline{r}_i$. The test then quantifies the compatibility of the experimental data with the reference distribution. We leave to future work the study of more elaborate approaches that can further optimize this procedure, for instance by feeding the information on $\hat d_i$, $\overline{r}_i$, or related quantities into machine-learning algorithms.

For concreteness, let us focus on the NNID. We denote by $\{\bm{d}_k\}_{k=1}^n$ a reference set of $m$-dimensional \ID vectors computed from $n$ datasets, where $n>m$, sampled from the same reference distribution under identical conditions, and by $\bm{d}'$ an $m$-dimensional vector computed from the experimental data, whose compatibility with the reference set we want to test. The $m$ entries of the vectors are populated by varying $i/N$ in the \ID computation.

In Section~\ref{sec:ID} we assumed the probability density of points to be locally constant and then derived from this assumption a series of properties for the \ID estimator. This was useful to set the stage for the new physics search presented here. However it is important to keep in mind that even when this initial assumption breaks down, our new physics test can still be performed as described in this Section.

Even if $\rho$ is not constant across the dataset, the random vector $\bm{d}$ can still be distributed as a multivariate normal random variable with mean $\bm{\mu}$ and covariance matrix $\bm{\Sigma}$, $\bm{d} \sim \mathcal{N}(\bm{\mu}, \bm{\Sigma})$. Call  $\bm{\hat \mu}, \bm{\hat{\Sigma}}$ the parameters empirically estimated employing the $n$ datasets associated to the reference distribution, and call $\hat T^2$ the would-be $\chi^2$,
\begin{align}
    \hat T^2 = \frac{n}{n+1}( \bm{d}' - \bm{ \hat \mu}) ^T \bm{\hat{\Sigma}}  \inv ( \bm{d}' - \bm{ \hat \mu}).
    \label{eq:T2emp}
\end{align}
Under the null hypothesis, namely $\bm{d'} \sim \mathcal{N}(\bm{\mu}, \bm{\Sigma})$, the variable $ F \equiv T^2 (n-m)/(m(n-1))$ follows an $F$-distribution with $(m,n-m)$ degrees of freedom, $F \sim F_{m,n-m}$ \cite{anderson2003introduction}.\footnote{Formally, $\hat T^2$ follows a \emph{Hotelling} distribution \cite{Hotelling1931TheGO}, $\hat T^2\sim T^2_{n-1,m}$.} This can be regarded as a generalization of the familiar Student's $t$-distribution. Note that in the limit $n\rightarrow \infty$, $T_{n,m}^2 \rightarrow \chi^2_m$.

A $p$-value for the compatibility between the reference model and the new data can be computed simply as
\begin{align}
p=P \left( F \geq \hat F  \right) \qquad F\sim F_{m,n-m}.
\label{eq:s2:pval}
\end{align}
and then converted in the more familiar significance $Z = \Phi^{-1}(1-p)$.

This procedure can be seen as a model-independent test of the agreement between the reference distribution and the experimental data. 
The biggest numerical burden is the computation of the distance matrix between events, which for a two-dimensional optimal transport problem, such as the EMD computation, is of $\mathcal O(N^2 D^3)$ for a dataset with $N$ points and $D$ features.

In Section~\ref{sec:applications} we also consider for the test the averages of the distances $\overline{r}_i$, which by the Central Limit Theorem are approximately normally distributed. In practice, this amounts to repeating the same procedure above but for the $\{\bm{ \overline{r}}_k\}_{k=1}^n$ vectors, where $\bm{\overline{r}}$ denotes the vector of the various $\overline{r}_i$ for a given dataset.  We choose to carry out the two tests, on \ID and $\overline{r}_i$, separately, to discuss their different physical properties, but to maximizes performances one should concatenate the two vectors, for instance as outlined in \cref{app:QQ}.

We stress that the reliability of the resulting $p$-value depends on the validity of the assumptions underlying the test; most crucially, that \cref{eq:T2emp} follows an $F_{m,n-m}$ distribution under the null hypothesis. In practice, this approximation can break down when the \ID vector components become highly correlated, or when their number $m$ approaches the number of reference samples $n$. In these regimes, the empirical covariance matrix $\bm{\hat \Sigma}$ acquires small eigenvalues, making its inversion numerically unstable and causing the test statistic to fluctuate more than expected from the $F$-distribution.

A standard approach to mitigate this problem is \emph{Principal Component Analysis} (PCA). We project the \ID vectors onto the subspace spanned by the first $M$ principal components, i.e. the directions with largest eigenvalues, and discard components associated with small eigenvalues that are most prone to numerical noise. The optimal number of retained components $M$ is determined by performing a Cramér–von Mises (CvM) test on the empirical distribution of $T^2$ obtained via a leave-one-out procedure under the null hypothesis: we choose the largest $M$ for which the CvM statistic indicates acceptable agreement with the theoretical $F$ distribution. As an additional qualitative cross-check, we inspect quantile–quantile (Q–Q) plots of the resulting $T^2$ values. Full details of this procedure are given in \cref{app:QQ}, where we show its application to the examples of \cref{sec:applications}.

In the next Section we perform this novel new physics test on a series of benchmark models and verify its main properties: 
\begin{enumerate}
    \item {\bf Flexibility}: Sensitivity to diverse Beyond the Standard Model scenarios with different kinematics.
    \item {\bf Robustness} (to showering and detector effects): The performances do not change appreciably when going from the parton-level to the detector-level simulation. 
    \item {\bf Robustness to Systematics}: Systematic errors can be incorporated in a statistically well-defined way and the test is insensitive to energy scale errors.
    \item {\bf Protection from the Curse of Dimensionality}: This feature is already present by construction in our test, but we demonstrate it explicitly in two examples, going up to $\sim 100$ dimensions. 
\end{enumerate}

\section{Flexibility, Robustness, Systematics and No Curse of Dimensionality}
\label{sec:applications}
In this Section we show the performances of our new physics search strategy on six benchmark models. The results are a first proof-of-principle test of our methodology and therefore we only consider the main  background in each search. In Section~\ref{sec:summary}, we summarize the significance of these results for each of the four properties listed at the end of the previous Section (Flexibility, Robustness to Detector Effects, Robustness to Systematics, No Curse of Dimensionality).

To test the performances of our methodology in a controlled setting, we apply our new physics test following a procedure that is not optimal as far as finding new physics goes, but it is the most transparent for testing. More concretely, we start by choosing a final state, for example two muons and one electron or a $W$ and a $Z$ boson. Then we identify the irreducible SM backgrounds and design a set of cuts to eliminate all other backgrounds. Only then we select a signal that can produce the chosen final state. The cuts are designed on the SM backgrounds, assuming no knowledge of the signal. Our cuts are chosen to simplify the analysis, i.e. to select a single SM background process. For this reason they often penalize our signal. An example of this is our main benchmark (called MSSM-3 in what follows), that leads to very boosted particles in the final state. The efficiency of our cuts on MSSM-3 is not optimal, as they were designed on $WZ$ production in the SM. Additionally, the cuts are designed to be exclusive, since their goal is to reduce the list of backgrounds relevant to the analysis, but the real purpose of our techniques is to find unexpected new signals and it should be applied on final states that are as inclusive as possible. We leave this to future work in collaboration with LHC experiments. Our choice simplifies the task of interpreting and validating the \ID and $\overline{r}_i$ curves and develop some physics intuition on them.

\subsection{Leptonic Decays of Charginos and Neutralinos}
\label{sec:MSSM_lep}

As a first example, we focus on a purely leptonic final state arising from the production of new weakly coupled particles, namely a set of charginos and neutralinos. The model consists of an electrically charged fermion $\chi_1 ^{\pm}$ and two neutral fermions $\chi^0_2, \chi^0_1$, with $R$-parity conserving couplings to the SM. We take $\chi^0_1$, to be the lightest fermion in the spectrum, so $R$-parity makes $\chi^0_1$ stable. 

\begin{figure}
\centering
\begin{tikzpicture}
\begin{feynman}
    \vertex (blob) at (0,0) {};

    \vertex (chi20_v)  at (2.5,  1.0);   
    \vertex (chi1pm_v) at (2.5, -1.0);   

    \vertex (Z_v)      at (4.5,  1.6);
    \vertex (chi10_u)  at (4.5,  0.4);
    \vertex (muplus)   at (6.2,  2.0) {\(\mu^+\)};
    \vertex (muminus)  at (6.2,  1.2) {\(\mu^-\)};

    \vertex (W_v)      at (4.5, -1.6);
    \vertex (chi10_d)  at (4.5, -0.4);
    \vertex (eplus)    at (6.2, -1.0) {\(e^+\)};
    \vertex (nue)      at (6.2, -2.2) {\(\nu_e\)};

    \diagram*{
        (blob) -- [plain] (chi20_v),
        (blob) -- [plain] (chi1pm_v),

        (chi20_v) -- [boson, edge label=\(Z\)] (Z_v),
        (chi20_v) -- [plain] (chi10_u),

        (Z_v) -- [anti fermion] (muplus),
        (Z_v) -- [fermion]      (muminus),

        (chi1pm_v) -- [boson, edge label'=\(W^+\)] (W_v),
        (chi1pm_v) -- [plain] (chi10_d),

        (W_v) -- [anti fermion] (eplus),
        (W_v) -- [fermion]      (nue),
    };
\end{feynman}


\draw (-2, 0.92) -- (-0.4, 0.05);
\draw (-2, 1) -- (-0.3, 0.13 - 0.54375*0.1);
\draw (-2, 1.08) -- (-0.3, 0.21 - 0.54375*0.1);

\draw (-2, -0.92) -- (-0.4, -0.05);
\draw (-2, -1) -- (-0.3, -0.13 + 0.54375*0.1);
\draw (-2, -1.08) -- (-0.3, -0.21 + 0.54375*0.1);

\node[left] at (-2.3,  1.0) {\(p\)};
\node[left] at (-2.3, -1.0) {\(p\)};


\filldraw[fill=gray!30,draw=black] (0,0) circle (0.4);


\node[above] at (1.4,  0.85) {\(\tilde\chi^0_2\)};
\node[below] at (1.4, -0.85) {\(\tilde\chi^\pm_1\)};

\node[right] at (chi10_u) {\(\tilde\chi^0_1\)};
\node[right] at (chi10_d) {\(\tilde\chi^0_1\)};

\end{tikzpicture}
\caption{Leading diagram for the leptonic MSSM signal in Section~\ref{sec:MSSM_lep}.}
\label{fig:mssm_sketch}
\end{figure}
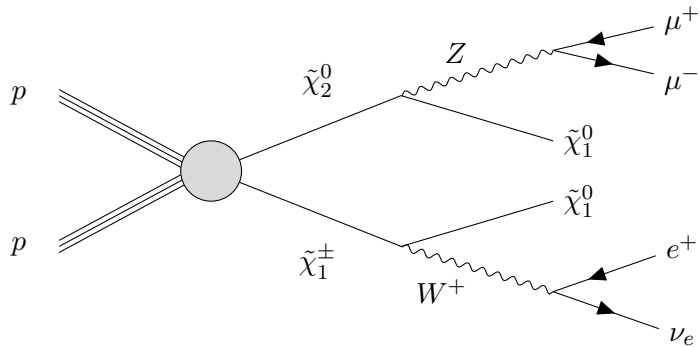

We analyze the process $ p p \rightarrow \chi_1 ^{+} \chi^0_2$, with on-shell $\chi^{+}_1$ and $\chi^0_2$ decaying as $\chi_1 ^{+} \rightarrow W^{+} \chi^0_1$ and $\chi^0_2 \rightarrow Z \chi^0_1$, respectively. The final state we consider contains a pair of $\chi^0_1$'s, a positron-neutrino pair from $W^{+} \rightarrow e^+\nu_e  $, and a muon-antimuon pair from $Z\rightarrow \mu^+ \mu^-$; the full diagram is depicted in \cref{fig:mssm_sketch}. The leading SM background consists of $p p \rightarrow W^+ Z$ events, with the same $W^+$ and $Z$ decays as above.

This is a MSSM-inspired simplified model, in which, however, we let the total cross section vary freely to study the behavior of the \ID with respect to the number of BSM events. Therefore the detailed admixture of Wino, Higgsino and Bino that makes up our particles is not relevant to the search.

\paragraph{Simulation and Event Selection.} 
We simulate the process with $\sqrt{s}=13.6$ TeV center of mass energy and choose a reference luminosity of 200 fb$^{-1}$. The simulation of the signal and background is carried out at parton level using {\small \textsc{MadGraph5\_aMC@NLO}} \cite{Alwall:2014hca} with the $\texttt{MSSM\_SLHA2}$ model \cite{Allanach:2008qq}. Hadronization and showering are performed with {\small \textsc{Pythia 8}} \cite{Sjostrand:2014zea}, and detector effects are simulated with {\small \textsc{Delphes}} \cite{deFavereau:2013fsa}.

The data is split into several samples (see Results paragraph below), and for each the \ID is obtained as described in \cref{sec:ID}. Specifically, we first compute the EMDs among events and then obtain the \ID by minimizing \cref{eq:logL}. The EMDs are computed with the Python package \texttt{EnergyFlow}~\cite{Komiske:2017aww, Komiske:2018cqr, Komiske:2019fks, Komiske:2019jim, Komiske:2019asc, Andreassen:2019cjw, Komiske:2020qhg}, using $\beta = 1$ and $R = 3$, and taking as input the $(p_T, \eta, \phi)$ of each particle (all treated as massless).

Since our new physics test will be applied at both parton and detector level to assess the difference in performance, we report here the details of the event simulation, reconstruction, and selection in both cases: 
\begin{enumerate}
    \item[$i)$] \emph{Parton level:} For each event, the muon, antimuon, and positron from the hard collision simulated in {\small \textsc{MadGraph}} are directly used in the computation of the NNID. The leptons are required to satisfy $p_T > 10$ GeV and $|\eta|<2.5$. Furthermore, we select isolated leptons by requiring the total $p_T$ of other particles within a cone of radius $R=0.3$ around each lepton to be smaller than 0.12 of the lepton's $p_T$ for electrons and 0.25 for muons. After all cuts we find an overall efficiency $\epsilon \simeq 0.44$ for the SM. In terms of the number of events, taking the results of \cite{CMS:2021icx, CMS:2024ild} for the total cross section and $W^{\pm}$ polarization, and the $W^+,Z$ branching ratios from \cite{ParticleDataGroup:2024cfk}, about 12,000 SM events are expected;
    \item[$ii)$] \emph{Detector level:} We employ the default CMS card \cite{CMScard} in {\small \textsc{Delphes}}, with the only modification being an adjusted lepton isolation requirement. Specifically, we apply the same isolation criteria as at parton level. Jets are clustered with the anti-$k_T$ algorithm with radius $R=0.5$.  Events selected for the analysis must contain exactly a muon, an antimuon, and a positron with $p_T>10$ GeV and $|\eta|<2.5$. Additional jets (from initial and final state radiation) are allowed if $p_{T,j}<40$ GeV. For the EMD computation only the three leptons are used. These criteria lead to approximately $4800$ selected SM events; the efficiency relative to the parton level selection is about $43\%$ for the SM.
\end{enumerate}

\paragraph{Results.} To get some intuition on the physics of the $\overline{r}_i$ and \ID search strategy, we first inspect the \ID curve as a function of $i/N$ and $\overline{r}_i$ in pure SM and BSM datasets at parton level. 
For the BSM model, here and in the following we fix $m_{\chi^0_1} = 150$ GeV and $m_{\chi^0_2} = m_{\chi_1 ^+}$, with $m_{\chi^0_2, \chi_1 ^+} = 250, 500, 1000$ GeV, which we refer to as MSSM-1,2,3, respectively.

\begin{figure}[!t]
\begin{center}
    \includegraphics[width=0.49\textwidth]{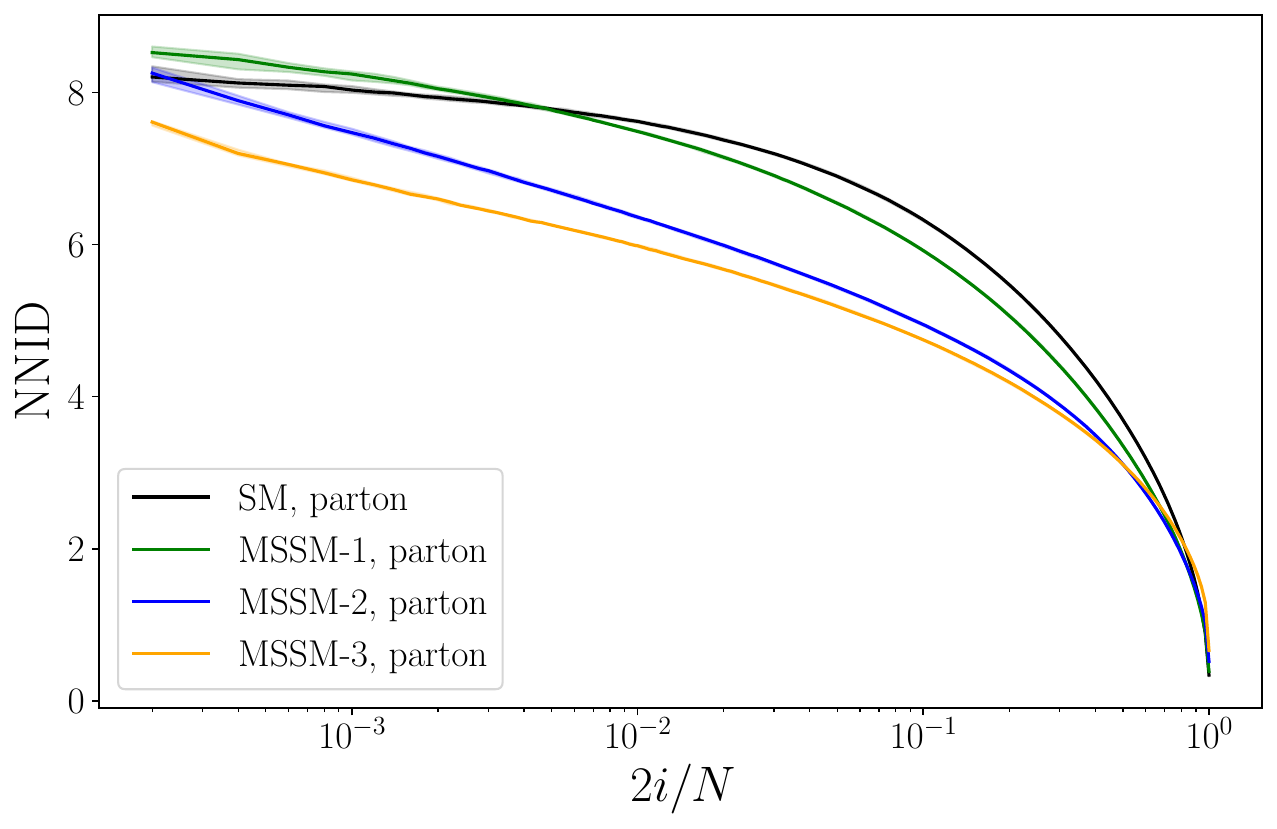}
    \hfill
    \includegraphics[width=0.49\textwidth]{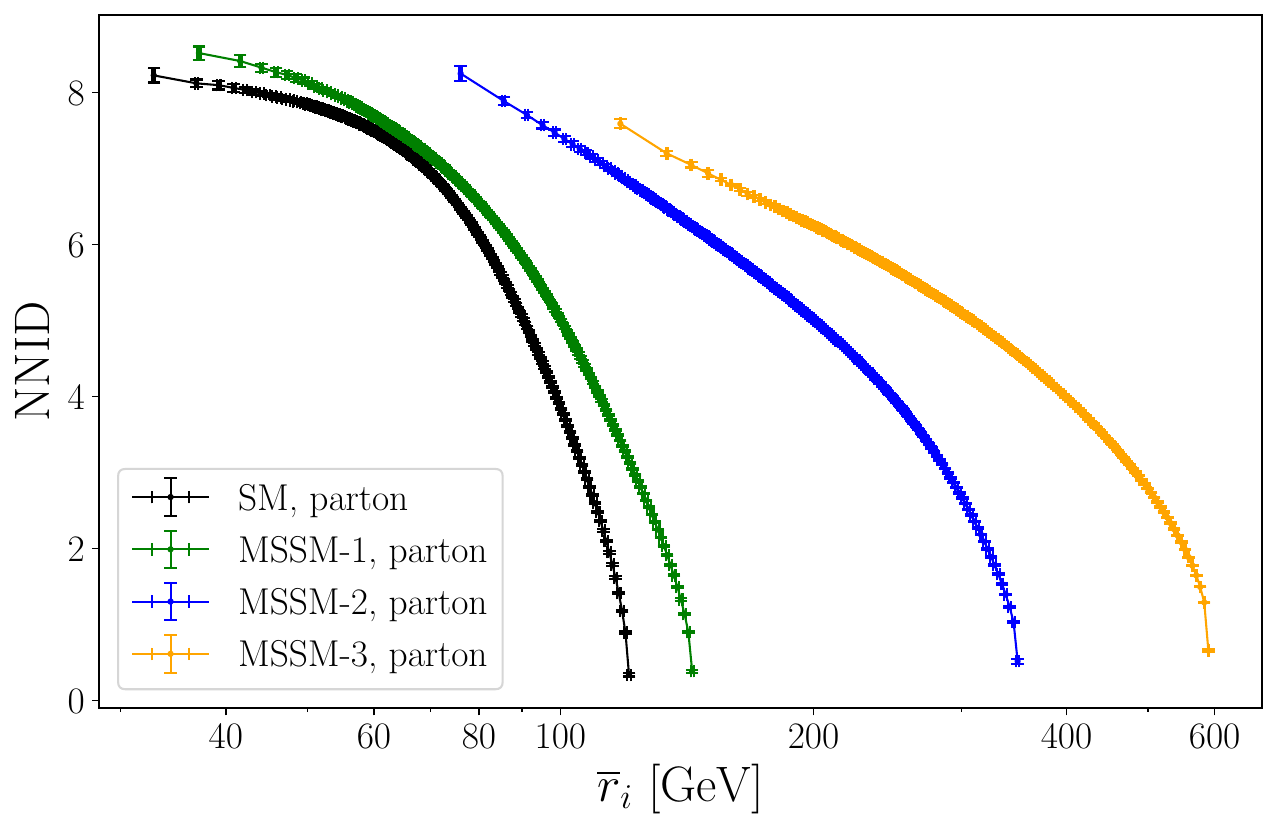}
    \hfill
    
    \end{center}
\caption{Parton level \ID curves as a function of $2i/N$ (left panel) and the average distance to the $i$-th neighbor (right panel) 
in the SM (black) and in three MSSM-like benchmark models described in the text. The curves are obtained from 100 samples. In the left panel the central line represents the median, while the band is the $68\%$ confidence interval. In the right panel we show the $68\%$ confidence interval for both axes as error bars.}
\label{fig:MSSM_IDcurves}
\end{figure}

The \ID curve is computed taking $j=2i$ and $m=200$ points logarithmically spaced between $i=1$ and $i=\lfloor N/2\rfloor $, with $N=10^4$ points.\footnote{At this stage we are only interested in the qualitative behavior of the \ID and use a large number of events for illustrative purposes.}
This operation is repeated for 100 signal and background samples, to obtain confidence level intervals for the NNID. We also compute the average distances $\overline{r}_i$ for each $i$ and plot the \ID against it. 
The results are given in Fig.~\ref{fig:MSSM_IDcurves}.

In the SM we expect the \ID estimate to asymptote to either 8 or 9 for small values of $i/N$, depending on whether we can resolve the $Z$ boson width or not. The theoretical counting of the independent degrees of freedom that justifies our expectation goes as follows. 
We are only including the two muons and the positron in the calculation of the NNID, and we treat them as massless, so we have at our disposal 9 variables.
The $2$ kinematical constraints from momentum conservation in the plane transverse to the collision do not hold because we are not including the neutrino in the calculation. The only additional kinematical constraint that we can impose comes from the $Z$ being on-shell. The $W$ does not count because again we are not including the neutrino in our calculation. 

The result in Fig.~\ref{fig:MSSM_IDcurves} (left) respects perfectly this estimate, falling between 8 and 9, but closer to 8, meaning that we can barely resolve the $Z$ width with the number of events at our disposal. 

For the MSSM models, we also ignore missing energy and only consider particles to which we can associate a three-component momentum. With this assumption the counting proceeds in exactly the same manner as in the SM. Hence, also here we expect a saturation of the \ID between $8$ an $9$. 
In Fig.~\ref{fig:MSSM_IDcurves}, we show that this counting is approximately respected for the MSSM-1 benchmark. The MSSM-2 and MSSM-3 benchmarks give a very different result; the latter, in particular, shows the biggest deviation compared to our theoretical counting. The muons and positron-neutrino pairs arising from the $Z$ and $W$ decays are very boosted, since $m_{\chi^0_2} \gg m_{\chi^0_1}, m_{Z,W}$. This leads to a strong collimation of their momenta, which may render it difficult for the estimator to disentangle them, thereby reducing the measured dimension of the manifold. This is evident from the right panel of \cref{fig:MSSM_IDcurves}, in which we can appreciate the boost in the form of much bigger typical distances $\overline{r}_i$ for the MSSM-2,3 cases.

Overall, the \ID shows a remarkable sensitivity to small kinematical changes in the models. For example MSSM-1 and the SM are quite close kinematically, but have different \ID curves. Interestingly, as the proportion $i/N$ increases, the difference persists or increases. This qualitative behavior shows that we can use the \ID as a test for new physics, at least for these specific models.

Fig.~\ref{fig:MSSM_IDcurves} also highlights the topological nature of the \ID (i.e. its insensitivity to absolute scales). The distances $\overline{r}_i$ are sensitive to the typical energy scale of the events and show clearly that MSSM-2 and MSSM-3 live at much higher scales compared to the SM. This is only partially reflected in the \ID which registers the collapse of some of the dimensions due to the MSSM events' boost, but is also loosing some of the kinematical information contained in the $\overline{r}_i$, as one can see by comparing the two panels of the Figure. We see this both as an issue and a positive feature. It is an issue because of the loss of information, but it is a feature because the \ID is completely orthogonal to traditional strategies that target high-energy tails or bumps. Additionally the topological nature of the \ID makes it much less sensitive to a series of effects that we describe below (for example errors on $p_T$ scale and resolution, extra input dimensions identical for signal and background). 

To solve the issue we enlarge our new physics test to include also the {\it average} distance $\overline{r}_i$ between neighbors, as discussed in the previous Section. We choose the average over the median because of asymptotic normality and to give a bigger weight to the small number of highly energetic new physics events as opposed to the large number of low-energy events in the SM. We build a vector of $\overline{r}_i$'s and treat it as the vector of \ID's in the test. We also performed a combined test concatenating the two vectors, but here we only show the results of the two separate tests to better understand their strengths and weaknesses.

To proceed with the new physics test outlined in \cref{sec:statTest}, we simulate $200$ SM \ID curves, now with 20 logarithmically spaced $i/N$ points, to speed up computations. We collect in \cref{app:QQ} the comparison of our test statistic under the null hypothesis with the theoretical $T^2$ distribution. 

Each SM sample has a variable number of events extracted from a Poisson distribution whose mean $\mu_\text{SM}$ is itself extracted from a Gaussian distribution with variance $ \delta \mu_{\rm SM}$. Here $\delta \mu_{\rm SM}$ accounts for the systematic error on the the total number of SM events, given by the sum in quadrature of the error on the luminosity~\cite{CMS:2021xjt} and that on the cross section of our main background~\cite{CMS:2021icx}. The error on the cross section $\delta \sigma_{WZ}/\sigma_{WZ} \simeq 4\%$ dominates over the error on the luminosity $\delta \mathcal{L}/\mathcal{L}\simeq 1.6\%$ and the Poisson statistical error, that for our choice of luminosity is around 1\%. 

Similarly, we compute 50 SM+BSM curves, with the number of SM events determined as above. 
On top of the SM events, we add $\text{Poisson}(\mu_{\text{BSM}})$ BSM events whose 
``nominal'' rate is fixed by the total cross-section ratio 
$\sigma_{\text{BSM}}/\sigma_{\text{SM}}$. 
Note that in the new physics test the effective rate used is the nominal one rescaled according to the SM and BSM efficiencies; in particular,  the efficiency of our parton-level cuts on these three benchmarks is $\simeq 0.7, 0.82, 0.73$ for MSSM-1,2,3, and roughly 33\% pass also the detector-level cuts for MSSM-3. This means that, for example, the effective $S/B$ at parton level for MSSM-3 is about 1.7 times larger than the nominal rate.
We then compare the \ID and $\overline{r}_i$ curves using the statistical test described in Section~\ref{sec:NP}, including the selection of the best number of points using PCA (see \cref{app:QQ}). This set of SM+BSM samples allows us to obtain confidence level intervals for the significance of our statistical test.

\begin{figure}[!t]
\centering
\includegraphics[width=0.49\textwidth]{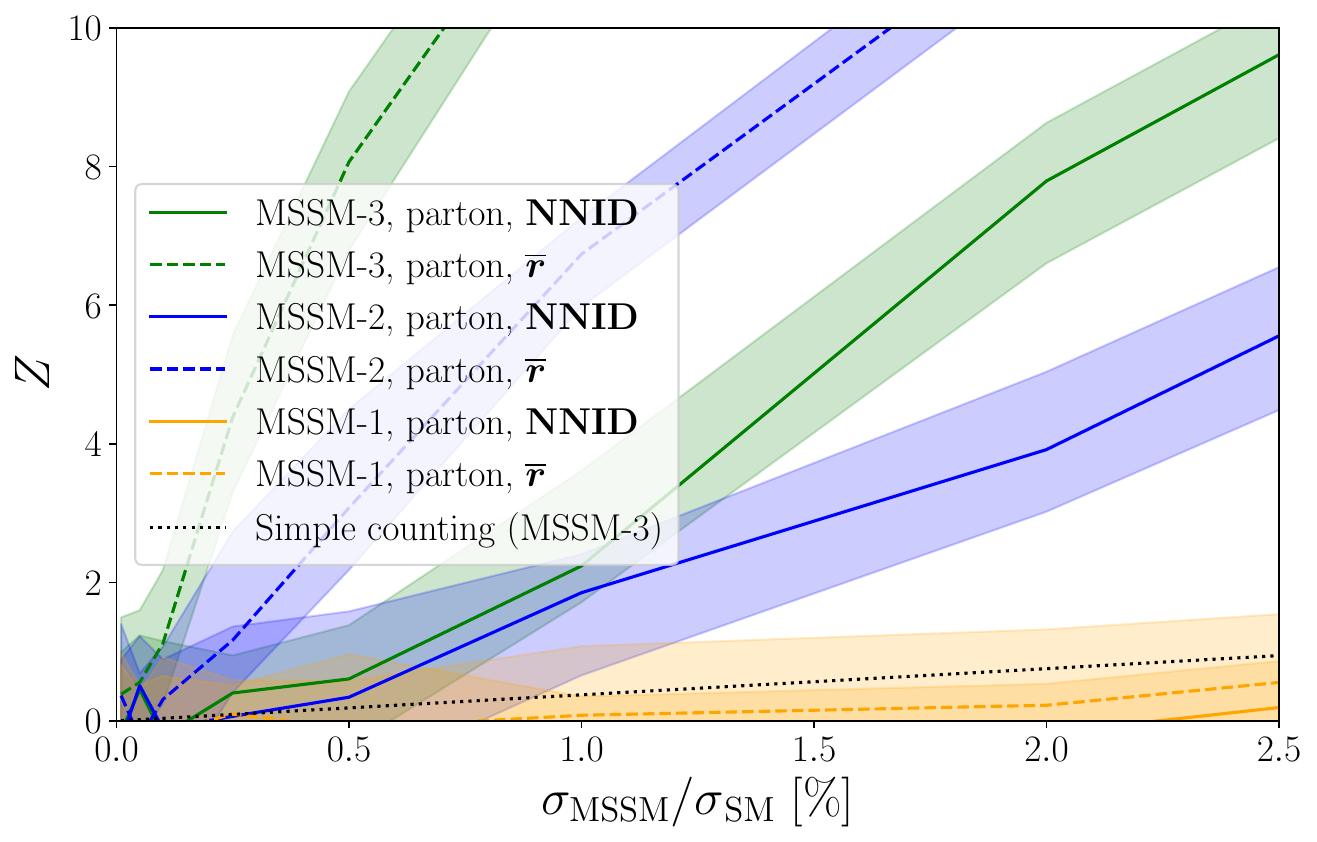}
\includegraphics[width=0.49\textwidth]{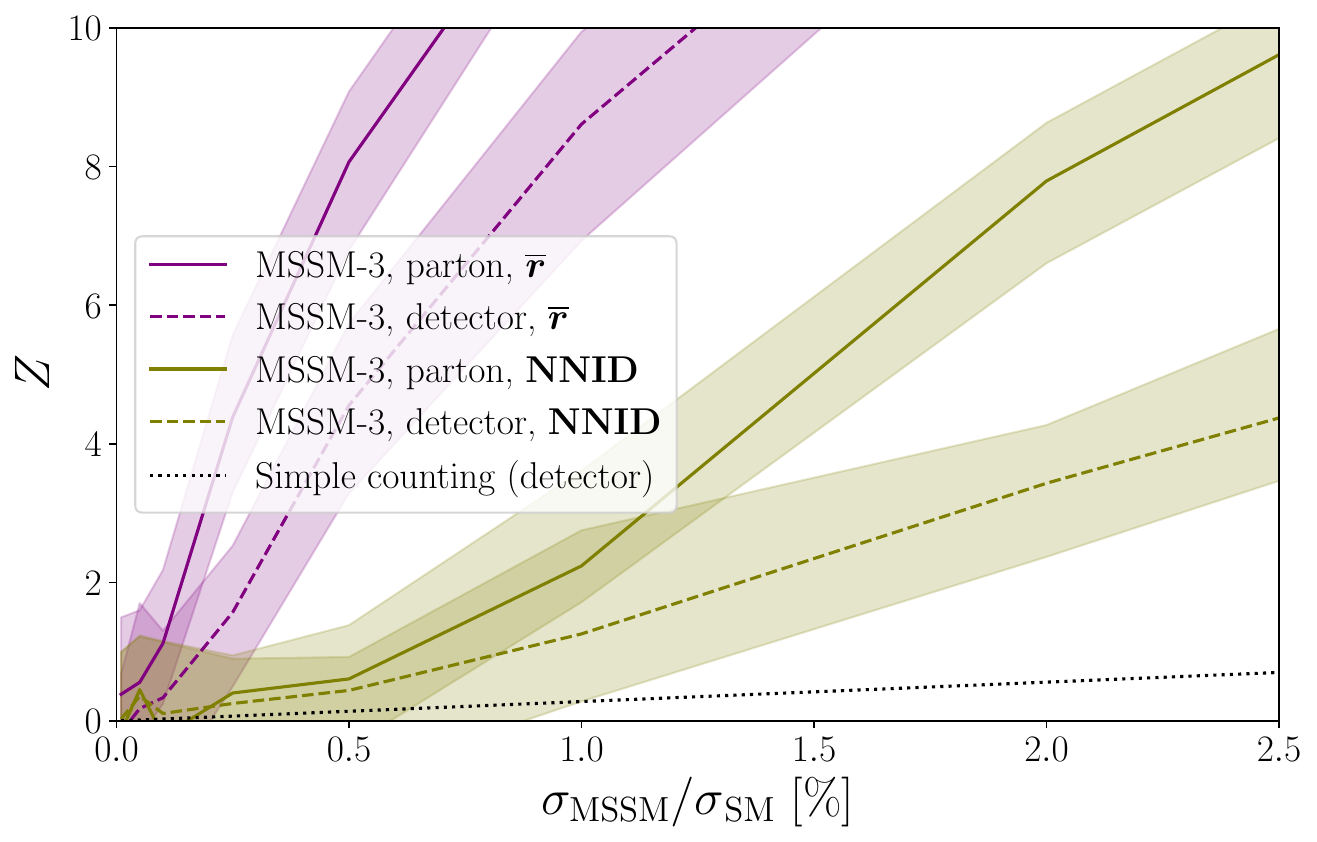}
\caption{Left Panel: Parton level Significance of the \ID and $\overline{r}_i$ new physics search versus the fraction of BSM events in the sample, for the three MSSM leptonic benchmarks described in Section~\ref{sec:MSSM_lep}. The solid and dashed lines are the median significances and are shown with the corresponding $68 \%$ CL band. The black dotted line corresponds to the simple counting significance, $Z \sim S/\delta B$. Right Panel: Comparison of the parton level and detector level significances for the MSSM-3 benchmark model.}
\label{fig:MSSM_IDvProp}
\end{figure}

In Fig.~\ref{fig:MSSM_IDvProp} we show the result of the new physics test on the benchmark models that we have just described. Both of our signal-agnostic tests outperform a simple counting experiment by orders of magnitude for MSSM-3, while they cannot meaningfully distinguish MSSM-1 from the SM. In this setup, the test based on the $\overline{r}_i$ variables achieves significantly better sensitivity than the one using the intrinsic dimension: the strong boost of the MSSM events amplifies the $\overline{r}_i$ values, leading to a much clearer separation between SM and MSSM. However this is not always the case as we are going to see in the next Section for a different type of signal.

Detector effects reduce the performance of the tests, as shown in the right panel of Fig.~\ref{fig:MSSM_IDvProp}.
This reduction is mainly due to the lower event selection efficiency at detector level, which effectively halves the available events in the reference samples and consequently increases the statistical fluctuations of the points along the \ID and $\overline{r}_i$ curves. As expected from general significance tests, this loss of statistics weakens the overall discriminating power.

For comparison, the dashed line in the figures shows the result of the ``trivial'' counting test, in which the $p$-value is obtained solely from the total number of events $N$ and the corresponding significance is computed as $Z=\mu_\text{BSM}/\sqrt{ \mu_\text{SM}+\delta \mu_\text{SM}^2}$, where $\delta \mu _{\rm SM}$ takes into account the uncertainty in the SM cross section and luminosity knowledge mentioned in the previous paragraphs.


\paragraph{No Curse of Dimensionality.}
Our methodology is inherently free from a look-elsewhere effect that grows exponentially with the number of dimensions. This is because we apply to each dataset a fixed analysis strategy, independent of its phase-space dimensionality. Consequently, we expect the performance of our method to remain stable when adding “spectator” particles that are distributed identically in the SM and in the new physics signal.

\begin{figure}[!t]
    \centering
    \includegraphics[width=0.49\textwidth]{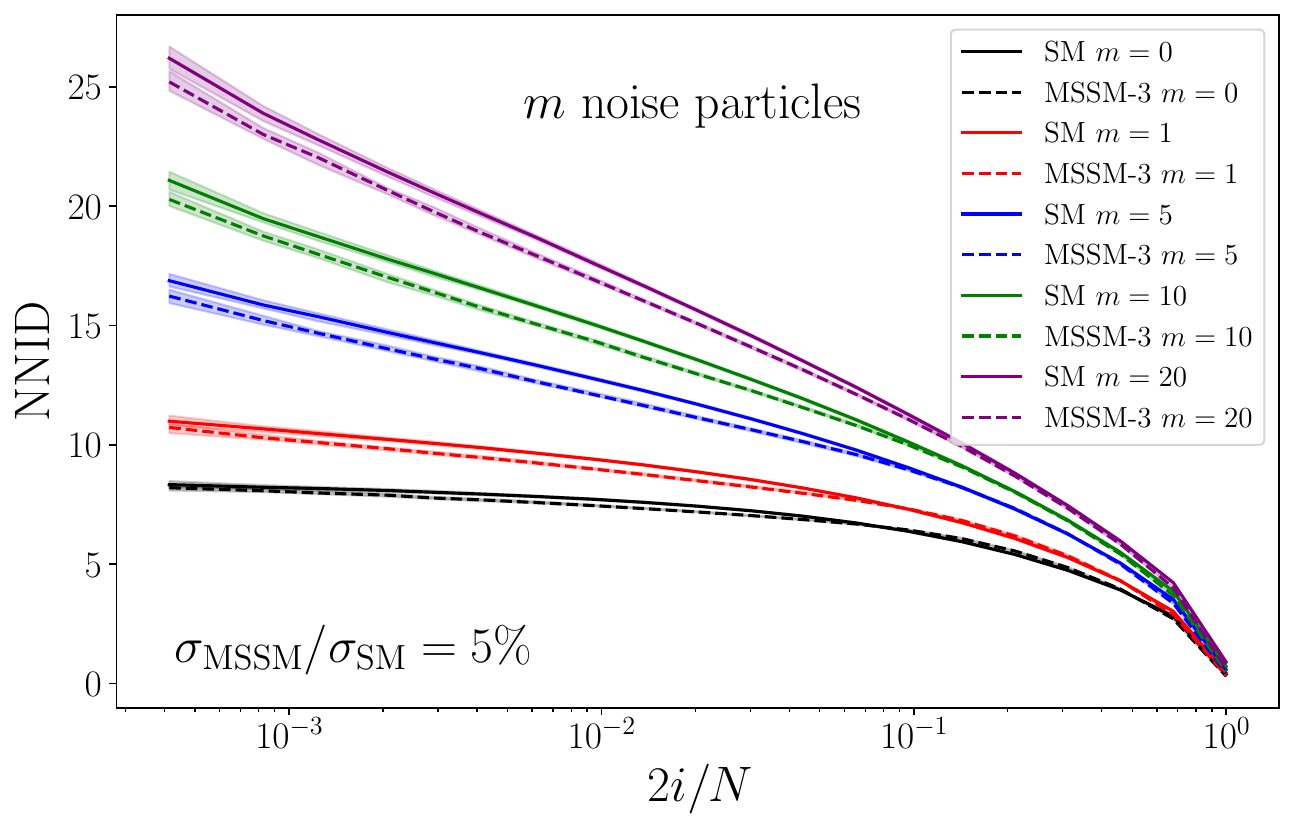}
    \includegraphics[width=0.49\textwidth]{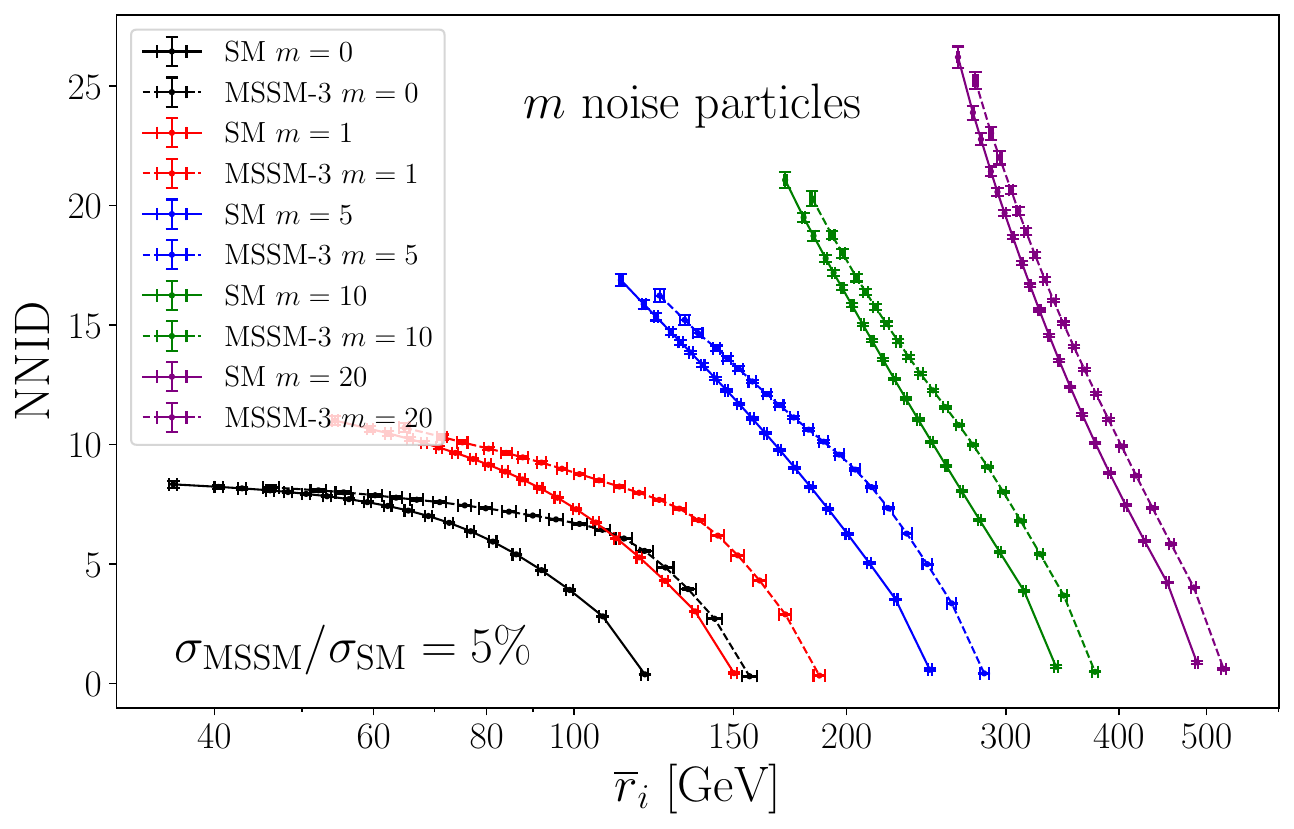}
    \includegraphics[width=0.49\textwidth]{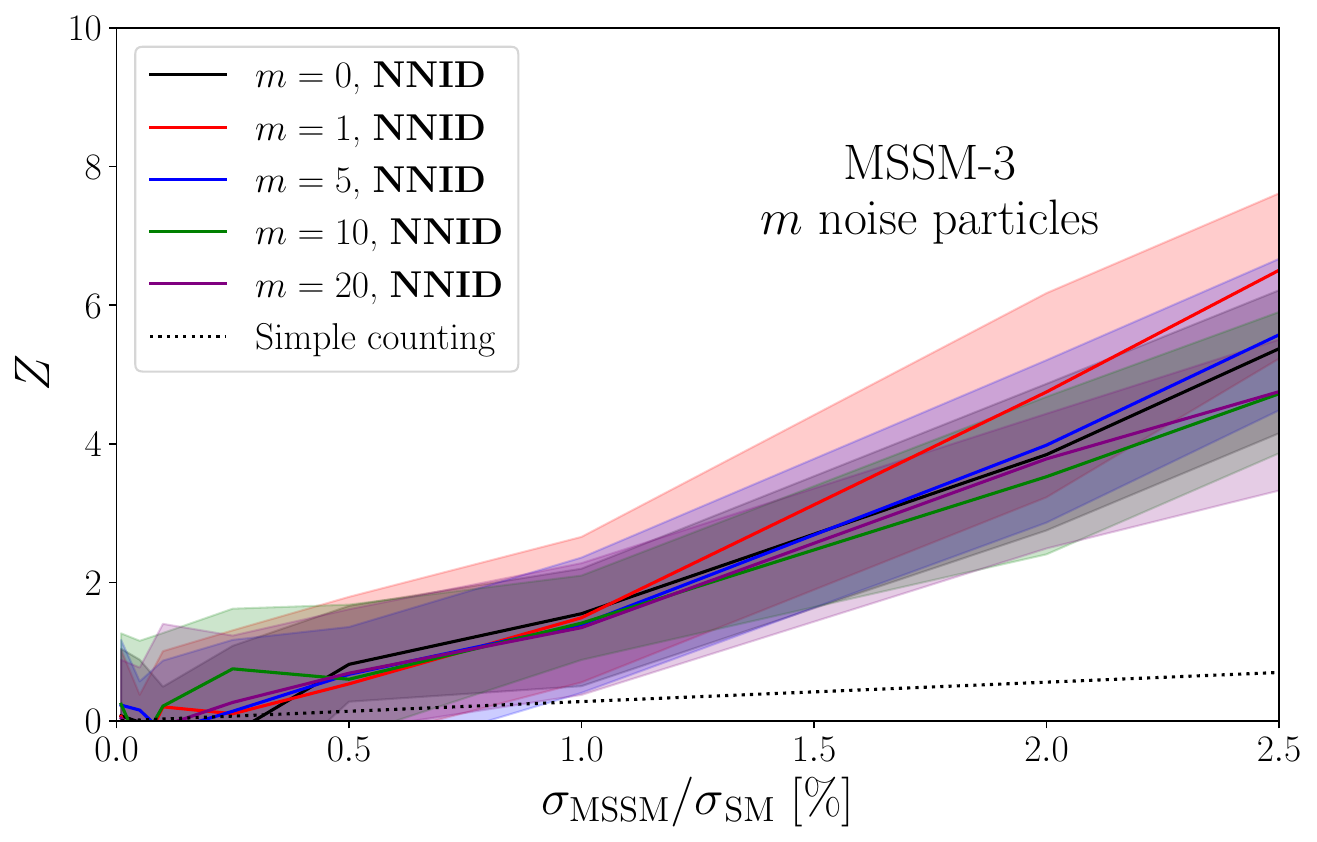}
    \includegraphics[width=0.49\textwidth]{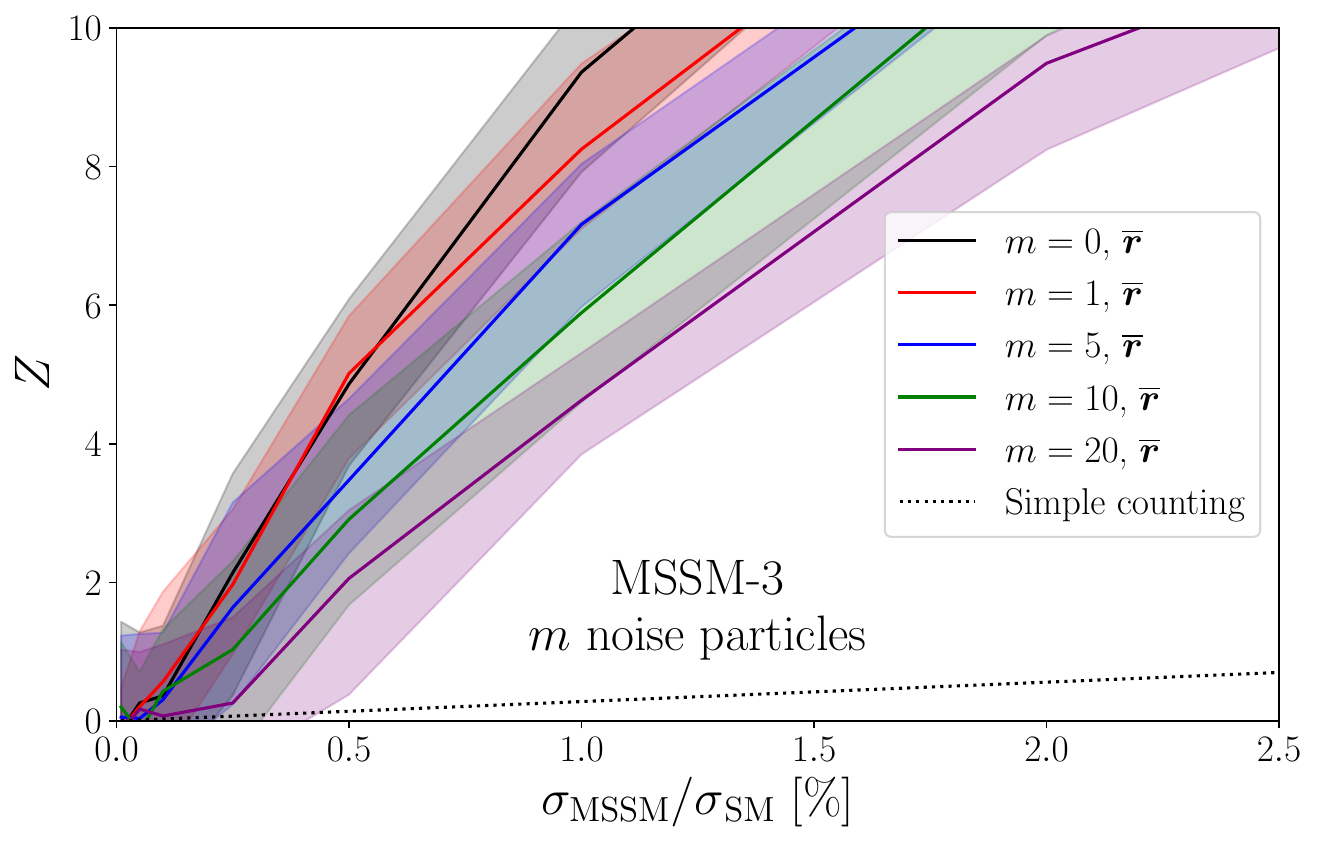}
\caption{Top row: \ID curves at detector level in the SM and MSSM-3 benchmark models described in Section~\ref{sec:MSSM_lep} with a variable number $m$ of spectator particles (i.e. fictitious particles that are identical in the MSSM-3 and SM samples). Bottom row: the significance of the \ID new physics test for the MSSM-3 benchmark model remains stable if the number $m$ of spectator particles is increased, while the $\overline{r}_i$ test suffers more.} 
\label{fig:MSSM_IDvRateNoise_det}
\end{figure}

We demonstrate this in Fig.~\ref{fig:MSSM_IDvRateNoise_det}, where we extend the MSSM-3 benchmark by including up to 60 additional dimensions corresponding to spectator particles. For each such particle, the values of $(p_T, \eta, \phi)$ are randomly and independently sampled from the distributions of SM muons. The results show that the performance of the \ID test on detector-level events remains unchanged when up to 15 spectator dimensions are added. By contrast, for other model-independent searches the $p$-value would degrade exponentially with dimensionality~\cite{DAgnolo:2019vbw}.

A performance decrease becomes visible only when adding 30 ($m=10$) or 60 ($m=20$) extra dimensions, though still at the level of about one sigma. This behavior is expected: in high dimensions, the distance between nearest neighbors grows with $d$ (the growth scaling as $\sqrt{d}$ for an Euclidean distance, for example), and once the typical inter-event distance exceeds the smallest intrinsic scale of the dataset, the search can no longer resolve it. For example, in comparing SM and MSSM-3 events, if $d$ is increased sufficiently, even the two SM muons may appear overlapping, effectively reducing the number of informative dimensions in the \ID curves below six. At that point, distinguishing the more boosted MSSM-3 events becomes impossible unless the total number of events $N$ is increased.

In contrast, the test based on $\overline{r}_i$ suffers a much stronger degradation when spectator particles are added. The addition of new dimensions distorts the $\overline{r}_i$ distribution, and since $\overline{r}_i$ enters the test directly, rather than through a topological quantity as the NNID, these distortions immediately wash out the distinctive MSSM contribution. As a result, it becomes significantly harder to identify a new effect hidden in the background.

\paragraph{Robustness to Systematics.}
As a first assessment of the robustness of our method against systematic errors, we concentrate on scale and resolution uncertainties in the charged particles' $p_T$. Once again, we take as a benchmark the detector-level MSSM-3 model.

Any global rescaling of $p_{T}$ uniformly rescales the distances $\overline{r}_i$, while leaving the \ID completely unchanged. Indeed, from the definition of the EMD in \cref{eq:EMD}, it follows directly that all distances scale linearly under dilations, as expected from their scaling dimension
\be
p_T \to \lambda p_T \quad \Rightarrow \quad r_{ki} \to \lambda r_{ki} .
\label{eq:pt_rescaling}
\ee

\begin{figure}[!t]
\centering
    \includegraphics[width=0.49\textwidth]{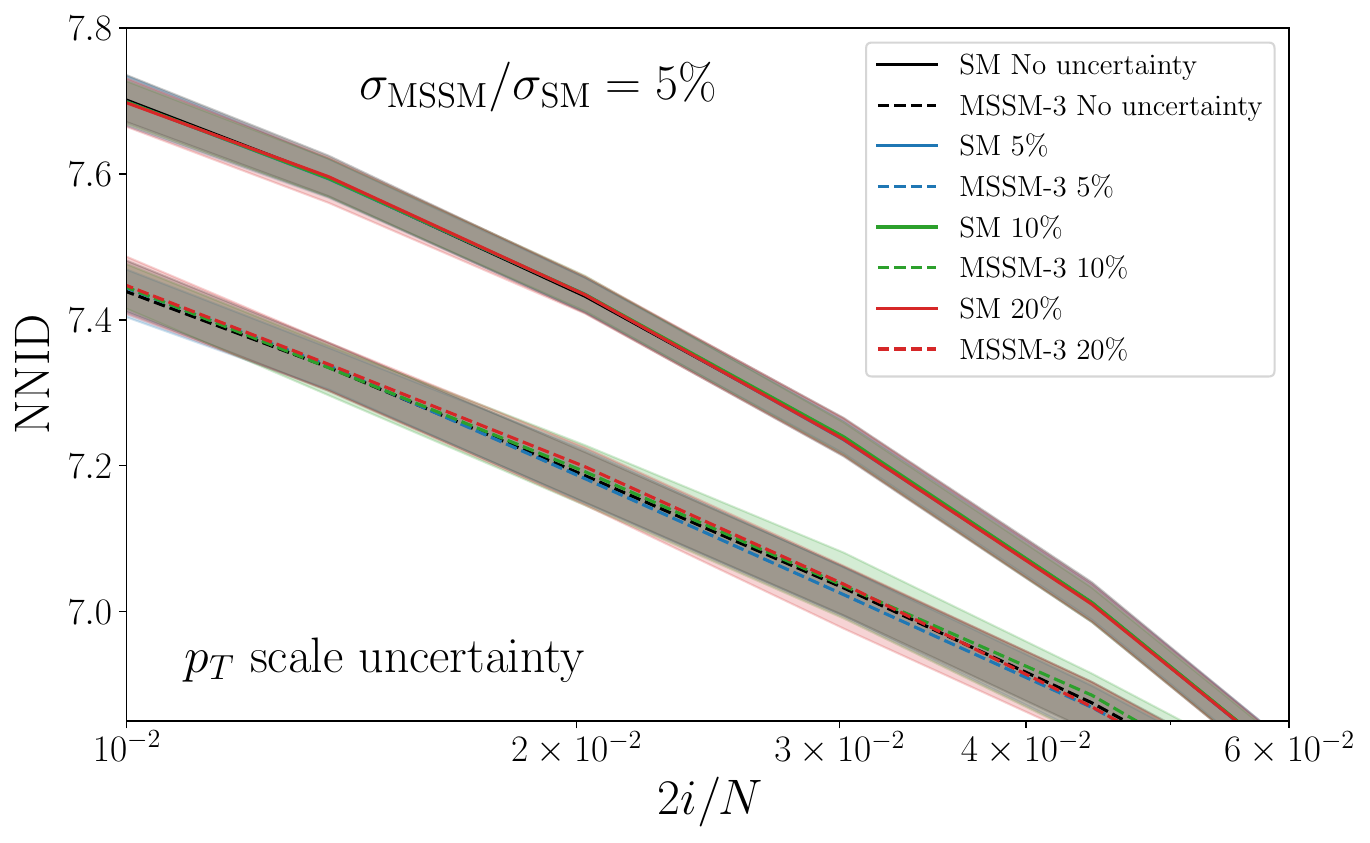}
    \hfill
    \includegraphics[width=0.49\textwidth]{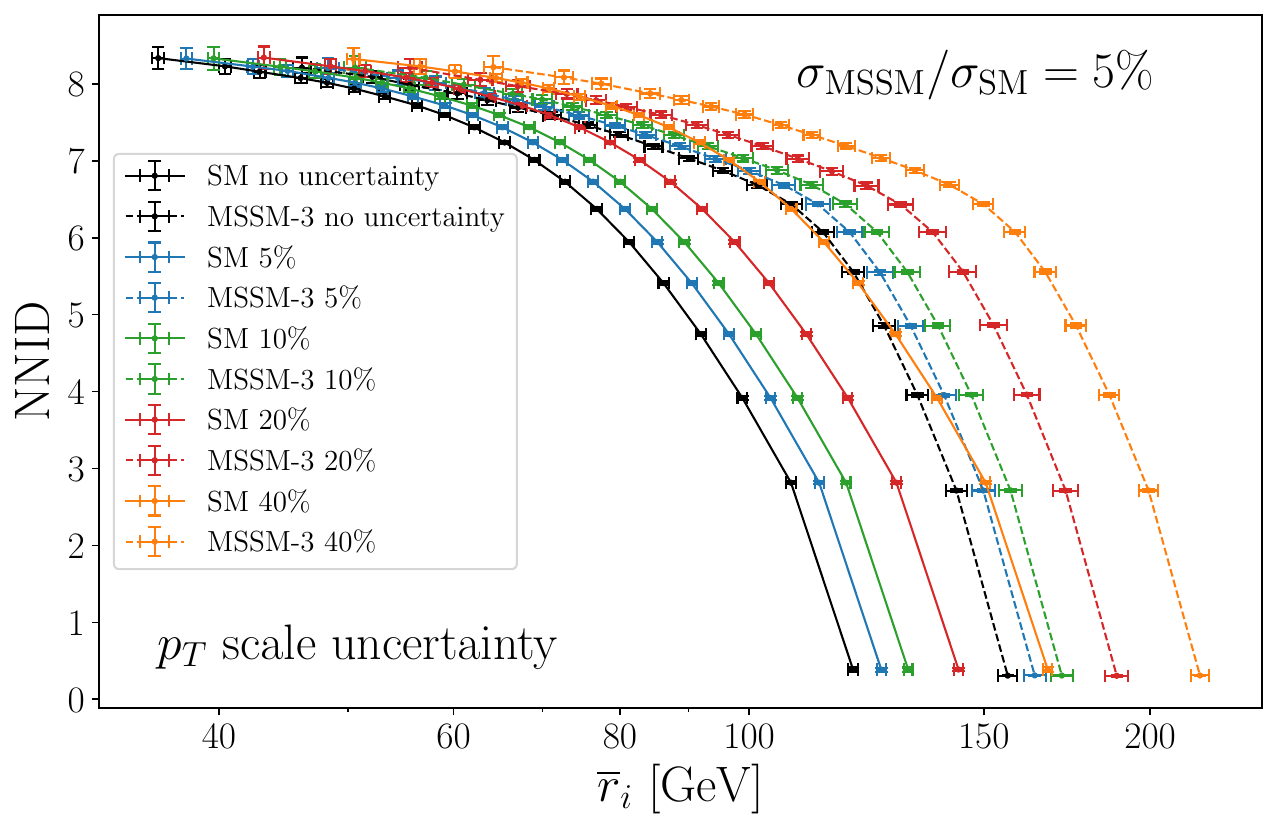}
    \includegraphics[width=0.49\textwidth]{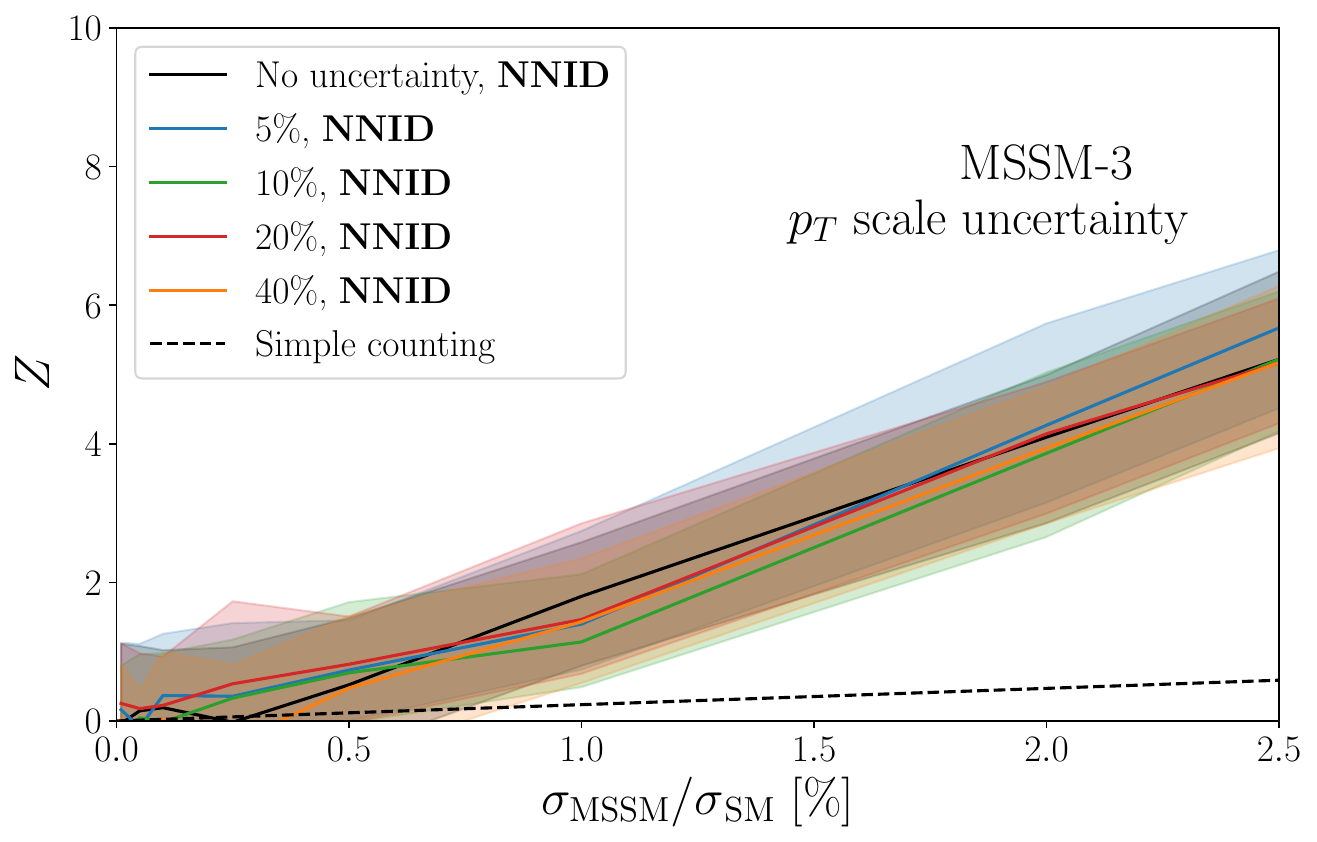}
    \hfill
    \includegraphics[width=0.49\textwidth]{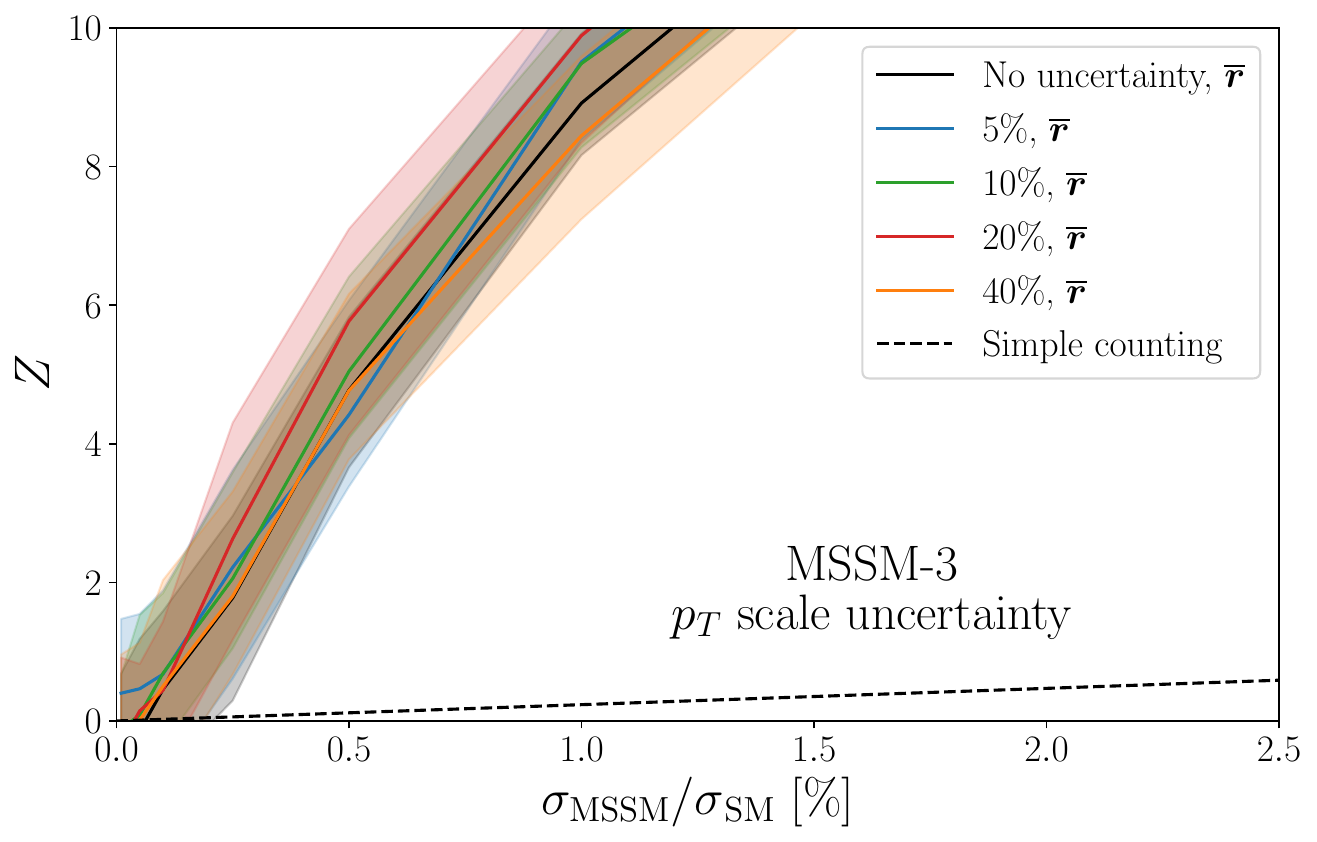}
\caption{A constant, uniform rescaling of all particles' $p_T$ leaves the \ID invariant, and thus the significance remains unchanged (left panles). The distances $\overline{r}_i$ get uniformly rescaled, but the significance is still unaltered (right panels). We rescale the $p_T$ by up to $40\%$ in the Figure, but we could have further increased this factor with no visible effect. The Figure demonstrates the insensitivity of our new physics search to energy scale uncertainties.}
\label{fig:MSSM_pt_det}
\end{figure}

Since the \ID depends only on ratios of these distances, it is exactly invariant under $p_T$ rescalings. We demonstrate this in \cref{fig:MSSM_pt_det}, where we apply a constant rescaling that increases all particles' $p_T$ values, mimicking the procedure used to estimate detector energy-scale uncertainties. We set $\lambda=(1+\alpha)$ in Eq.~\eqref{eq:pt_rescaling} with $\alpha=\{5\%, 10\%, 20\%, 40\%\}$. Both the \ID curve (zoomed in here to highlight the 68\% confidence band) and the corresponding significance remain unchanged compared to the unrescaled case. The same conclusion holds, of course, for rescalings that decrease $p_T$.
For the $\overline{r}_i$, while the rescaling seems to alter the distribution of distances, the uniform transformation in \cref{eq:pt_rescaling} ensures that the test statistic in \cref{eq:T2emp} is invariant. Consequently, the significance is unaffected, as is evident from the curves shown in \cref{fig:MSSM_pt_det}.

The impact of the finite $p_{T}$ resolution of the detector on the value of the \ID is suppressed by the sample size  ($N\approx5000$ in the Figure). While the CMS {\small \textsc{delphes}} card that we used already incorporates a few-percent $p_T$ smearing for leptons, we deliberately amplify this effect by applying an additional $5\%, 10\%, 20\%$ or $40\%$ smearing to the selected final state events. 

\begin{figure}[!t]
\centering
    \includegraphics[width=0.49\textwidth]{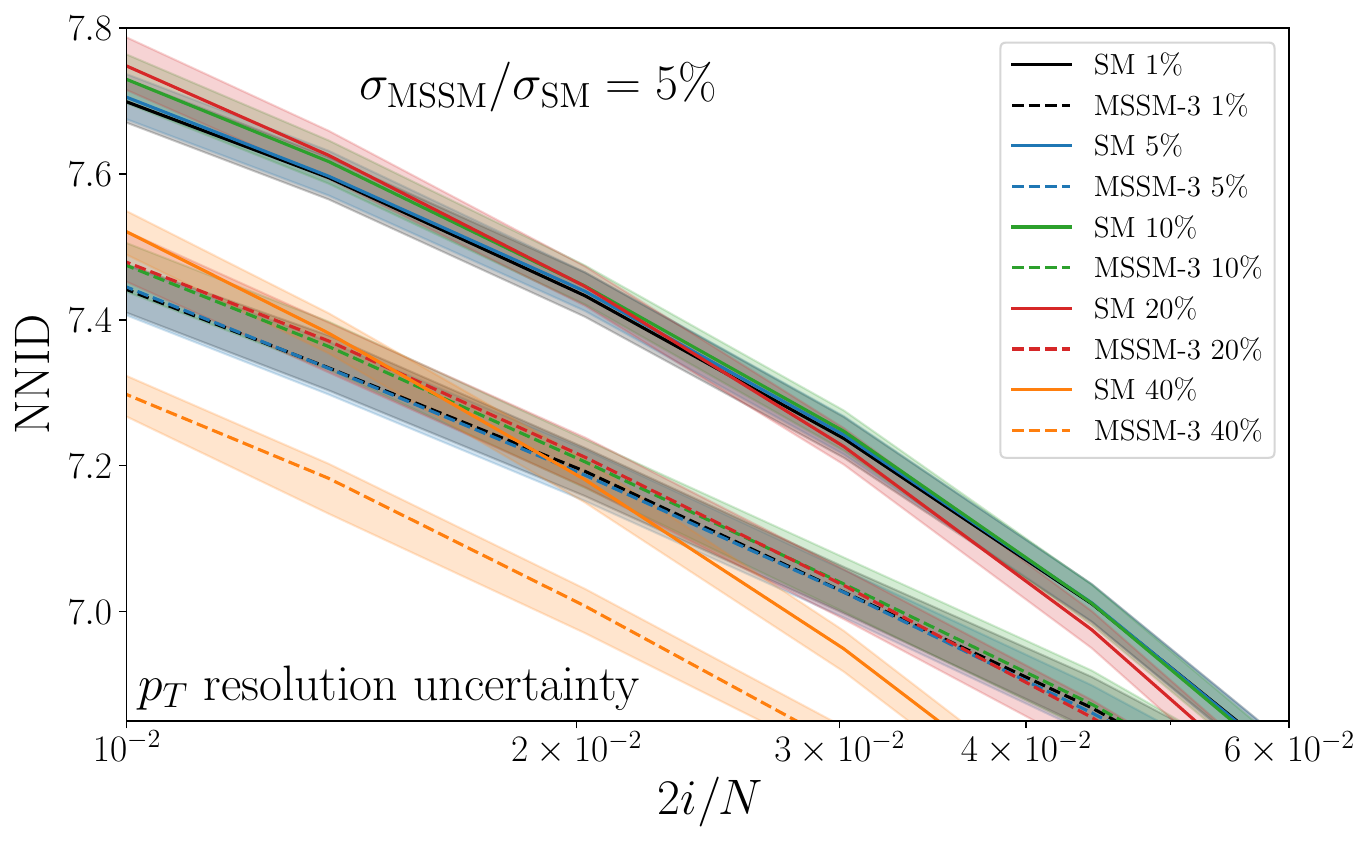}
    \hfill
    \includegraphics[width=0.49\textwidth]{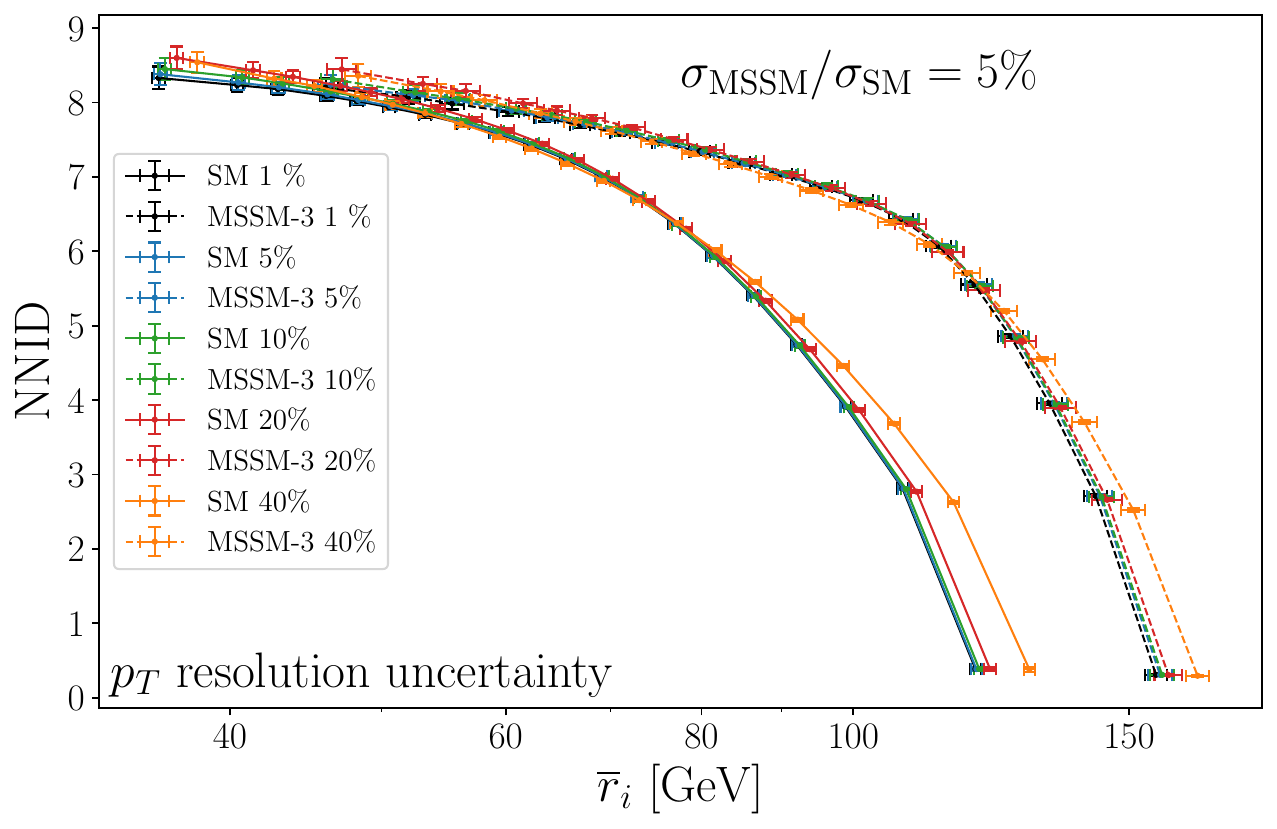}
    \includegraphics[width=0.49\textwidth]{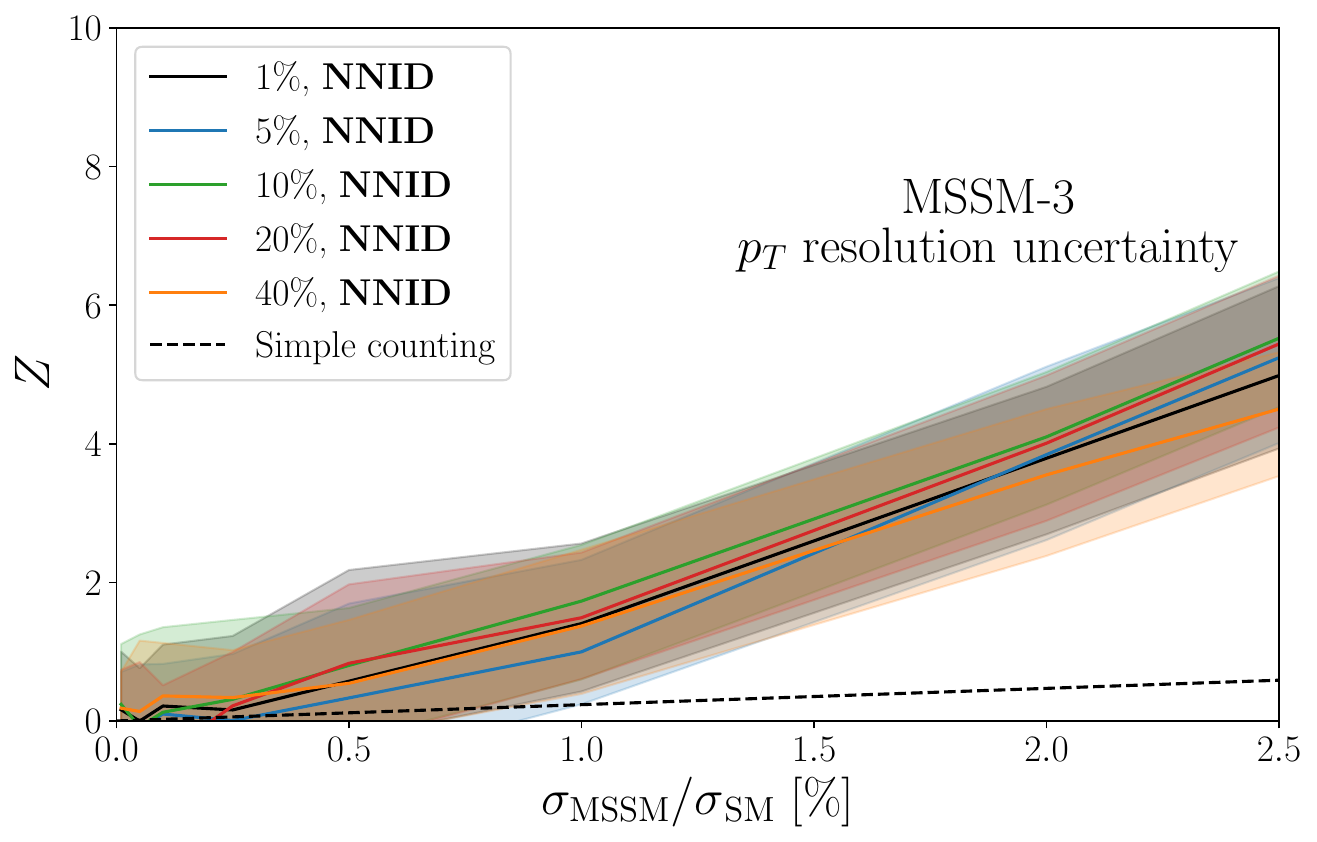}
    \hfill
    \includegraphics[width=0.49\textwidth]{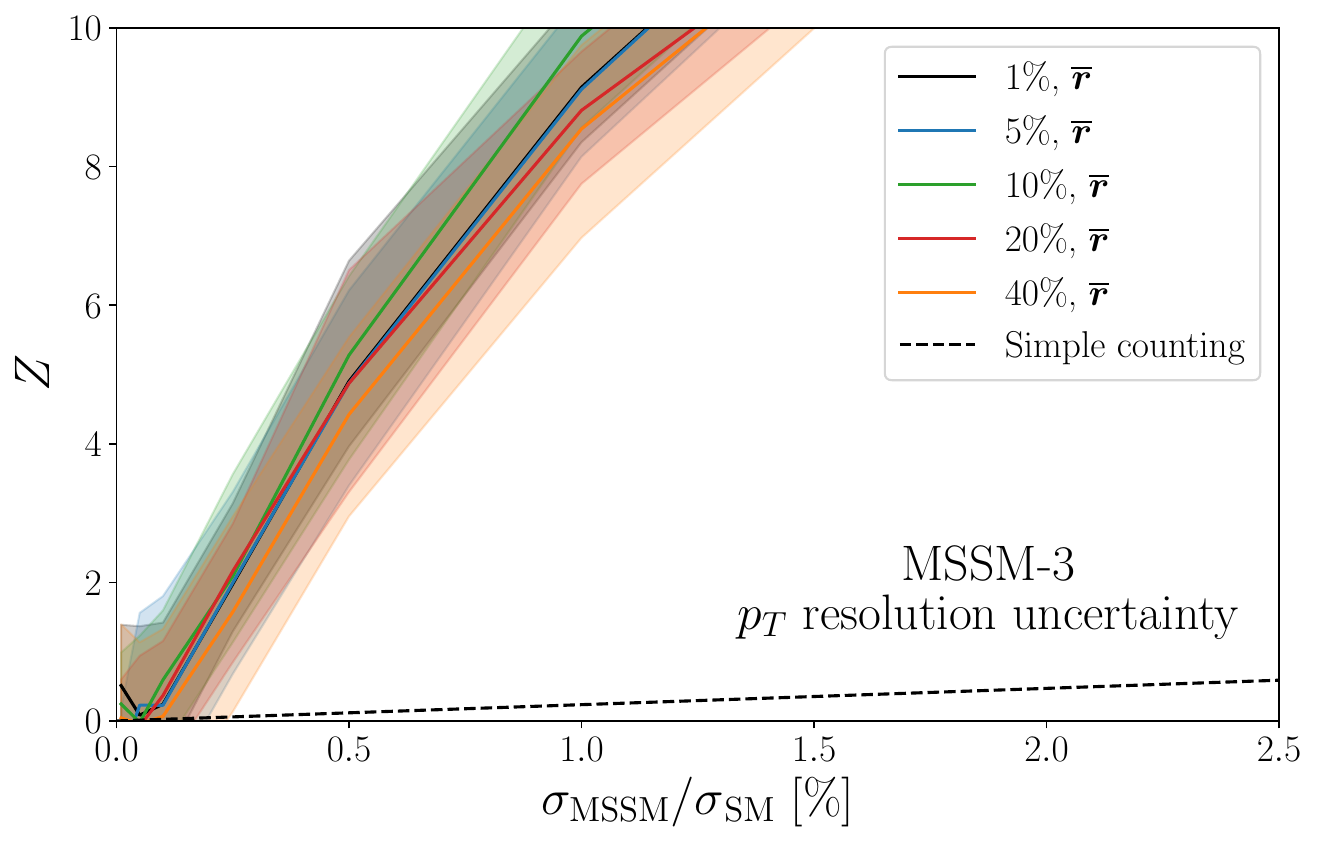}
\caption{Robustness of the \ID and the associated statistical test to a Gaussian smearing of all particles' $p_T$. The Figure demonstrates the insensitivity of our new physics search to energy resolution uncertainties.}
\label{fig:MSSM_IDvRate_pt_gauss_det}
\end{figure}

By standard M-estimator theory, the variance of the NNID, $\sigma_d^2$, scales as $1/N$.  An additional per-particle smearing of relative size $\sigma_{p_T}$ inflates the variance of the distances, and hence of each log-ratio $\mu_{k,ij} =r_{k,j}/r_{k,i}$, by an $\mathcal O(\sigma_{p_T}^2)$ term, but its contribution to $\sigma_d^2$ is suppressed by $1/N$.  Consequently, even a $\mathcal O(10\%)$ smearing in $p_{T}$ produces only a few-percent increase in the standard error of $d$, which is already at the percent level. From the top-left panel of \cref{fig:MSSM_IDvRate_pt_gauss_det}, we can indeed appreciate how the 68\% confidence level bands for the \ID maintain the same width. The most notable differences are the shifts in the median, particularly for $\sigma_{p_T} = 20\%,40\%$, associated to the hard cut that we impose to keep $p_T>0$ which distorts the distribution of the distances, as shown in the top-right panel of \cref{fig:MSSM_IDvRate_pt_gauss_det}.
The same argument applies to the variance of $\overline{r}_i$, which is also already at percent level and hence is essentially unaffected even for big additional smearing.

Since the significance depends on the relative distance between the \ID and $\overline{r}_i$ curves in units of their standard deviation, it should not be appreciably affected by the smearing, given its small impact on the standard deviation of their distribution. Indeed this is what we can see from the bottom panels of \cref{fig:MSSM_IDvRate_pt_gauss_det}, in which the significances remain essentially unchanged, except for the biggest smearing values which induce a small decrease of $Z$.  

Overall, this demonstrates that our $\overline{r}_i$ and NNID-based significance is both exactly insensitive to global scale shifts and highly resilient to large random $p_{T}$ variations, making the method very robust to detector effects on particles' energies and momentum. We expect similar results to hold for errors on the particles' positions ($\eta$ and $\phi$), given that their contribution to the EMD is similar to that of the particle $p_T$.

In principle, such a transformation would also affect which events pass the $p_T$ selection cuts, since their absolute thresholds are not invariant under rescaling. In our analysis, the rescaling (and later the smearing) is applied only after event selection, so this effect is not included. In Appendix~\ref{app:scale} we rescale the $p_T$ of all particles before event selection and we find a non-zero effect, as expected, but still a very modest impact on the significance as shown in Fig.~\ref{fig:MSSM_lep_significance_pT_rescaling_cut} and explained in the Appendix.

\subsection{Leptonic $Z'$}\label{sec:Zp}

\begin{figure}[ht]
    \centering

\begin{subfigure}{0.32\textwidth}
\centering
\begin{tikzpicture}
\begin{feynman}

    \vertex (q)  at (-1, 1.5) {\(q\)};
    \vertex (qb) at (-1,-1.5) {\(\bar q\)};
    \vertex (v1) at (-0.5,  0);
    \vertex (v2) at ( 0.5,  0);

    \vertex (mu1) at ( 1,  1.5) {\(\mu^-\)};
    \vertex (mu2) at ( 1, -1.5) {\(\mu^+\)};

    \vertex (v3)  at (0.75,-0.75);
    \vertex (v4)  at (1.5,-0.5);
    \vertex (mu3) at (2.5,  0) {\(\mu^-\)};
    \vertex (mu4) at (2.5, -1) {\(\mu^+\)};

    \vertex (label) at (1.2, -0.3) {$Z'$};

    \diagram*{

        (q)  -- [fermion]      (v1),
        (qb) -- [anti fermion] (v1),
        (v1) -- [boson, edge label=\(Z\)] (v2),

        (v2) -- [fermion]      (mu1),
        (v2) -- [ anti fermion]      (v3) -- [anti fermion] (mu2),

        (v3) -- [boson] (v4),
        (v4) -- [fermion]      (mu3),
        (mu4) -- [fermion] (v4),
    };
\end{feynman}
\end{tikzpicture}
\caption{}
\end{subfigure}
    \hfill
\begin{subfigure}{0.32\textwidth}
\centering
\begin{tikzpicture}
\begin{feynman}

    \vertex (q)  at (-1, 1.5) {\(q\)};
    \vertex (qb) at (-1,-1.5) {\(\bar q\)};
    \vertex (v1) at (-0.5,  0);
    \vertex (v2) at ( 0.5,  0);

    \vertex (mu1) at ( 1,  1.5) {\(\mu^-\)};
    \vertex (mu2) at ( 1, -1.5) {\(\mu^+\)};

    \vertex (v3)  at (0.75,-0.75);
    \vertex (v4)  at (1.5,-0.5);
    \vertex (mu3) at (2.5,  0) {\(\mu^-\)};
    \vertex (mu4) at (2.5, -1) {\(\mu^+\)};

    \vertex (label) at (1.1, -0.3) {$\gamma/Z$};

    \diagram*{

        (q)  -- [fermion]      (v1),
        (qb) -- [anti fermion] (v1),
        (v1) -- [boson, edge label=\(Z\)] (v2),

        (v2) -- [fermion]      (mu1),
        (v2) -- [ anti fermion]      (v3) -- [anti fermion] (mu2),

        (v3) -- [boson] (v4),
        (v4) -- [fermion]      (mu3),
        (mu4) -- [fermion] (v4),
    };
\end{feynman}
\end{tikzpicture}
\caption{}
\end{subfigure}
    \hfill
    \begin{subfigure}{0.32\textwidth}
        \centering
        \begin{tikzpicture}
            \begin{feynman}
                \vertex (q)   at (-1.8,  1.5) {\(q\)};
                \vertex (qb)  at (-1.8, -1.5) {\(\bar q\)};
                \vertex (v1)  at (-0.6,  0.7);
                \vertex (v2)  at (-0.6, -0.7);

                \vertex (w1)  at ( 1.0,  0.9);
                \vertex (mu1) at ( 2.0,  1.5) {\(\mu^-\)};
                \vertex (mu2) at ( 2.0,  0.3) {\(\mu^+\)};

                \vertex (w2)  at ( 1.0, -0.9);
                \vertex (mu3) at ( 2.0, -0.3) {\(\mu^-\)};
                \vertex (mu4) at ( 2.0, -1.5) {\(\mu^+\)};

                \vertex (label1) at ( 0.2, 1.2) {$\gamma/Z$};
                \vertex (label2) at ( 0.2, -1.2) {$\gamma/Z$};

                \diagram*{
                    (q)  -- [fermion]      (v1)
                         -- [fermion]      (v2)
                         -- [fermion]      (qb),

                    (v1) -- [boson] (w1),
                    (w1) -- [fermion]      (mu1),
                    (mu2) -- [ fermion] (w1),

                    (v2) -- [boson] (w2),
                    (w2) -- [fermion]      (mu3),
                    (mu4) -- [ fermion] (w2),
                };
            \end{feynman}
        \end{tikzpicture}
        \caption{}
    \end{subfigure}

\caption{Dominant production mechanism for the $Z'$ signal in Section~\ref{sec:Zp} (a) and leading SM backgrounds (b), (c).}
\label{fig:Zp_sketch}
\end{figure}
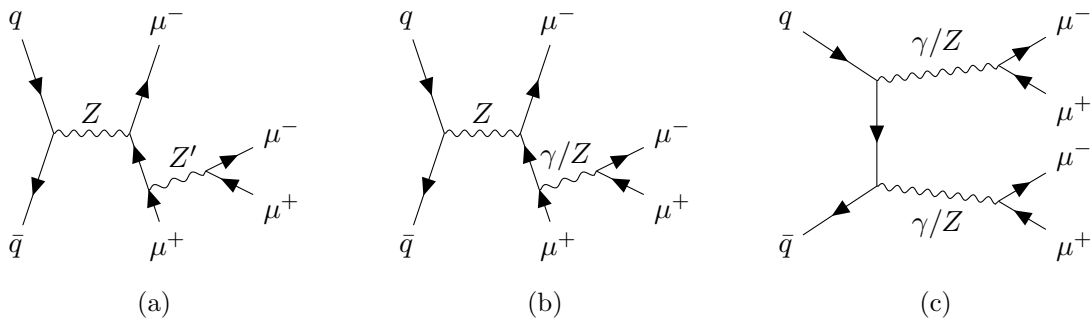


To test our approach as a model-independent search, we now consider a completely different signal.
Specifically, we focus on a light $Z'$ model arising from gauging the $U(1)_{L_\mu - L_\tau}$ symmetry of the SM,  where $L_{\mu, \tau}$ are the muon and tau lepton numbers. The process we study is $pp \to 2\mu^+ 2\mu^-$, where a $Z'$ boson is radiated by a final-state muon and subsequently decays into a $\mu^+\mu^-$ pair (the characteristic topology is depicted in Fig.~\ref{fig:Zp_sketch} a). This setup is quite different compared to the previous example, where the final-state particles originating from the decay of heavy new physics were highly energetic. Here we do not have boosted particles in the final state and, additionally, we have a relatively low mass resonance (in what follows we take $m_{Z'}< m_Z$).

The leading irreducible Standard Model background consists of diagrams similar to that of the $Z'$, i.e. with a radiatively emitted $Z$ boson, along with other contributions, that include $Z$ bosons and virtual photons as intermediate states, as shown in \cref{fig:Zp_sketch}. We include all relevant tree-level processes in our background simulation (but not the 1-loop box diagram with virtual quarks, considered in~\cite{ATLAS:2023vxg}).

\paragraph{Simulation and Event Selection.} Events are generated at parton level with a custom {\small \textsc{MadGraph}} implementation of the $Z'$ model~\cite{delAguila:2014soa}. We set $\sqrt{s}=13.6$ TeV and a reference luminosity of 400 fb$^{-1}$, doubled with respect to \cref{sec:MSSM_lep} to provide a sizeable amount of events. We then run the data through {\small \textsc{pythia} 8} and {\small \textsc{delphes}} to simulate hadronization and detector effects. The procedure for computing $\overline{r}_i$ and the \ID is the same as in Section~\ref{sec:MSSM_lep}, and it is carried out again for two sets of events:
\begin{enumerate}
    \item[$i)$] \emph{Parton level}: for each event, the 2 muons and 2 antimuons from {\small \textsc{MadGraph}} are directly employed in the computation of distances, subject to the same cuts as in \cref{sec:MSSM_lep} regarding rapidty, $p_T$ and muon isolation. This leads to datasets of $\approx 2600$ SM events with the reference luminosity of $400$ fb$^{-1}$;
    \item[$ii)$] \emph{Detector level}: we employ again the custom CMS card, and select only events with at least two isolated muons and two isolated antimuons with $p_T>10$ GeV, $|\eta|<2.5$. As before, jets are allowed if $p_{T,j}<40$ GeV, although they are not employed for the computation of distances. The efficiency of the selection after the detector simulation retains only about $50\%$ of the events compared to the parton case.
\end{enumerate}

\begin{figure}[!ht]
    \includegraphics[width=0.49\textwidth]{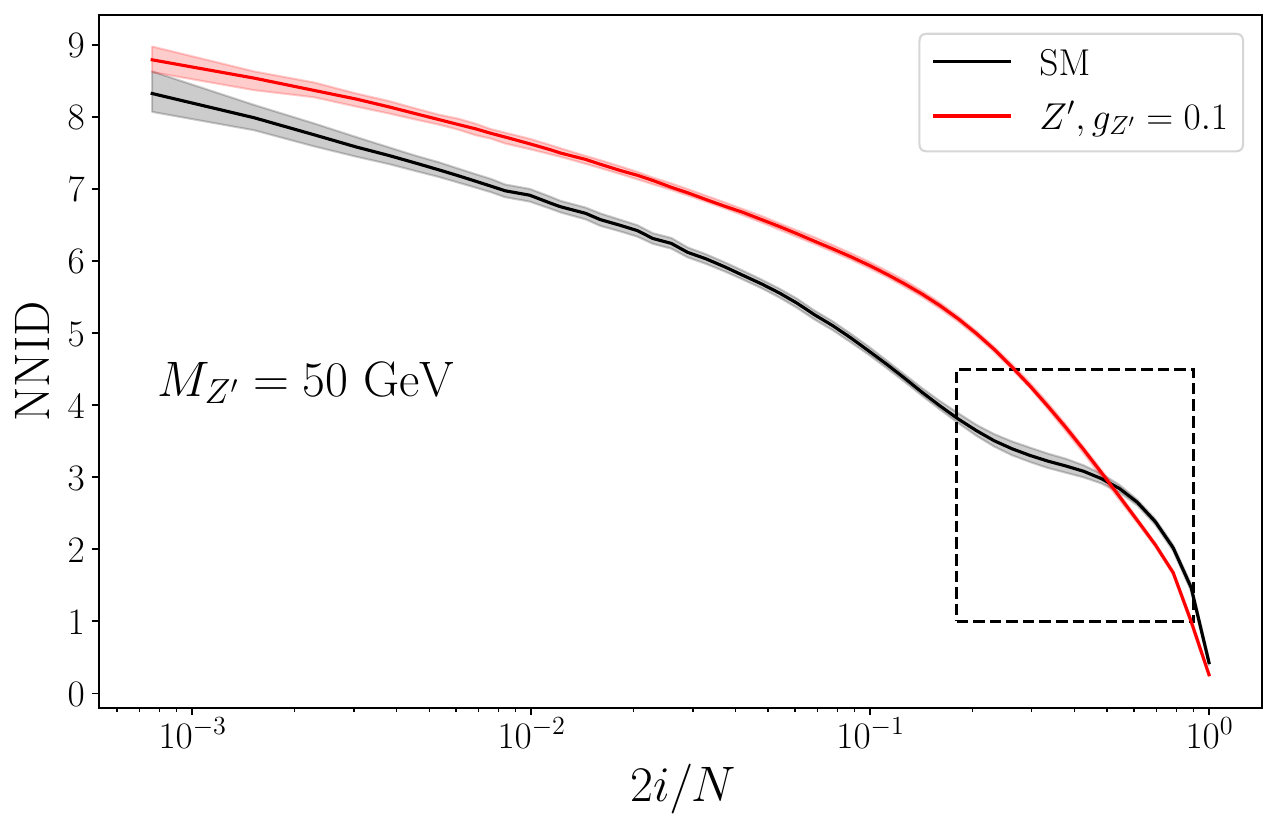}\hfill
    \includegraphics[width=0.49\textwidth]{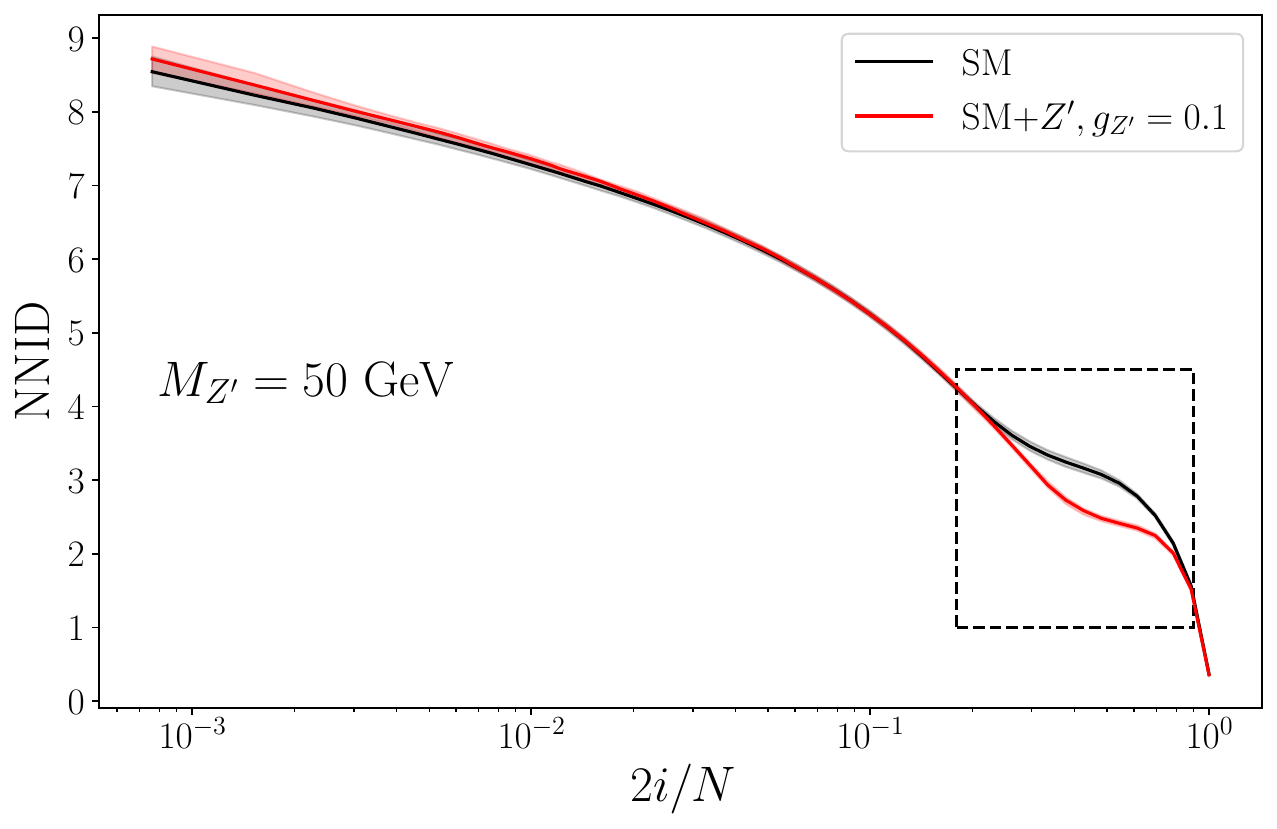}\hfill
    \includegraphics[width=0.49\textwidth]{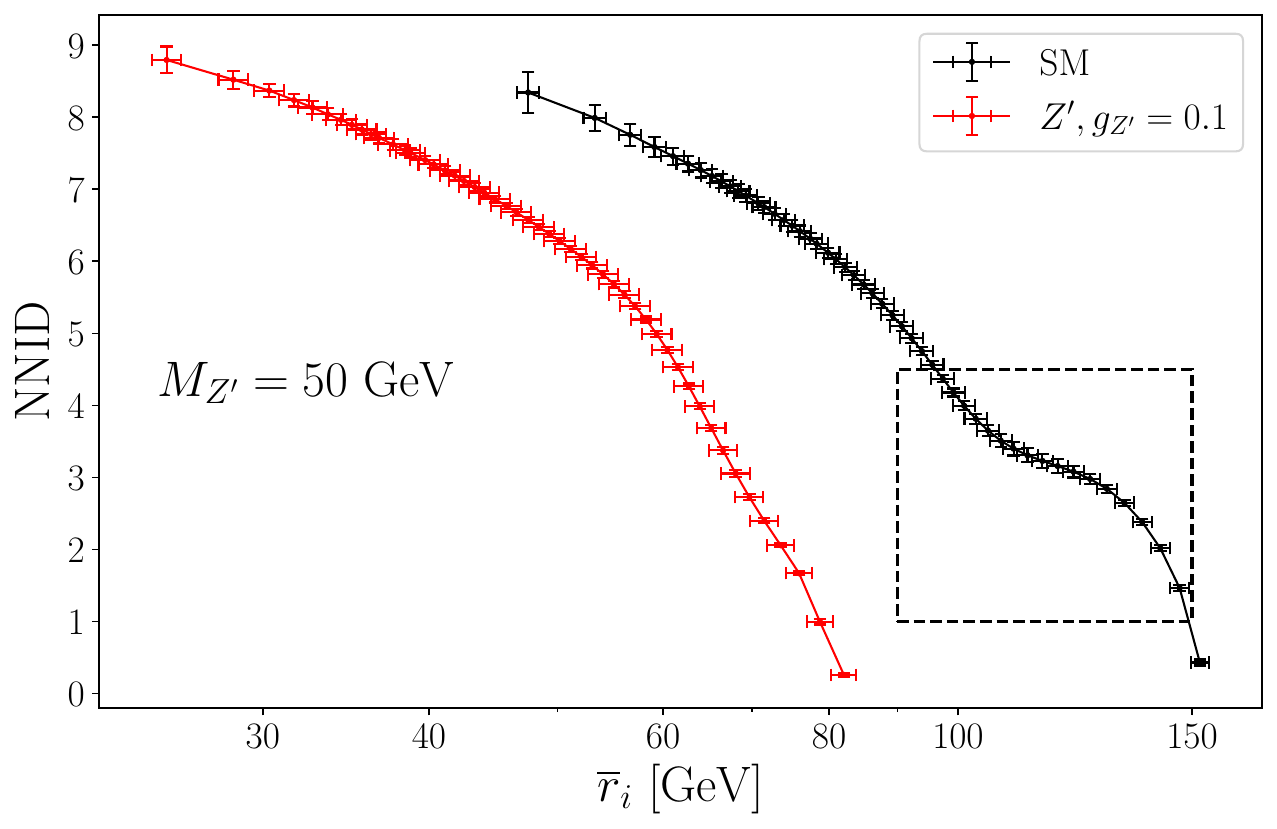}\hfill
    \includegraphics[width=0.49\textwidth]{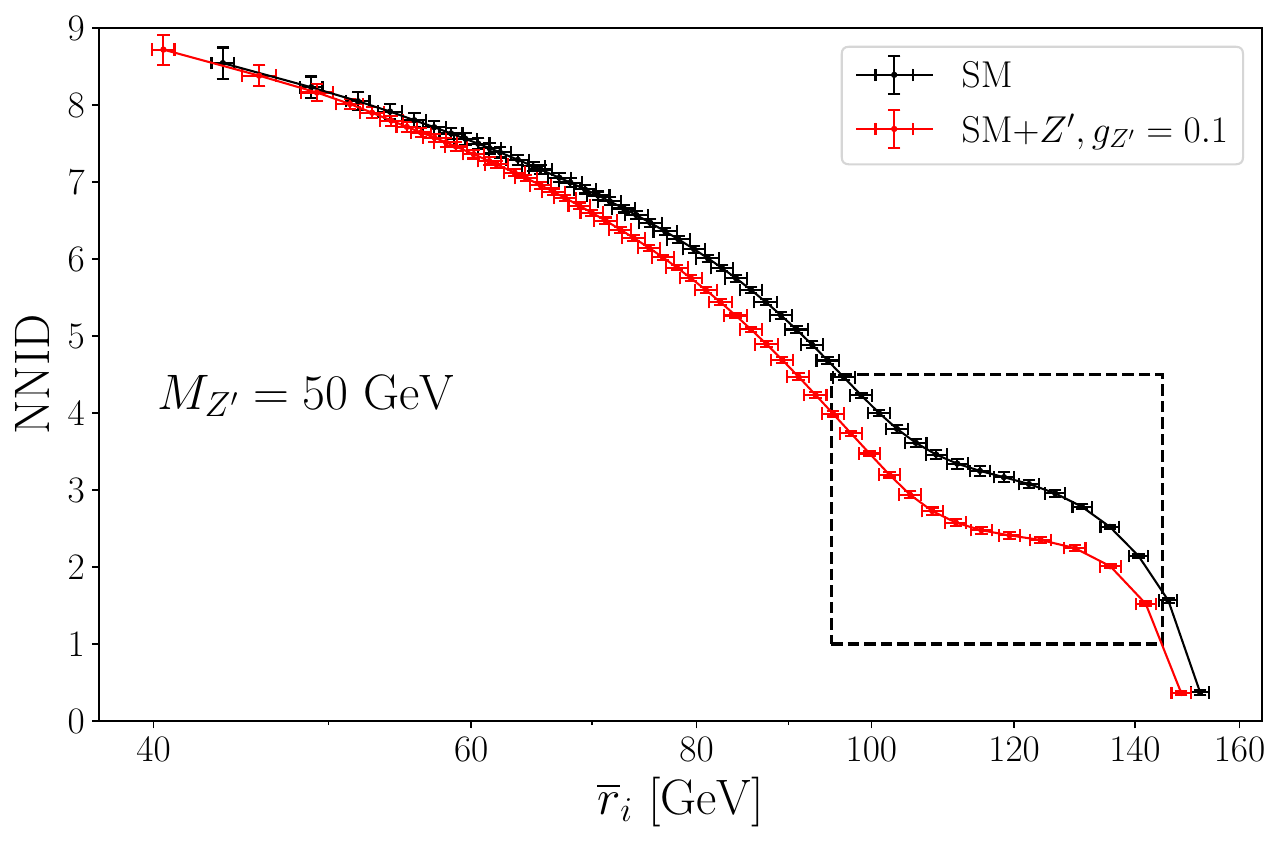}\hfill
    \includegraphics[width=0.49\textwidth]{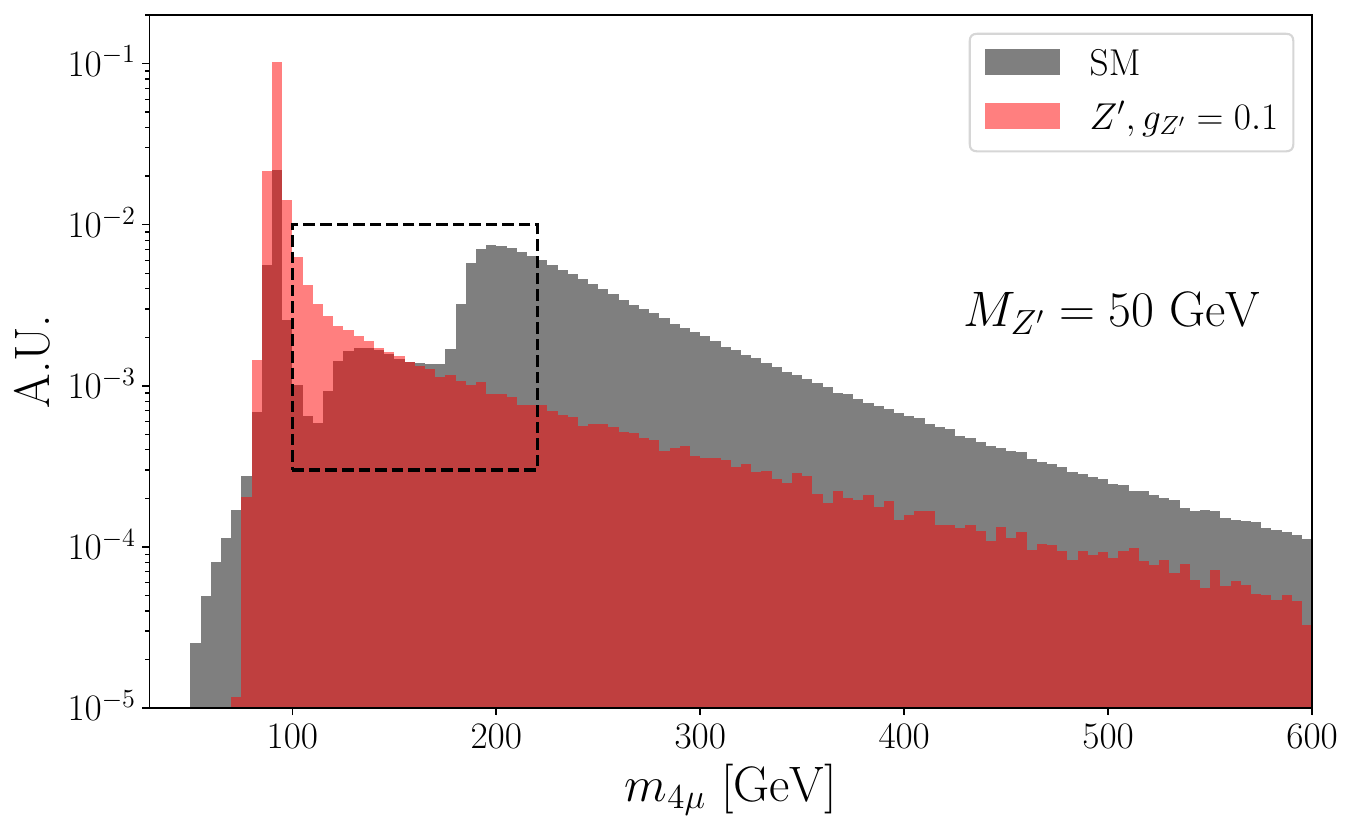}\hfill
    \includegraphics[width=0.49\textwidth]{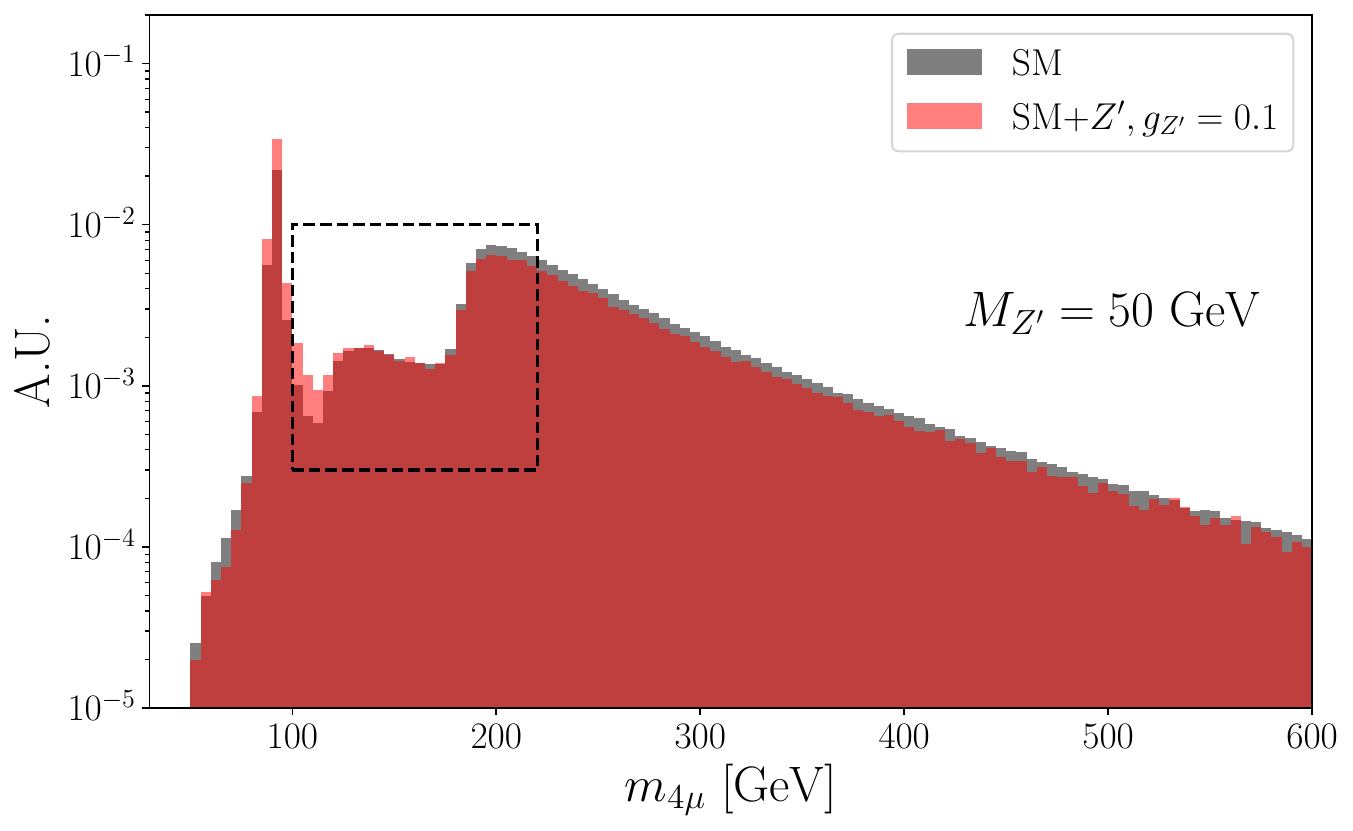}\hfill
\caption{\ID curves at parton level as a function of $2i/N$ and $\overline{r}_i$ in the pure SM and pure $Z'$ samples (left). In the right panels we show a comparison between the SM and the SM+$Z'$ samples. The mass of the $Z'$ is fixed to  $M_{Z'}=50$ GeV with coupling $g_{Z'}=0.1$, and the curves are obtained with a fixed number of SM events $N=2613$ for illustrative purposes. The bottom row shows the invariant mass distribution of the four muons in the event. We highlight with a dashed black box the region of the \ID curves corresponding to the two shoulders in the invariant mass distribution of the bottom panels.}
\label{fig:Zp_IDcurves}
\end{figure}

\paragraph{Results.} We generate 500 samples of the SM background, and compute the \ID and $\overline{r}_i$ with 50 logarithmically spaced values of the normalized index $i/N$. The granularity is improved compared to the model in \cref{sec:MSSM_lep} to compensate the smaller number of events per sample in this benchmark. We collect in \cref{app:QQ} the comparison of our test statistic under the null hypothesis with the theoretical $T^2$ distribution.

For the $Z'$ signal, we fix the mass to $M_{Z'} = 50$~GeV and vary the gauge coupling to muons and taus, denoted as $g_{Z'}$, in the range $[0.01, 0.1]$ in steps of 0.01. As a reference, the number of  $Z'$ events passing the cuts and selection at detector level ranges from a few ($g_{Z'} = 0.01$) to about 100 ($g_{Z'}=0.1$). To obtain confidence level bands, we generate 50 samples for each value of the coupling.

The \ID curves for both the pure and the combined background+signal samples are shown in \cref{fig:Zp_IDcurves}. Interestingly, both the pure SM and SM+$Z'$ cases exhibit a distinct ``plateau'' at large values of $2i/N$, corresponding to scales of $\mathcal{O}(100)$ GeV. A comparison with the invariant mass distributions of the four leptons $2\mu^+ 2\mu^-$, shown in the bottom panels of \cref{fig:Zp_IDcurves}, suggests a possible connection with the two shoulders highlighted by the dashed boxes. This is confirmed by the absence of such a bump in the \ID curves and invariant mass distributions in the leptonic MSSM case (see \cref{fig:MSSM_IDcurves}) and in the pure $Z'$ samples in Fig.~\ref{fig:Zp_IDcurves}. 

\begin{figure}[!t]
\centering
\includegraphics[width=0.8\textwidth]{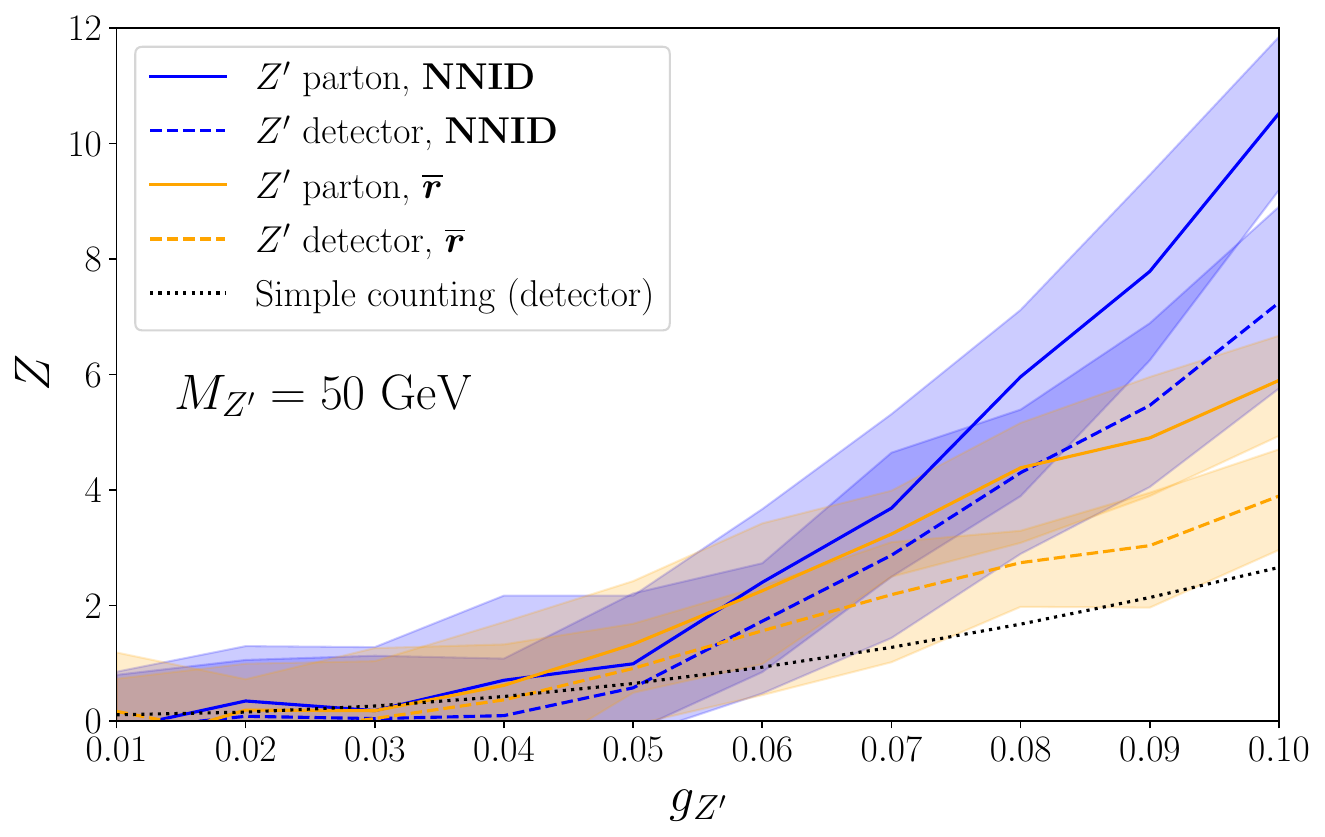}
\caption{Significance bands at 68\% CL of the new physics test at parton and detector level, as a function of the $Z^\prime$ coupling to leptons. The curves are obtain using the test described in \cref{sec:statTest}, either using information from the \ID (blue curves) or from $ \overline{r}_i$ (orange curves).}
\label{fig:Zp_ZvRate}
\end{figure}

The significances obtained from the \ID and $\overline{r}_i$ tests described in \cref{sec:statTest}, at both parton and detector level, are shown in \cref{fig:Zp_ZvRate} as a function of $g_{Z'}$. The sensitivities at parton and detector level are compatible within one sigma.

In this case, the test based on the \ID is more powerful than that based on $\overline{r}_i$, in contrast to the findings of \cref{sec:MSSM_lep}. The reduced sensitivity of the $\overline{r}_i$ test can be traced to the kinematics of the process: the events associated with $Z'$ production are not significantly more boosted than the corresponding SM background ones, since $m_{Z'}$ is close to the $Z$-boson mass. As a result, the distribution of $\overline{r}_i$ alone provides limited discrimination. The \ID, on the other hand, appears to enhance the separation between signal and background by effectively exploiting a non-linear function of $\overline{r}_i$, thereby leading to a more powerful test.

As in previous cases, the simple counting experiment (dashed black line in the figure) is dominated by systematic uncertainties ($4.4\%$ on the SM cross section, combined with a $1.6\%$ uncertainty on the luminosity~\cite{CMS:2020gtj}).

The main message of this Section is that our \ID test is sensitive also to the $Z'$ signal. The kinematics explored in this Section and the previous one are completely different from each other and support our statements on the flexibility of our new physics search strategy. Additionally, the performances at parton level and detector level remain comparable, further confirming the robustness of our test. 

\subsection{Semileptonic Decays of Charginos and Neutralinos}\label{sec:MSSMhad}

In this section we show that our techniques can find new physics without clustering jets in a $\sim$100-dimensional phase space.
To this end, we employ the MSSM-3 model introduced in \cref{sec:MSSM_lep}, but consider a semileptonic final state where the $W$ boson decays to jets: $W \rightarrow jj$.

\paragraph{Simulation and Event Selection.}
As in the previous sections, parton-level events, hadronization, and detector effects are simulated using {\small \textsc{MadGraph}}, {\small \textsc{Pythia}}, and {\small \textsc{Delphes}}, respectively, with the same configuration cards as in the leptonic study.
We take $\sqrt{s}=13.6$ TeV and generate batches of 10,000 clean parton level events, formally fixing the effective luminosity through the corresponding cross section (see below). This choice is motivated by the fact that the cross section of the relevant background process would otherwise produce datasets far too large for us to process with our available resources, if we were to match the luminosity used in \cref{sec:MSSM_lep}. We perform the analysis on two sets of events:

\begin{enumerate}
\item[$i)$] \emph{Parton level}: the four charged partons from the hard collision generated by {\small \textsc{MadGraph}} are used directly. The selection cuts are $p_T > 10$ GeV for muons and $p_T > 30$ GeV for jets, both within $|\eta| < 2.5$. The same muon isolation as in~\cref{sec:MSSM_lep} is imposed, and we further require $\Delta R_{jj} \geq 0.4$ to mimic the detector-level jet separation (since jets are clustered with anti-$k_T$ and radius $R=0.4$, see below). We also impose $70 \text{ GeV} \leq m_{jj} \leq 90\text{ GeV}$ and $76 \text{ GeV} \leq m_{\mu\mu} \leq 106\text{ GeV}$;

\item[$ii)$] \emph{Detector level}: we use the custom CMS card of \cref{sec:MSSM_lep,sec:Zp}, this time with a jet radius $R=0.4$. Events are required to contain an isolated muon and an isolated antimuon with $p_T > 10$ GeV and $76 \text{ GeV} \leq m_{\mu\mu} \leq 106\text{ GeV}$, as well as at least two jets with $p_T > 30$ GeV and $70 \text{ GeV} \leq m_{jj} \leq 90\text{ GeV}$.
\end{enumerate}

With these cuts, the leading SM background consists of $Z+$2 jets events; the cross section for $WZ$ is orders of magnitude smaller. The dimuon invariant mass window suppresses the $t\bar t$ background. We stress that additional cuts, for instance on the missing transverse momentum as in dedicated MSSM searches for this channel \cite{ATLAS:2022zwa}, could further improve the signal isolation. However, since our goal is to identify potential traces of new physics within a fixed final-state configuration (here $2\mu2j$), without prior knowledge of the signal, imposing such cuts would artificially enhance the significance. With the cuts above, the parton-level cross section for the SM background is $\approx 3$ pb; the relative efficiency from parton to detector level is about 45\% for both signal and background.\footnote{This cross section and our events are obtained simulating $Z+2j$ at LO with {\small \textsc{MadGraph}}. We do not generate an inclusive matched sample because we choose to neglect the corrections to the $p_T$ distribution of our jets with $p_T > 30$~GeV (that are of $\mathcal{O}(\alpha_s \log (m_Z/p_T))$ in this first proof-of-principle study of the methodology.}

To test the robustness of our search against the dimensionality of the input space, we compute distances and \ID at detector level in two setups: (a) using muons and clustered jets as inputs, and (b) using muons and the \emph{subcomponents} of the selected jets. The latter case results in an average of about 30 particles per event, corresponding to a search performed in an $O(100)$-dimensional space.

\paragraph{Results.}

We compute the \ID on 200 SM samples. We collect in \cref{app:QQ} the comparison of our test statistic under the null hypothesis with the theoretical $T^2$ distribution for the various sets of events described below (parton level, detector level clustered and unclustered). Then, we proceed as in the leptonic case: we include a variable fraction of MSSM events in the SM samples, and perform the new physics test. The fraction is set in a fiducial region defined by loose cuts:
$$ p_{T,j}>25 \text{ GeV}, \quad  p_{T,\mu^{\pm}}>10 \text{ GeV},\quad |\eta_{j, \mu^{\pm} } |<2.75 \, .$$
Since the efficiency for SM and MSSM with respect to our event selection is different (about $10\%$ for the SM and $60\%$ for the MSSM at parton level), the effective $S/B$ ratio in the actual computation is almost a factor of ten larger. This explains why in this case a smaller percentage of signal is needed to get sizeable significances compared to Sections \ref{sec:MSSM_lep} and \ref{sec:Zp}.

\begin{figure}[!t]
\begin{center}
    \includegraphics[width=0.49\textwidth]{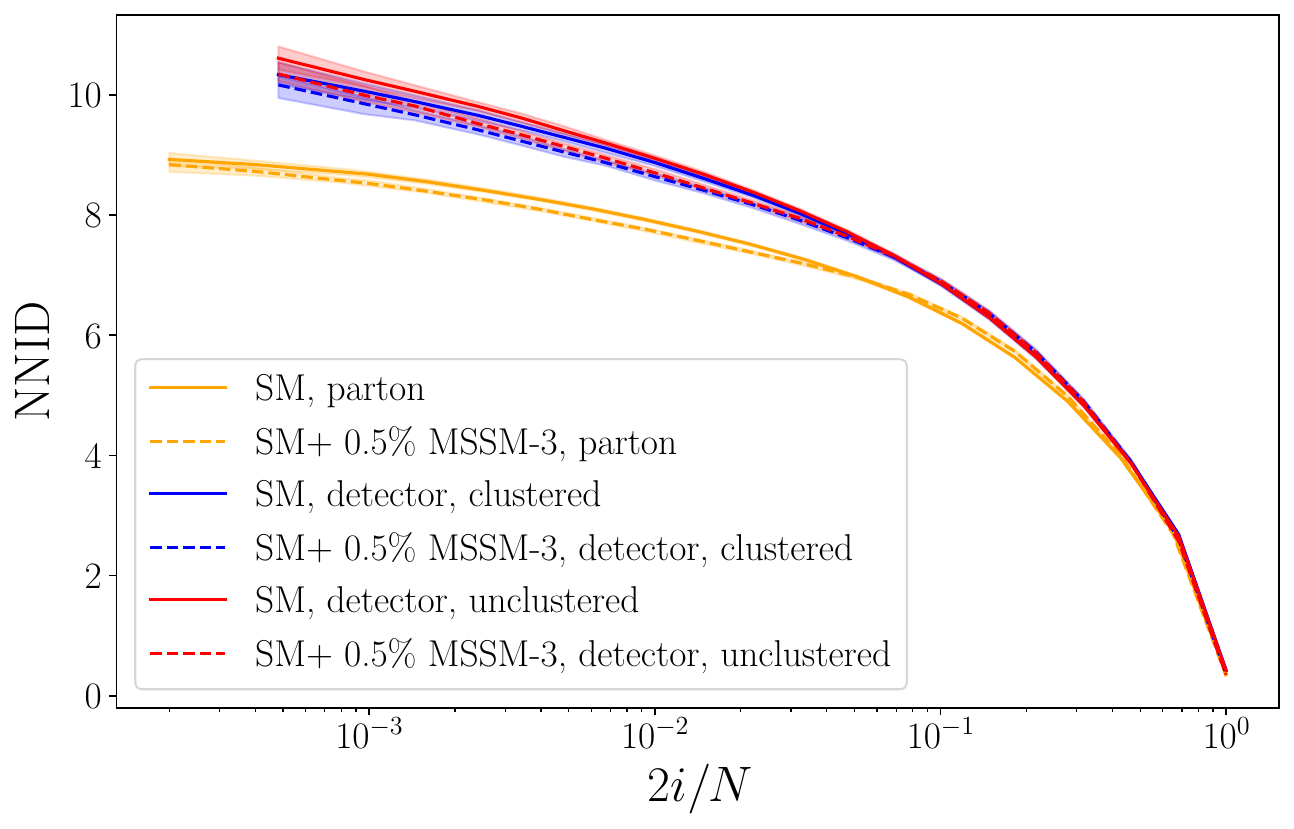}
    \hfill
    \includegraphics[width=0.49\textwidth]{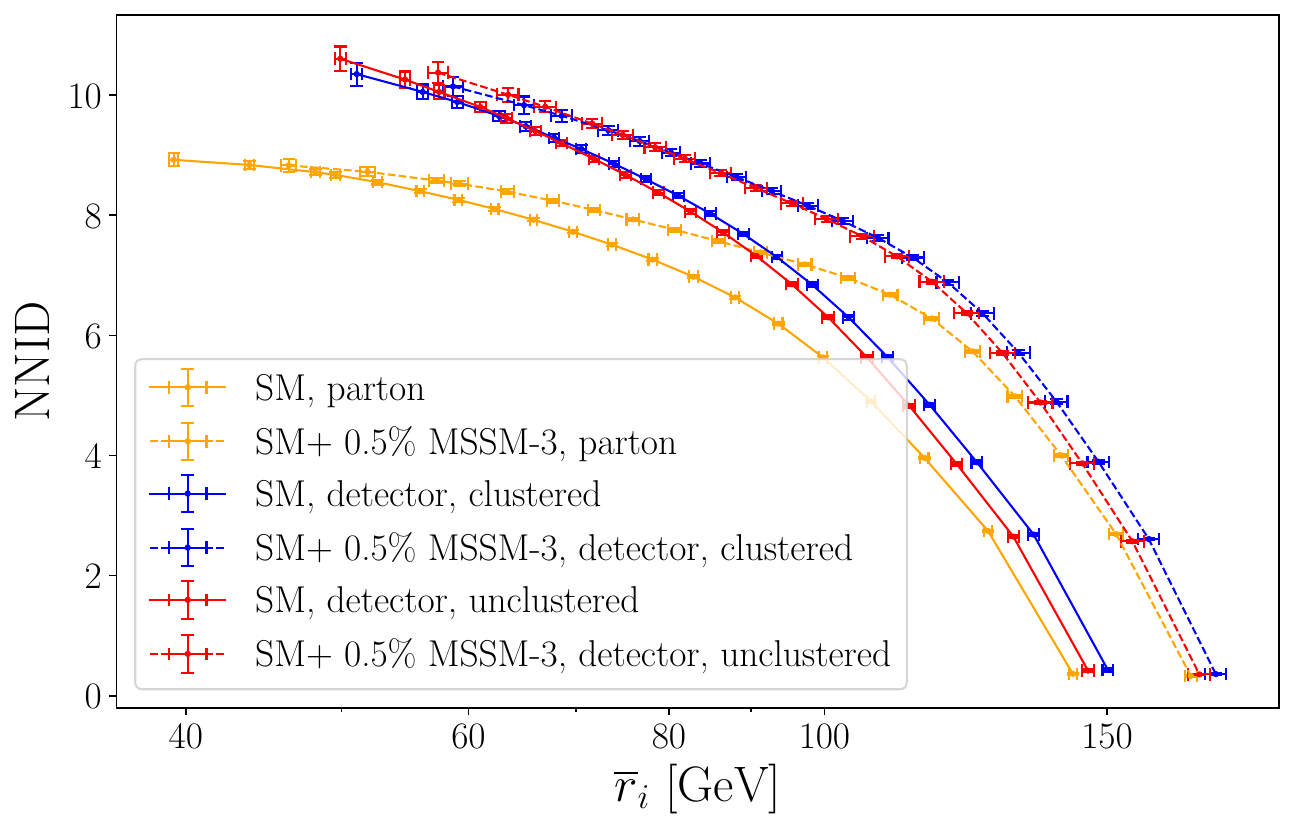}
    \hfill
    \end{center}
\caption{\ID curves at parton, clustered detector and unclustered detector level as a function of $2i/N$ (left panel) and the average distance to the $i$-th neighbor (right panel). The central line represents the median, while the band is the $68\%$ confidence interval.}
\label{fig:MSSM_semilep_IDcurves}
\end{figure}

In \cref{fig:MSSM_semilep_IDcurves} we show the \ID as a function of the proportionality vector $2i/N$ and of the average distance to the $i$-th neighbor, $\overline{r}_i$, for pure SM samples and SM + MSSM-3 events. The input to the \ID computation varies between 12 kinematical variables in the parton case, to more than 100 for detector-level unclustered events. Nonetheless, \ID curves do not vary dramatically between these two sets of events. The results in the left panel of Fig.~\ref{fig:MSSM_semilep_IDcurves} show that the transition between partons and jets encodes information, as the \ID grows at small $i/N$ between the two sets of events from roughly 9 to around 11. Additionally, the ID correctly recognizes that the 100 or so particles in the unclustered events are in reality encoding information about a much smaller set of degrees of freedom. Indeed we can see in the left panel of Fig.~\ref{fig:MSSM_semilep_IDcurves} that the \ID computed at detector level starting from jets (so overall 12 input variables) gives roughly the same result as the \ID computed from jet constituents (so around 100 input variables). In the right panel of the figure we show that also the typical distances between neighbors do not change appreciably between clustered and unclustered calculations, further demonstrating the robustness of these observables. The shorter distances probed in the parton-level dataset arise from the larger number of available events, which allows the nearest-neighbors estimator to explore smaller scales. At fixed $\overline{r}_i$, the intrinsic dimension extracted from detector-level data is indeed larger, as expected, because hadronization and detector smearing introduces additional noisy directions.

\begin{figure}[!t]
\centering
    \includegraphics[width=0.8\textwidth]{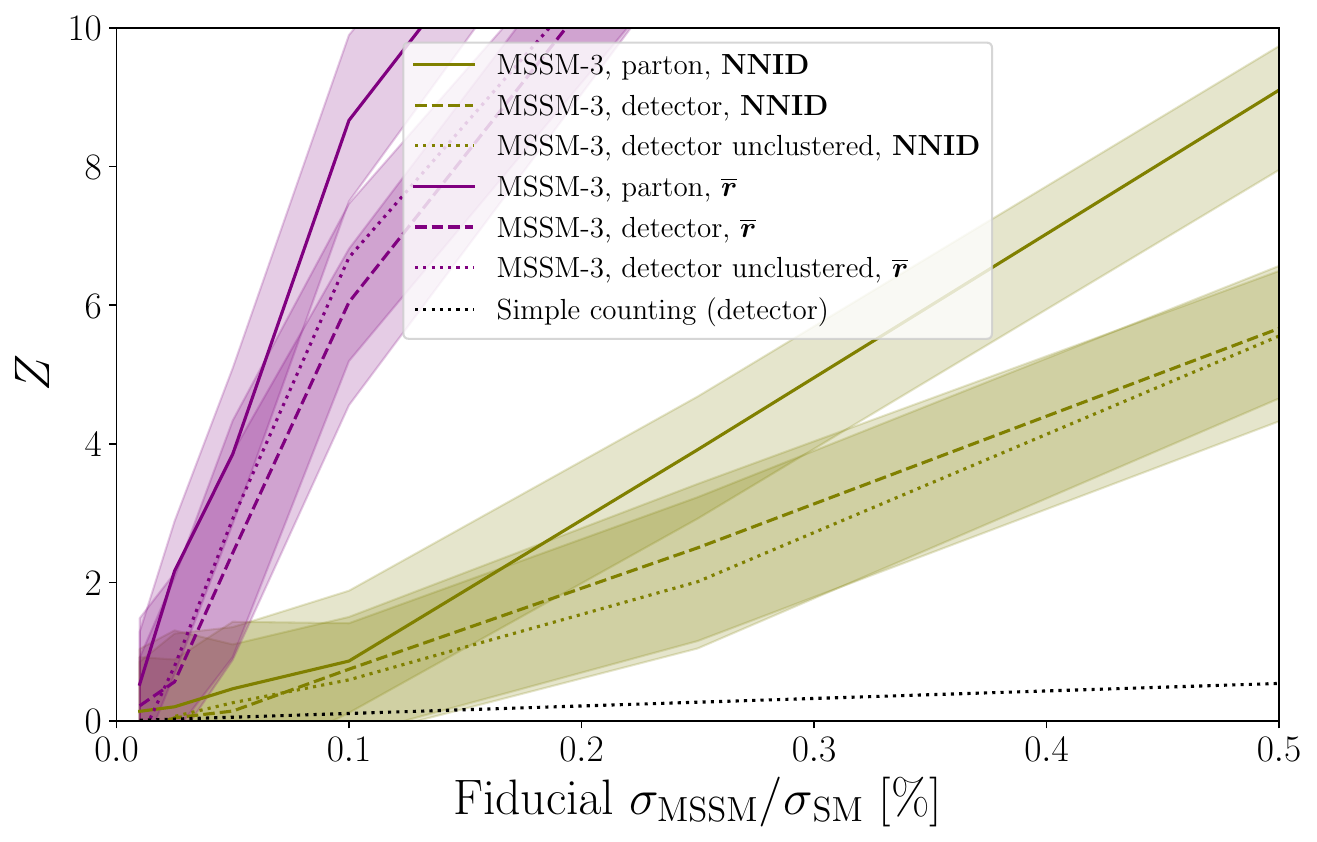}
\caption{Significance bands at 68\% CL of the new physics test in the semileptonic MSSM-3 model at parton level and detector level (with clustered and unclustered jets), as a function of the ratio between the BSM and SM cross section in the fiducial region.}
\label{fig:MSSM_semilep_ZvRate}
\end{figure}

In \cref{fig:MSSM_semilep_ZvRate} we display the usual significance plots as a function of a variable BSM cross section. In this case, for the reference ``naive'' counting experiment we take a total error of $\approx 9\%$ on the theory SM prediction for $Z+2j$ \cite{CMS:2018mdf}.
Quite remarkably clustered and unclustered events at detector level give the same median significance with also the error bar on the significance remaining stable. We emphasize that, in the unclustered case, this corresponds to a model-independent new physics search in a space of order $\mathcal O(100)$ dimensions. This demonstrates the robustness of our \ID and $\overline{r}_i$ new physics test and supports its application to real data.

The parton-level performances are instead higher than the detector-level ones. We can trace this difference to the different efficiency of our event selection cuts. The larger number of events at parton level reduces the statistical error on \ID and $\overline{r}_i$. 

In \cref{fig:MSSM_ZvSB} (\cref{app:scale}) we provide the same significance plot in terms of $S/B$, since the efficiency of the cuts going from the region of phase space where we define our fiducial cross sections to the signal region that we use to compute the \ID makes $S/B$ larger than the fiducial cross section ratio.

\subsection{Summary of Results}\label{sec:summary}

This work represents an initial proof-of-principle study in which we have included the Standard Model background most closely resembling the signal. Nevertheless, the results obtained here are already highly encouraging and clearly illustrate the capabilities and distinctive advantages of our new physics search strategy.

\begin{enumerate}
    \item {\bf Flexibility}: sensitivity to diverse Beyond the Standard Model scenarios with different kinematics. This is demonstrated by the variety of signals explored in the three previous Sections. We had boosted leptons from the MSSM-3 signal in Section~\ref{sec:MSSM_lep}, low-mass resonances from the $Z'$ in Section~\ref{sec:Zp}, boosted jets in Section~\ref{sec:MSSMhad} and heavy non-resonant new physics from the MSSM-2 signal in Section~\ref{sec:MSSM_lep}.
    \item {\bf Robustness} (to showering and detector effects): the performances are degraded when going from the parton-level to the detector-level simulation. We trace this effect to the lower efficiency of the detector-level event selection and the increased statistical error on \ID and $\overline{r}_i$. However, the performances remain unaltered going from clustered jets to jets' constituents and remain comparable between leptonic and hadronic decays of the $W$. We take this as a promising sign of the robustness of our methodology, leaving to future work the study of different showering algorithms. This is shown in Fig.s~\ref{fig:MSSM_IDvProp} and~\ref{fig:Zp_ZvRate} for the MSSM-3 leptonic and $Z'$ signals, respectively and in Fig.~\ref{fig:MSSM_semilep_ZvRate} for the MSSM-3 semileptonic signal. A comparison between leptonic and semileptonic MSSM in shown in Appendix~\ref{app:scale}.
    \item {\bf Robustness to Systematics}: systematic errors can be incorporated in a statistically well-defined way. We included $p_T$ scale and resolution uncertainties in the leptonic MSSM-3 benchmark in Fig.s~\ref{fig:MSSM_pt_det} and~\ref{fig:MSSM_IDvRate_pt_gauss_det} and found no sensitivity to momentum scale uncertainties and a mild sensitivity to momentum resolution uncertainties. We expect similar results for angular resolution uncertainties.
    \item {\bf Protection from the Curse of Dimensionality}: This feature is already present by construction in our definition of the NNID, but we demonstrate it explicitly in a toy example in Fig.~\ref{fig:MSSM_IDvRateNoise_det}, and by detecting new physics in 100 dimensions in Section~\ref{sec:MSSMhad}.
\end{enumerate}

\subsubsection{Qualitative Comparisons with Other Search Strategies} A quantitative comparison with other model-independent search strategies goes beyond the scope of this work. However, at least conceptually, it is easy to show that our technique is complementary to the two broad categories of model-independent searches identified in the introduction. Our methodology does not require a localized resonance in some kinematical variable and in this sense is complementary to the techniques put forward in~\cite{Collins:2018epr, Collins:2019jip, Benkendorfer:2020gek} that require one. 
The more flexible strategies in~\cite{DAgnolo:2018cun, DAgnolo:2019vbw, dAgnolo:2021aun} can in principle detect any of our signals, but only at much larger cross sections. Their global $p$-value decreases exponentially with the dimension of the input space, placing the $d\simeq 100$ and $d \simeq 60$ examples that we have discussed well beyond their reach with any future LHC luminosity. Complementing~\cite{DAgnolo:2018cun, DAgnolo:2019vbw, dAgnolo:2021aun} with a pre-training step of dimensionality reduction as in~\cite{BrightThonney2025, grosso2025sparseselforganizingensembleslocal} can solve the problem, but for now it has been done with event-by-event observables for localized signals. So our ``global" approach is conceptually quite different and it would be interesting to compare the performances of the two ideas in more detail.

It is, of course, possible to find examples where other techniques would shine and ours would underperform. It is the case, for instance, of the LHC Olympics dataset~\cite{Kasieczka_2021}, which contains only resonant signals. Those signals are easily detectable with~\cite{Collins:2018epr, Collins:2019jip, Benkendorfer:2020gek}, but require a larger cross section with our methodology. Similarly, any intricate low-dimensional signal that requires combining multiple final states each with kinematics close to the SM and a small cross section is better searched for using~\cite{DAgnolo:2018cun, DAgnolo:2019vbw, dAgnolo:2021aun} rather than what we proposed in this work. 

To conclude this brief and qualitative discussion, we would like to emphasize two aspects of model-independent searches that are obvious to the practitioners, but often underappreciated by the broader community.
We chose to compare our sensitivities to that of a counting experiment because this is the only option when one knows nothing about the signal and other model-independent search strategies fail by construction. Asking if a model-dependent search targeted to our $Z'$ signal, for instance, would do better, is not a honest question. It forces a comparison between what we do when we know nothing about the signal and what we do when we know everything about it. Obviously, there is always a better way to construct a search compared to what we did, if we know what we are looking for. 
So all the hundred (or so) model-independent search strategies developed so far are truly useful only in two situations. The first is when we suspect that signals lying outside our current theory priors might be hidden in the data. The second is when the range of reasonable possibilities is so large that we do not have the manpower to search for all of them individually. An example of the first case is an object that does not trace a helix in the magnetic field of the LHC detectors. An example of the second is a Higgs-like particle with sizeable branching ratios in many final states but a cross section small enough that we must combine all channels with the right weights in order to observe it.

The second aspect of model-independent searches that is obvious, but worth to state explicitly is that anomaly detection is only the first step in a new physics search. We can establish if an anomalous event is coming from a new physics phenomenon only if we can define a reference hypothesis to compare it with and output a $p$-value. After identifying a set of anomalous events, we still have to set up a statistical test to determine whether they are just rare SM events or truly new physics. If we consider for instance anomaly detection with autoencoders, the most natural way to do it (but not the only way) would be to use the autoencoder output as a test statistics and then perform a standard frequentist goodness-of-fit test.

This short discussion completes the analogy between the most common use of autoencoders in high energy physics and the second category of model-independent search strategies that we mentioned in the introduction. The ideas in those two sets of papers are just two different ways of obtaining a test statistic needed to perform a goodness-of-fit test. A good test statistic necessarily describes correctly the probability distribution of the reference data (i.e. the SM) against which we want to compare the experimental data, so an autoencoder needs a larger and larger number of parameters in the hidden layer as we increase the dimensionality $d$ of the training data. The increase in internal parameters must be comparable to that in~\cite{DAgnolo:2018cun, DAgnolo:2019vbw, dAgnolo:2021aun} since both methodologies aim at fitting the probability distribution of the reference data with a neural network. Hence we expect a look-elsewhere effect that increases exponentially in $d$ also for autoencoders.

The analogy between these two methodologies becomes an equality in the case of normalized autoencoders~\cite{CMS:2025lmn, yoon2023autoencodingnormalizationconstraints} where the autoencoder is built explicitly to learn the probability distribution of the training data, just as was done in~\cite{DAgnolo:2018cun, DAgnolo:2019vbw, dAgnolo:2021aun} with different loss functions. Note that normalized autoencoders are built precisely to correct what makes traditional autoencoders unsuitable for anomaly detection, i.e. that they are not learning correctly the probability distribution of their training data for outliers.

However, as stated in the introduction, this is not the only possible use of autoencoders or dimensionality reduction in general. If we keep the number of extracted features fixed and do not increase it with the input dimension of the dataset, we end up with an analysis strategy in the first category, with a smaller look-elsewhere effect, but a correspondingly smaller expressivity on the signal. In that case we aim at well-describing the SM only on a lower-dimensional subspace of the input space.

\section{Conclusion and Outlook}
\label{sec:conclusions}
We have discussed how to perform a data-driven measurement of the dimension of the phase space of collider events. In Sections~\ref{sec:NP} and~\ref{sec:applications} we have demonstrated that it can be used as a model-independent new physics search strategy. It allows to capture new physics precisely where it could still be hiding at the LHC, i.e. in subtle correlations between hundreds or thousands of observables that in existing searches are coarse-grained and reduced to a few high-level quantities. Our methodology, as is, already allows to search for new physics without clustering jets (see Section~\ref{sec:MSSMhad}) and we see it as a first step towards performing a search directly on raw data. There are of course a number of open questions on how to analyze raw data directly, so many of them that we will not even list them explicitly, but this is a first conceptual step in this direction.

The results presented here aim at illustrating the salient conceptual features of the methodology. We see this as a first paper that will be followed by more detailed studies which will include more background processes, comparisons between different showering algorithms and MC tunes, the full set of systematic uncertainties for each analysis and studies of even higher-dimensional spaces.

In addition to new physics searches, there is a number of other applications of the \ID that we have barely mentioned in the rest of the paper, but we would like to explore in future work. We highlight here those that we find most interesting:
\begin{enumerate}
    \item {\it What is the intrinsic dimension of a jet?} The answer depends on the energy scale (as noted also in~\cite{Komiske:2019fks, Komiske:2019jim, Komiske:2022vxg}) and encodes interesting information on the transition between the perturbative and non-perturbative regimes. Compared to existing studies~\cite{Komiske:2019fks, Komiske:2019jim, Komiske:2022vxg} the \ID is more robust experimentally at small EMDs for finite datasets, as illustrated by the toy model of Section~\ref{sec:toy}. This leaves open the question of its calculability in perturbative QCD, which is itself an interesting  problem for future work.
    \item {\it What is the right dimension for the latent space of an autoencoder or other dimensionality reduction techniques?} This question is usually answered heuristically in high energy physics, but the \ID provides a (rather standard) data-driven methodology to find the optimal size of these spaces.
     \item {\it Is this a quark or a gluon jet?} From preliminary studies we found that the \ID of quark and gluon jets, computed from their constituents, is appreciably different and could be used as an observable to improve their classification, as was done with the correlation dimension in~\cite{Komiske:2022vxg}.
     \item {\it Can the \ID improve model-dependent searches?} The application of neural networks (and older techniques such as boosted decision trees) to model-dependent analyses seems to have left very little margin of improvement to human ingenuity. However model-dependent searches are optimized using event-by-event observables and not global properties of the dataset. We would like to explore the possibility that the \ID curves computed in this work can be used to further increase the sensitivity of model-dependent searches.
     \item {\it More ``global'' observables for new physics searches.} As mentioned just above, new physics is often searched for using observables that can be defined on an event-by-event basis, but here we have discussed a way to do it using only the measurement of a property of the entire dataset. The \ID has no meaning for a single event. In this sense our strategy is orthogonal to anomaly detection and could be the first of many other applications of global properties of a dataset to detect new physics. We already took a first step in this direction by using the average distances $\overline{r}_i$ to detect new physics.
\end{enumerate}
In this work we have explored the intrinsic dimension of a dataset as a tool for model-independent new physics searches. As such, it allows to study very high-dimensional spaces, and could reveal signals still hiding in LHC data. We see this technique as something qualitatively new in the already vast landscape of model-independent searches. 

Furthermore, the \ID of a dataset can have a long list of other applications to the analysis of collider data, including the study of non-perturbative aspects of QCD and the optimization of model-dependent searches.

\section*{Acknowledgements}
We would like to thank L. Wang, J. Thaler, A. Larkoski and G. Grosso for useful discussions. RTD would like to thank P. Urbani for introducing him to the NNID. RTD and GR acknowledge ANR grant ANR-23-CE31-0024 EUHiggs for partial support. This research was supported in part by grant NSF PHY-2309135 to the Kavli Institute for Theoretical Physics (KITP). The work of AG was supported in part by the European Union - Next Generation EU under Italian MUR grant PRIN-2022-RXEZCJ.
The work of AV has
received funding from the Swiss National Science Foundation (SNF) through the Eccellenza Professorial Fellowship ``Flavor Physics at the High Energy Frontier'', project
number 186866. Some of the calculations were performed at sciCORE (\href{http://scicore.unibas.ch/}{http://scicore.unibas.ch/}) scientific computing center at University of Basel.

\appendix

\section{Empirical $T^2$ Distribution, Validation and PCA} 
\label{app:QQ}

In this Appendix, we expand upon and illustrate the procedure outlined in \cref{sec:statTest} for validating the $T^2$ statistic under the null hypothesis.  The primary source of discrepancy between the theoretical and empirical distributions arises when the \ID vector components are highly correlated or when the feature count $m$ becomes comparable to the number of reference samples $n$.  In these regimes, the empirical covariance matrix $\bm{\hat\Sigma}$ develops small eigenvalues, its inversion is numerically unstable and amplifies statistical noise, causing large fluctuations in the test statistic.

To diagnose this instability, we construct the empirical $T^2$ distribution via a leave-one-out approach under the null hypothesis: for each $j=1,\dots,n$, we remove the $j$-th reference vector, compute $\bm{\hat\mu}^{(-j)}$ and $\bm{\hat\Sigma}^{(-j)}$ from the remaining $n-1$ samples, then evaluate
\begin{align}
\hat T^2_{j} \;=\; \frac{n-1}{n}\,\bigl(\bm{d}_{j} - \bm{\hat\mu}^{(-j)}\bigr)^{T}\,\bigl[\bm{\hat\Sigma}^{(-j)}\bigr]^{-1}\,\bigl(\bm{d}_{j} - \bm{\hat\mu}^{(-j)}\bigr)\,.
\end{align}
This yields $n$ (approximately independent, up to $\mathcal O(1/n)$ correlations) realizations of $\hat T^2$ under the null hypothesis.

A qualitative check is provided by a Q–Q plot of the empirical quantiles of $\{\hat T^2_j\}$ against the theoretical quantiles of the $T^2_{n-1,m}$ distribution (equivalently proportional to $F_{m,(n-1)-m}$).  Deviations from the diagonal, beyond expected finite-sample fluctuations in the upper tail, signal a breakdown of the null hypothesis assumptions.

For a quantitative assessment, one may employ nonparametric goodness-of-fit tests such as Kolmogorov–Smirnov \cite{an1933sulla} or Cramér–von Mises (CvM) \cite{anderson1962distribution}.  We prefer the CvM statistic because it weights deviations in the tails and near the median equally, yielding a more balanced measure of agreement.

Since the root cause of the discrepancy is covariance‐matrix instability, a natural remedy is dimensionality reduction.  We implement a PCA-based selection of principal directions as follows.

\medskip
\noindent\textbf{PCA Selection Algorithm:}
\begin{enumerate}
  \item For each reference vector $\bm{d}_j$, remove it from the dataset and perform PCA on the remaining $n-1$ vectors.  Apply the same rotation induced by PCA to $\bm{d}_j$.
  \item For each $k=1,\dots,m$, compute $\hat T^2_j(k)$ using only the first $k$ principal components (ordered by descending covariance matrix eigenvalue).
\end{enumerate}
\noindent This produces an $n\times m$ matrix of $\hat T^2_j(k)$ values.  For each column $k$, we apply the CvM test to compare its empirical distribution to the theoretical $T^2_{n-1,k}$ distribution, yielding a $p$-value $p(k)$.  Plotting $p(k)$ versus $k$ reveals the largest $k$ for which the empirical and theoretical distributions remain in statistical agreement.  Finally, we inspect the Q–Q plot at the chosen $k$ to confirm visually the adequacy of the fit. This final step also helps identify and potentially remove remaining outliers, which are still visible in some plots.

The same procedure can be repeated for the $\{\bm{\overline r} \}$ vectors and for the joint $\{ (\bm{d}, \bm{\overline r} ) \}$ set. When combined, PCA should be performed globally on the concatenated and standardized vector. Standardizing the concatenated vector before applying PCA is important, as otherwise observables with big relative but small absolute variations may be obscured by those with larger scales.

\begin{figure}[!ht]
\centering
\begin{minipage}[c]{0.49\textwidth}
  \centering
  \includegraphics[width=\textwidth]{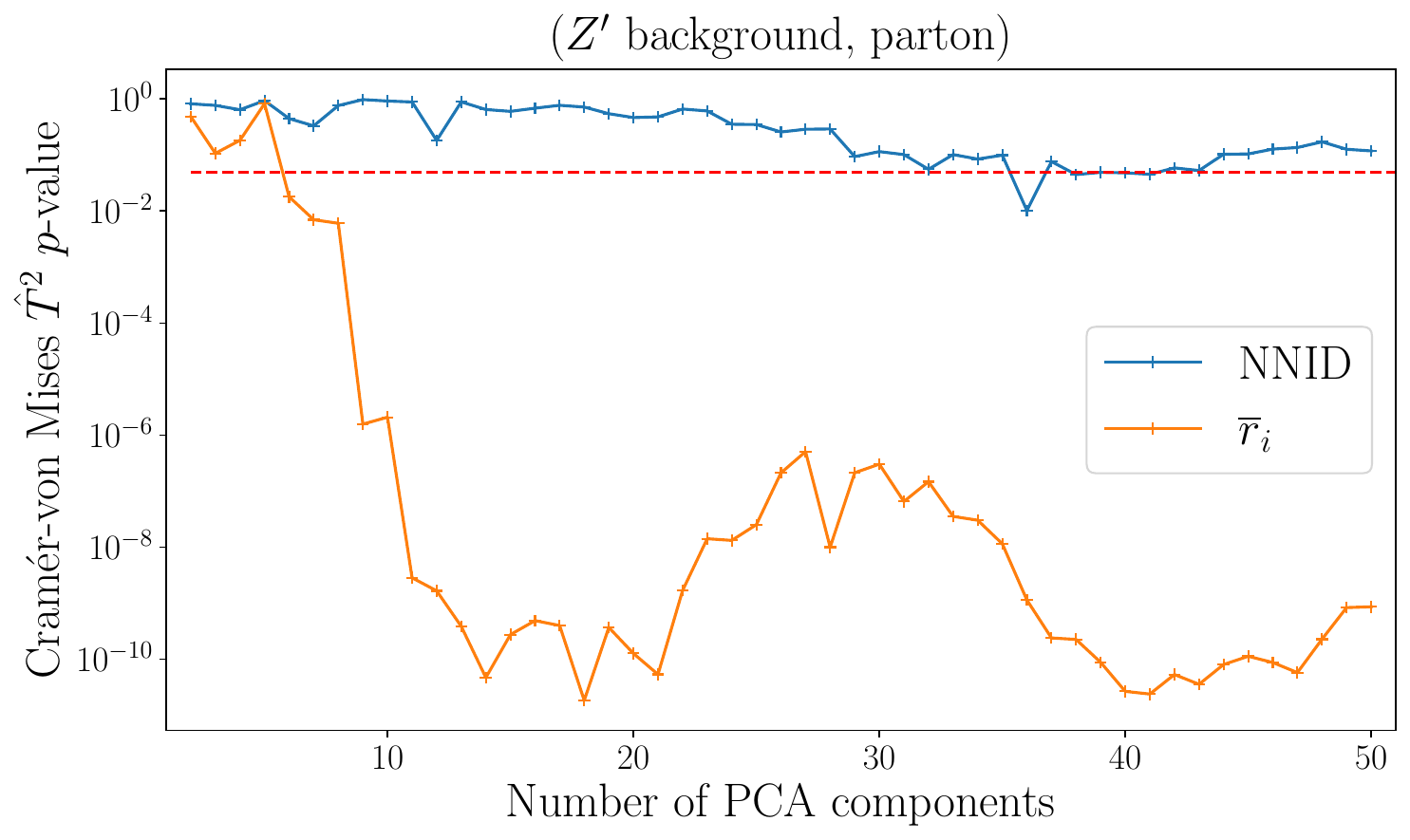}
\end{minipage}%
\hfill
\begin{minipage}[c]{0.49\textwidth}
  \centering
  \includegraphics[width=\textwidth]{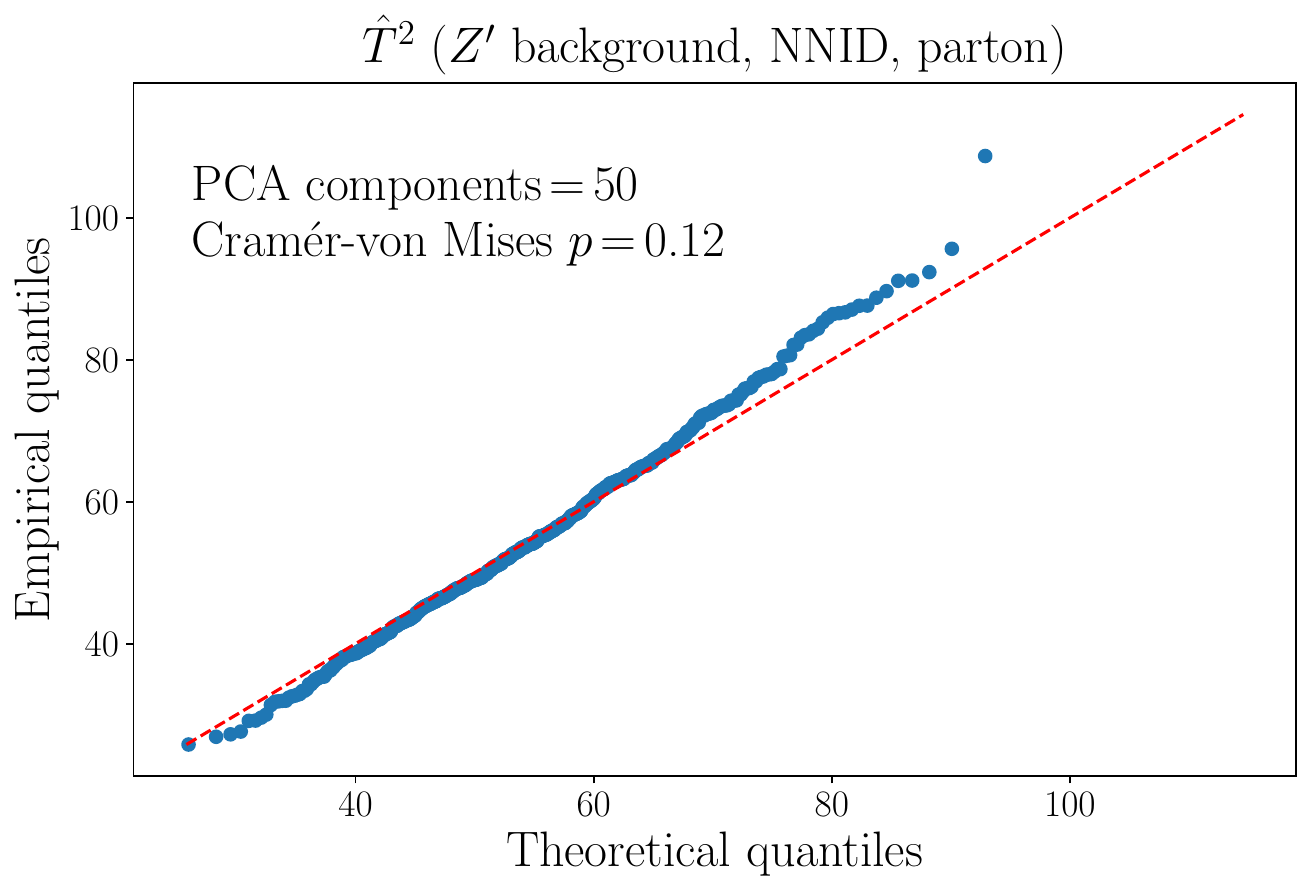}
\end{minipage}
\begin{minipage}[c]{0.49\textwidth}
	\centering
	\includegraphics[width=\textwidth]{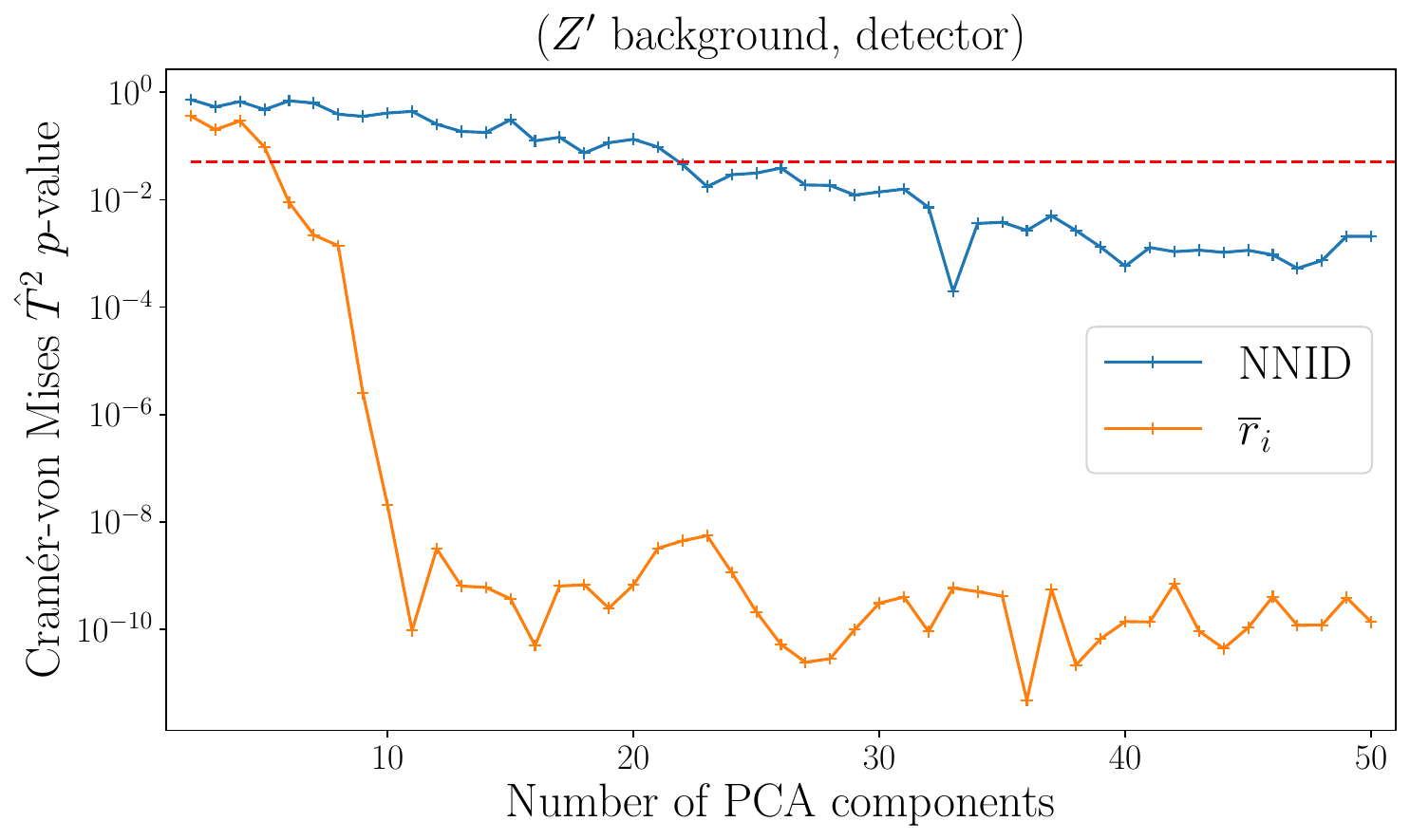}
\end{minipage}%
\hfill
\begin{minipage}[c]{0.49\textwidth}
	\centering
	\includegraphics[width=\textwidth]{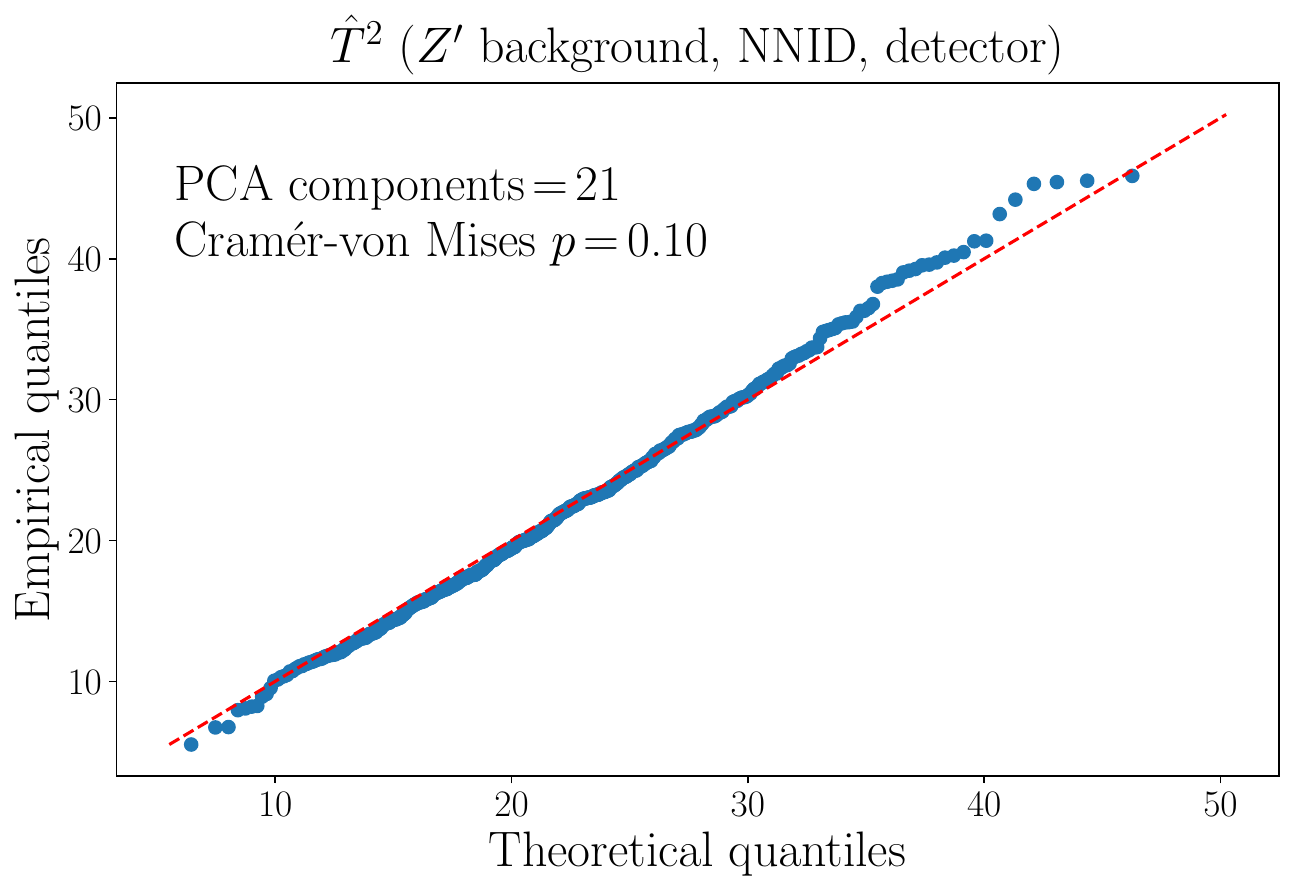}
\end{minipage}
\caption{CvM $p$-value as a function of the number of principal components retained for the SM background of the $Z'$ signal in Section~\ref{sec:Zp}. The right panel shows the qualitative agreement at the optimal number of components via a Q-Q plot for the \ID vector.}
\label{fig:systPCAexample}
\end{figure}

 \begin{figure}[!ht]
\centering
\includegraphics[width=0.49\textwidth]{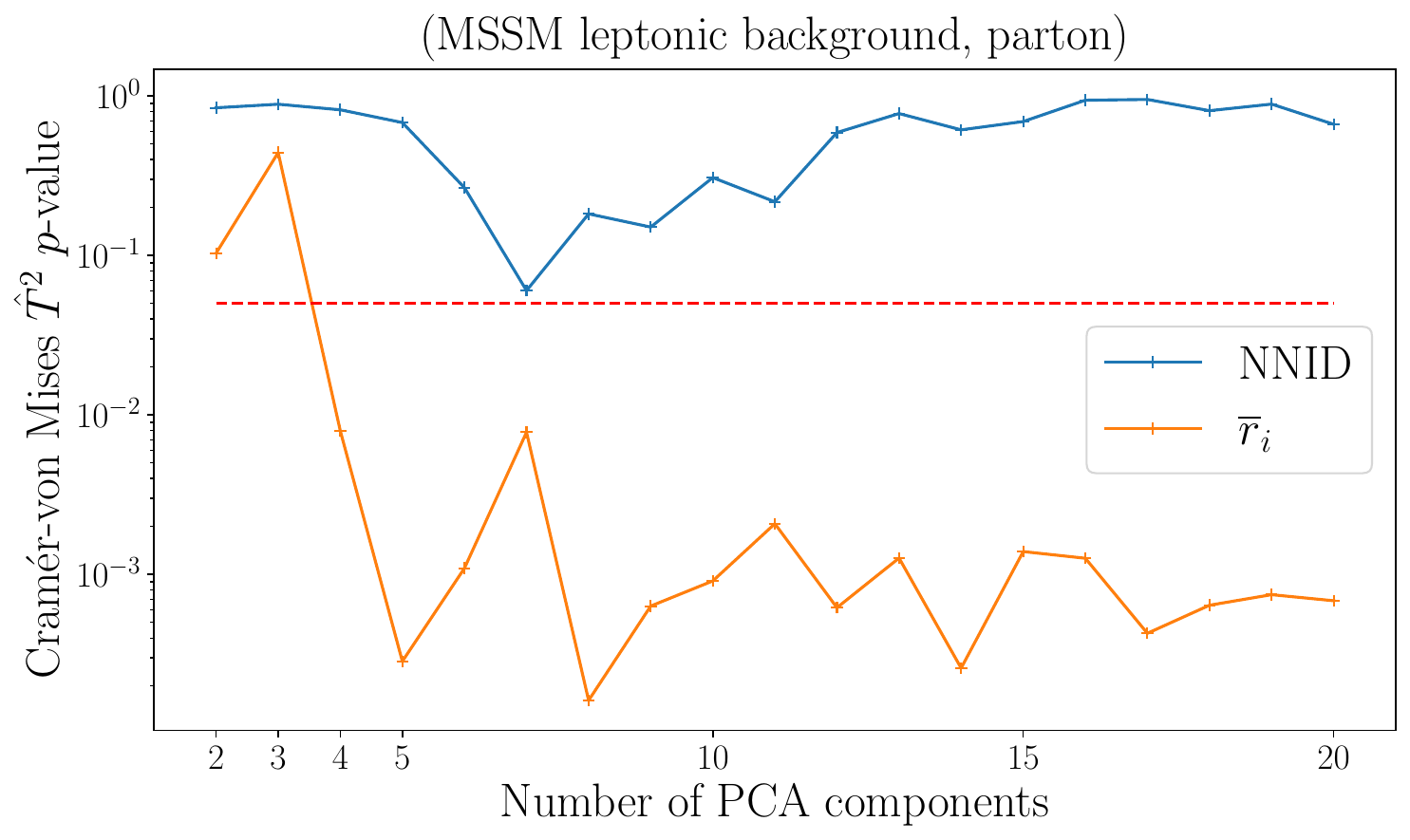}
\hfill
\includegraphics[width=0.49\textwidth]{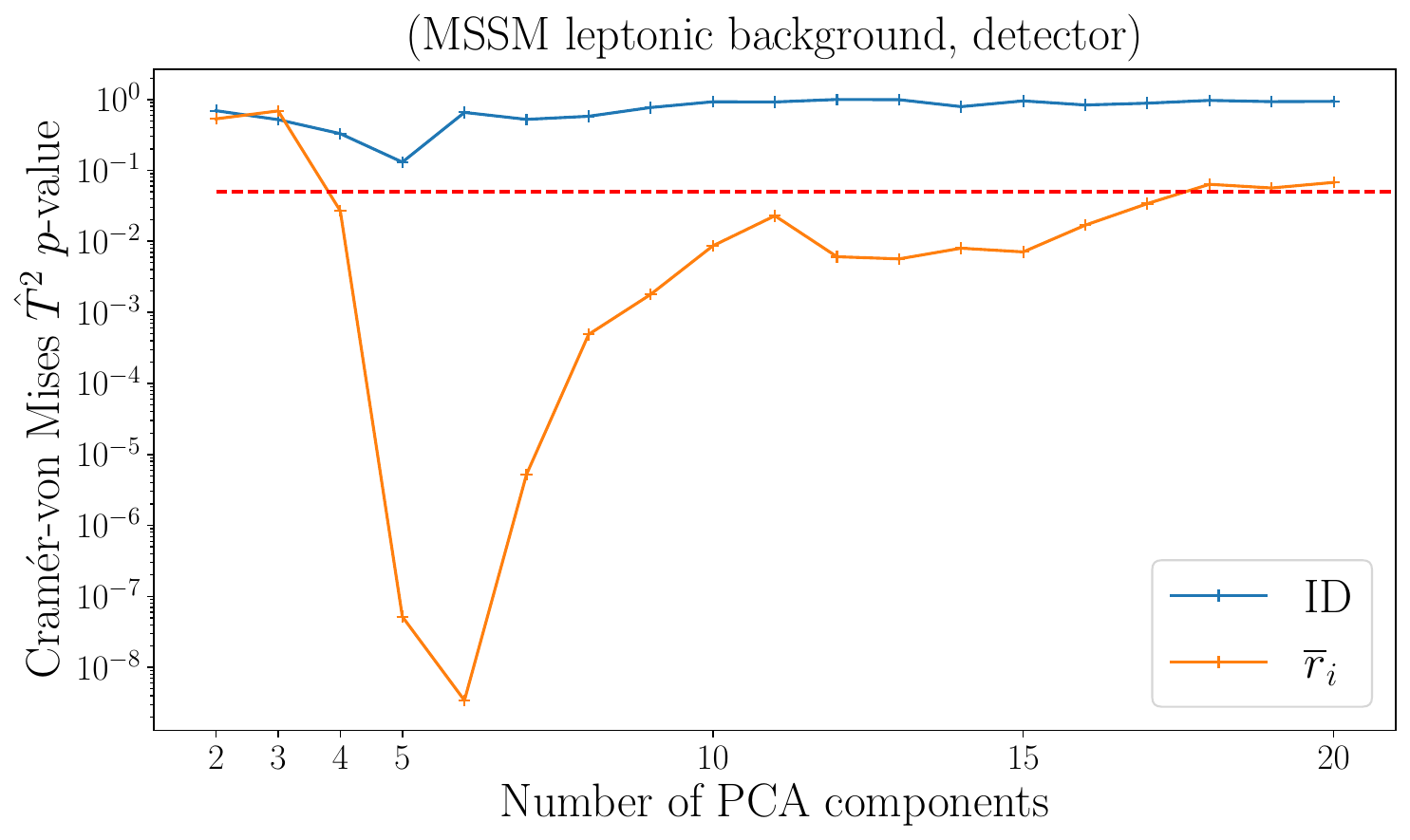}
\hfill
\caption{CvM $p$-values as a function of the number of principal components retained for the SM background of the MSSM leptonic signal in Section~\ref{sec:MSSM_lep} at parton and detector level.}
\label{fig:QQMSSMlep}
\end{figure}

\begin{figure}[!ht]
\centering
\begin{minipage}[c]{0.49\textwidth}
	\centering
\includegraphics[width=\textwidth]{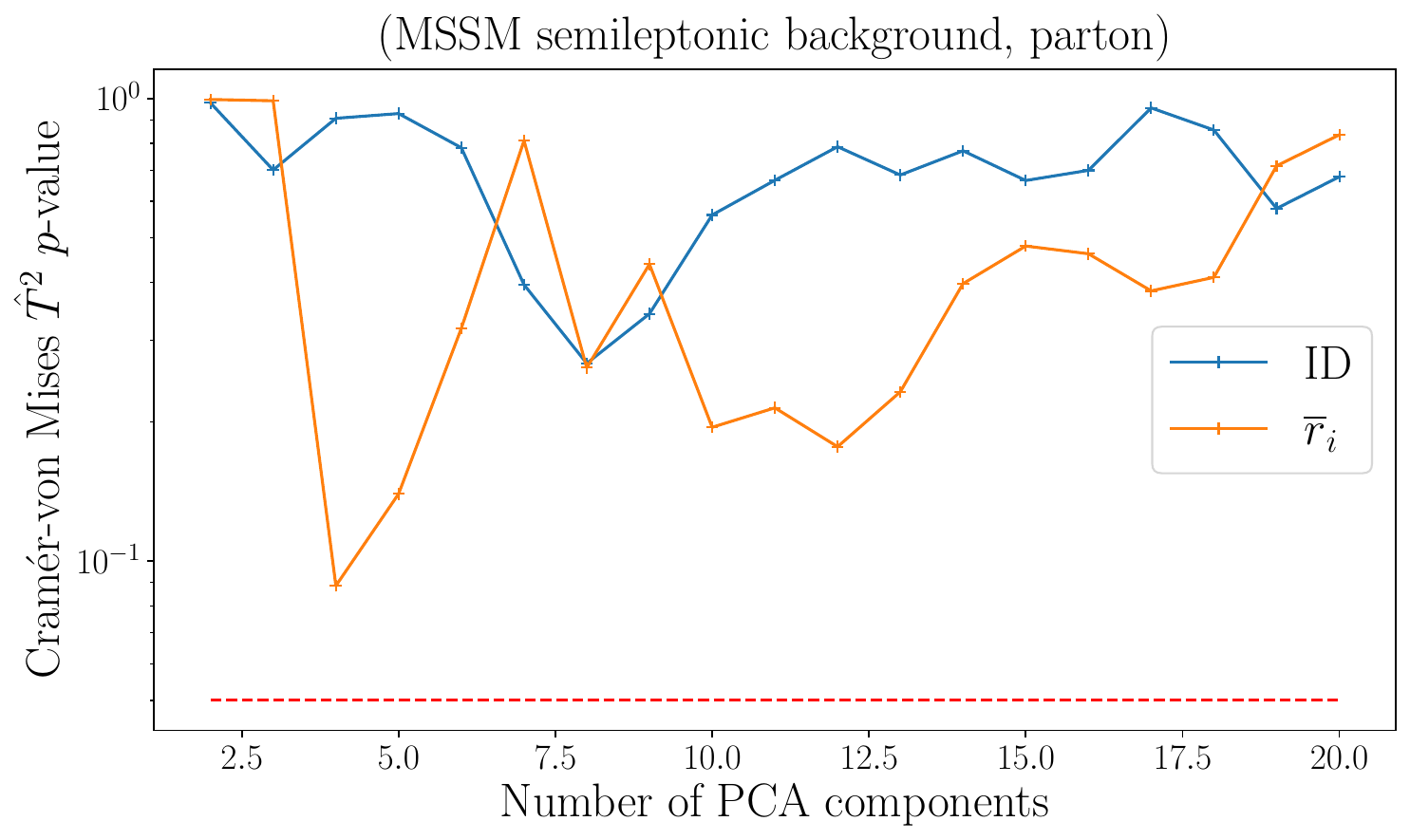}
\end{minipage}%
\hfill
\begin{minipage}[c]{0.49\textwidth}
	\centering
\includegraphics[width=\textwidth]{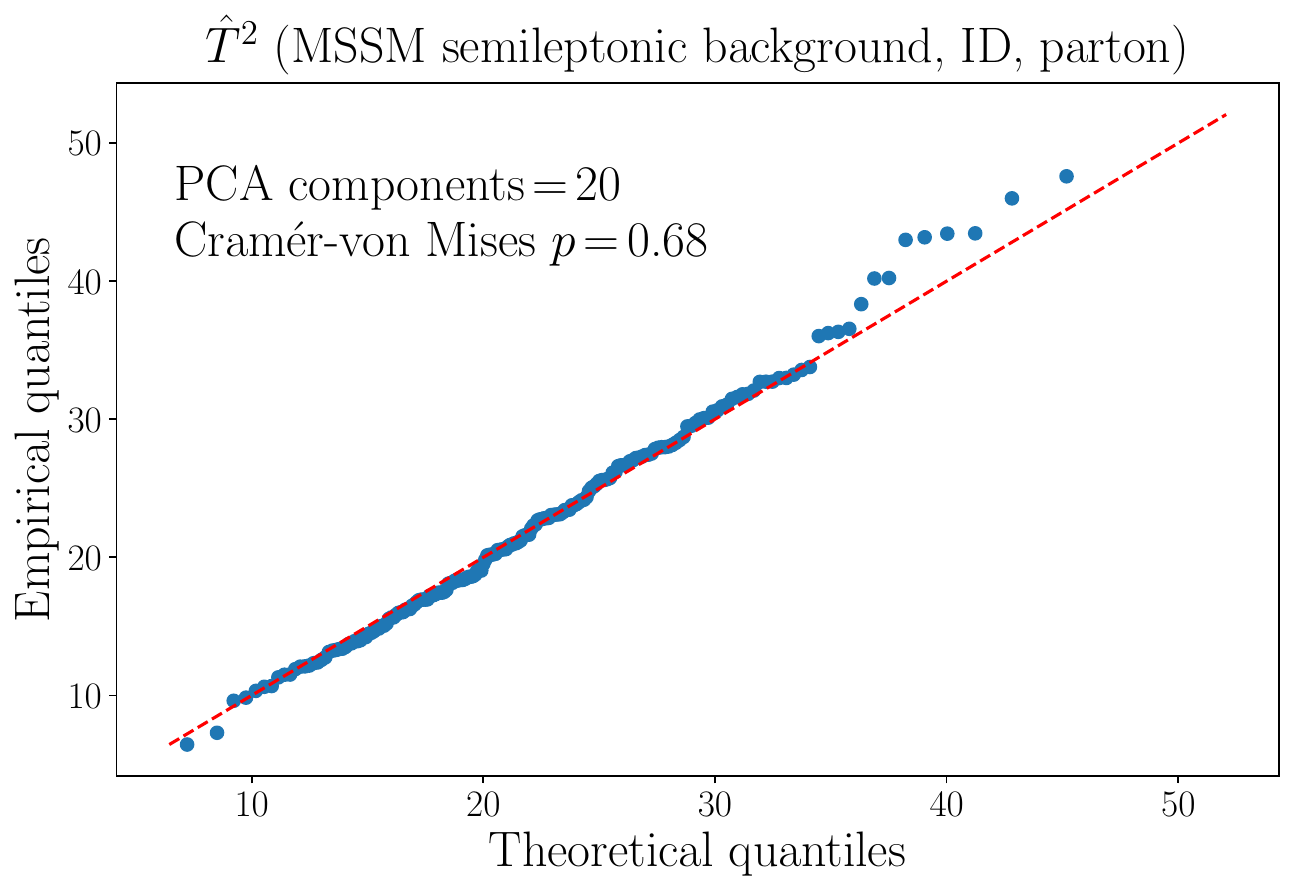}
\end{minipage}
\begin{minipage}[c]{0.49\textwidth}
	\centering
\includegraphics[width=\textwidth]{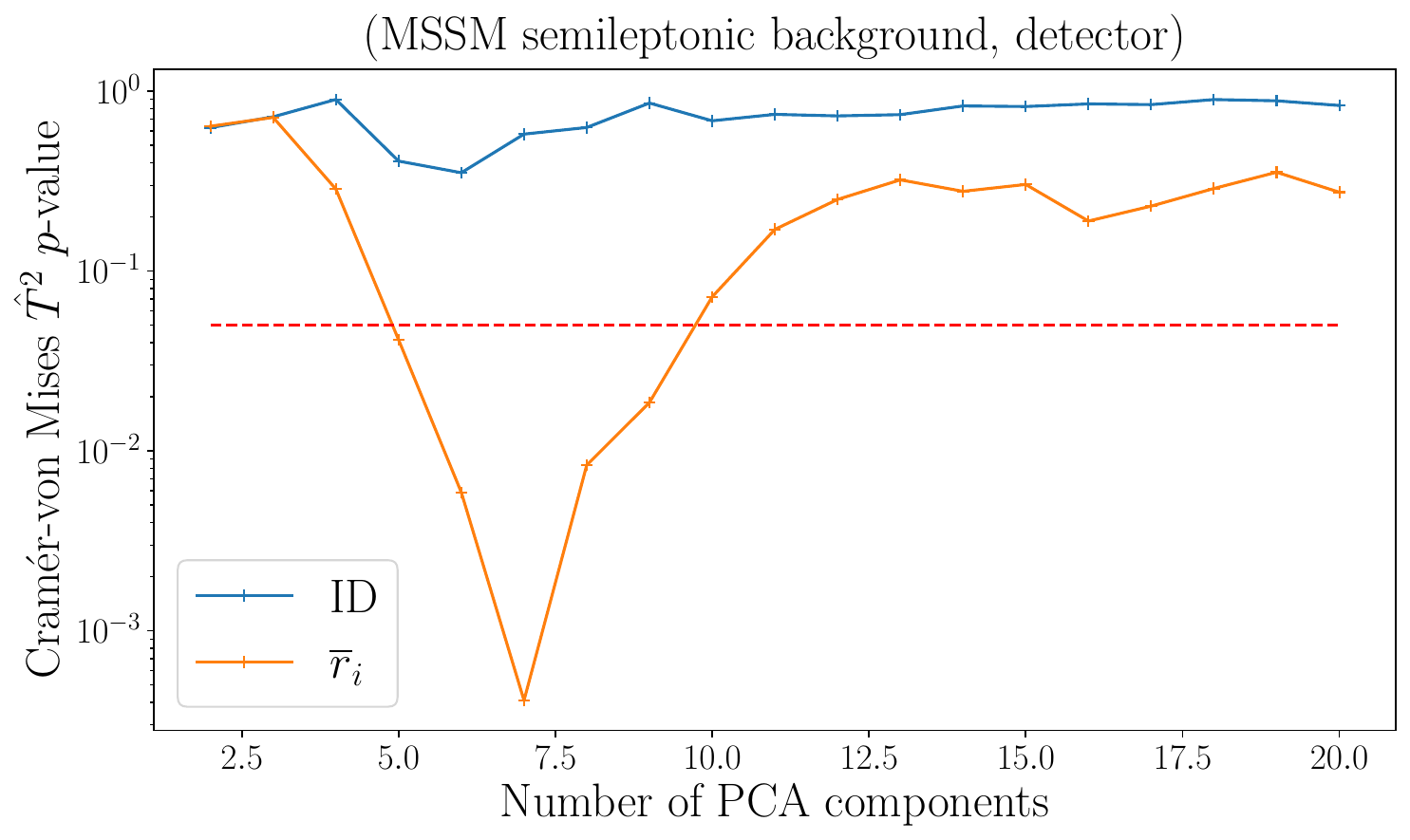}
\end{minipage}%
\hfill
\begin{minipage}[c]{0.49\textwidth}
	\centering
\includegraphics[width=\textwidth]{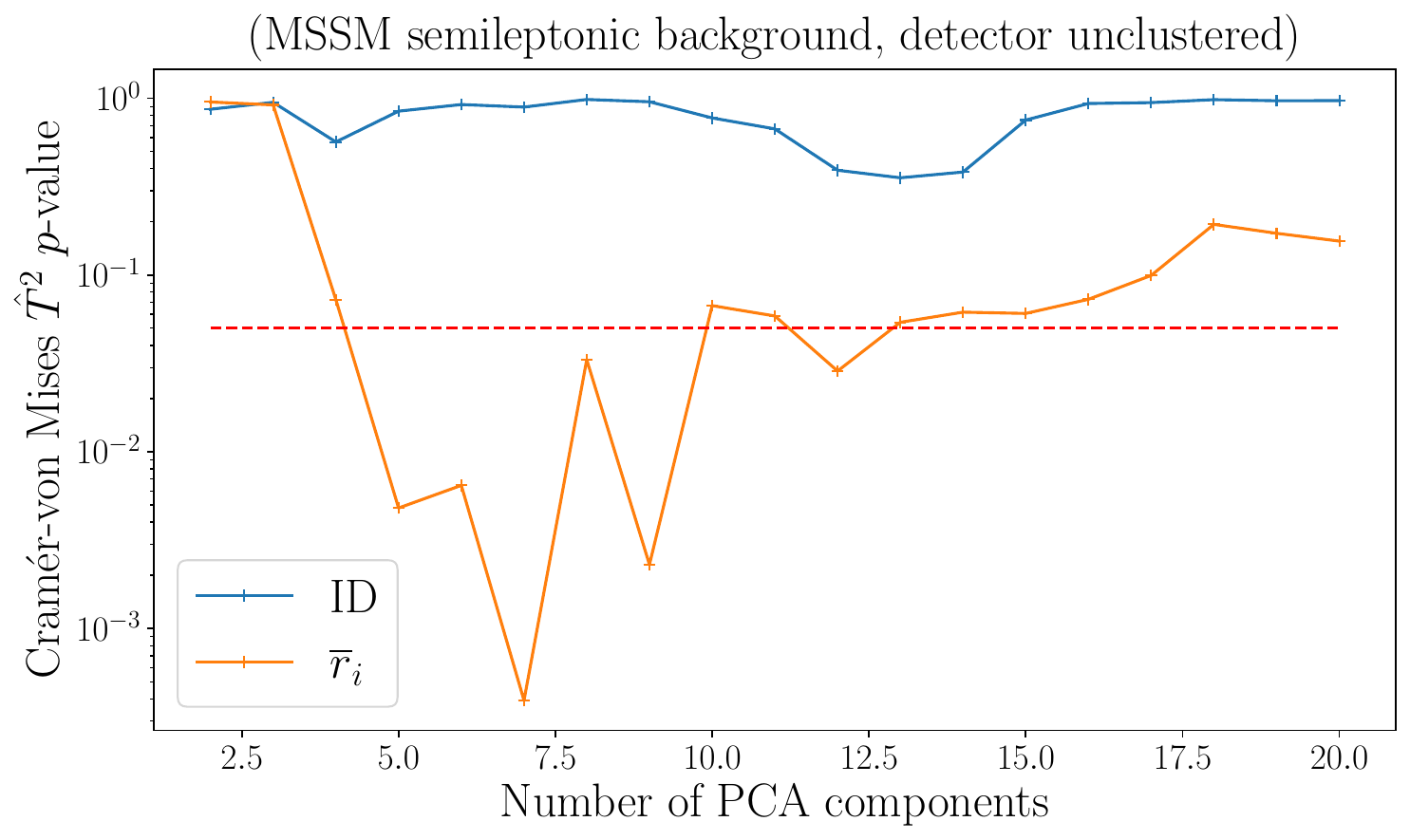}
\end{minipage}
\caption{Q-Q plot corresponding to the optimal number of PCA components for the \ID vectors at parton level (top right), and CvM $p$-value as a function of the number of principal components retained for the SM background of the MSSM semileptonic signal in Section~\ref{sec:MSSMhad}.}
\label{fig:QQMSSMhad}
\end{figure}

A representative example is shown in \cref{fig:systPCAexample}, applied to the Standard-Model background of the $Z'$ signal in \cref{sec:Zp}.  The left panel displays the CvM $p$-value curves for the \ID vectors and the $\overline{r}_i$ vectors. For small $k$, the CvM tests shows good compatibility between the theoretical and empirical reference distributions. As $k$ increases, small eigenvalues introduce instability and the $p$-values fall.  Setting a 5\% cutoff selects, for instance, $k=21$ components for the \ID vector; the corresponding Q–Q plot in the right panel confirms the adequacy of this choice.

For completeness, we include in the following pages the diagnostic plots for the other benchmark models analyzed in \cref{sec:applications}, following the same procedure used in \cref{fig:systPCAexample}. In Figs.~\ref{fig:QQMSSMlep} and \ref{fig:QQMSSMhad} we present the corresponding results for the leptonic and semileptonic MSSM backgrounds discussed in Section~\ref{sec:MSSM_lep}, \ref{sec:MSSMhad}.

Interestingly, at very large $k$ the $p$-value curves sometimes recover,  particularly in the case of $\overline{r}_i$ for the MSSM background (Fig.s \ref{fig:QQMSSMlep}, \ref{fig:QQMSSMhad}), wrongly suggesting a good fit.  We attribute this to the diminishing discriminating power of the $F_{k,(n-1)-k}$ distribution when $k$ is large: noise amplifications fail to produce statistically significant deviations, giving an illusion of agreement.  This regime, however, is unsuitable for new-physics discovery and should be avoided.

\section{More Details on the MSSM Analyses}
\label{app:scale}

In \cref{sec:MSSM_lep} we showed the impact on the significance test based on the \ID\ and $\overline{r}_i$ when uniformly rescaling the particle transverse momenta ($p_T$) \emph{after} the event selection had been carried out. Here we examine how the results change if the rescaling is applied \emph{before} the event selection. In this case, we downscale $p_T$ by $5$, $10$, and $20\%$ with respect to the true values.

\begin{figure}[!htb]
    \centering
    \includegraphics[width=0.49\linewidth]{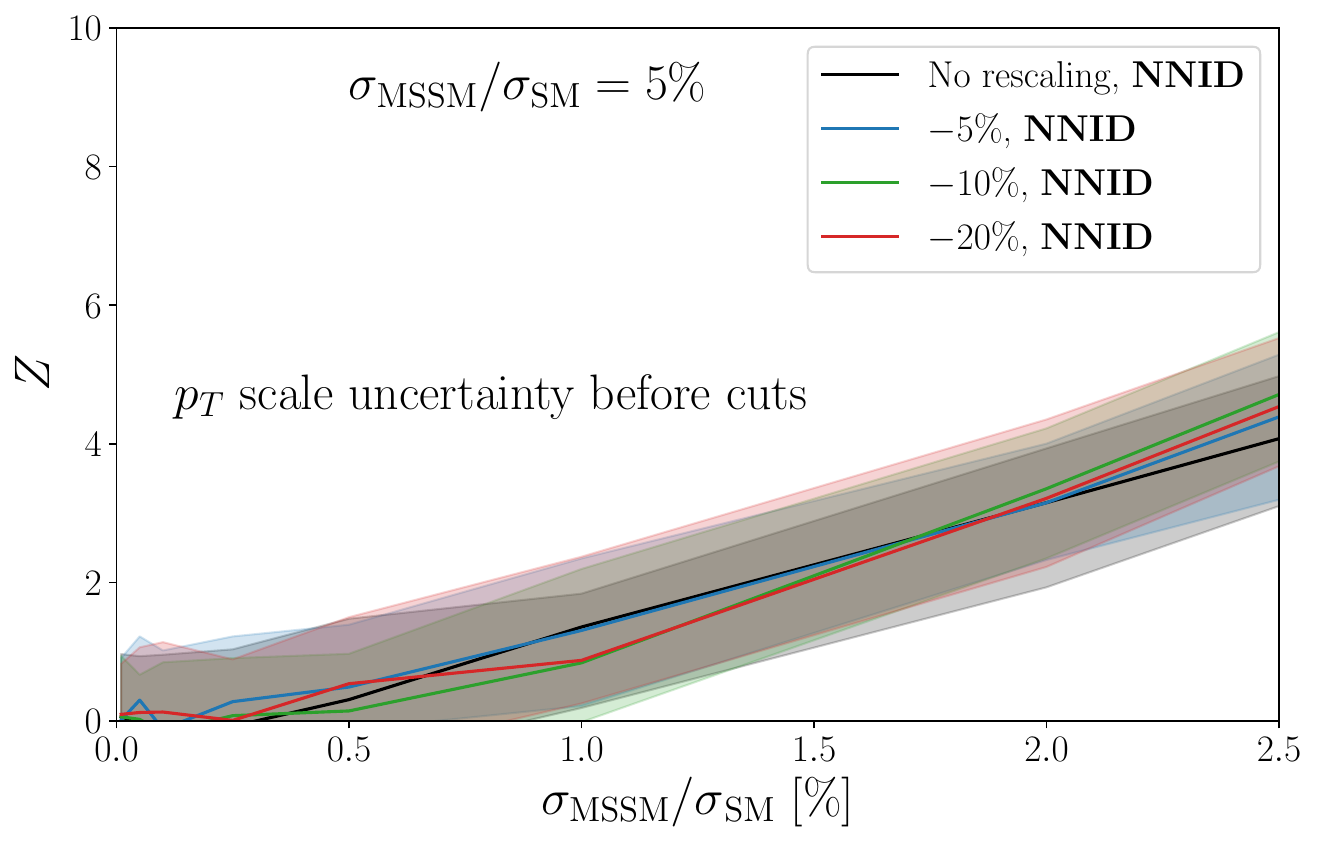}
    \includegraphics[width=0.49\linewidth]{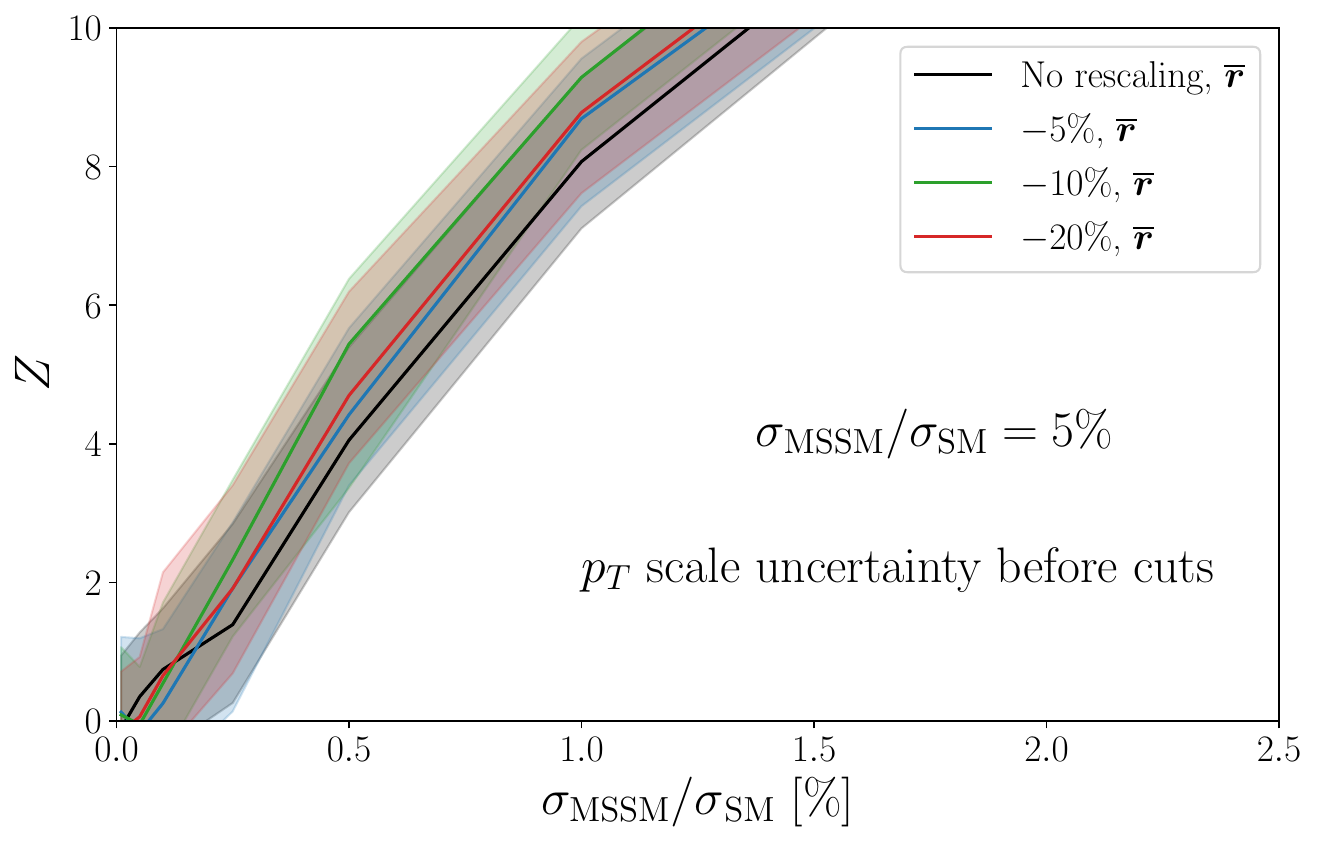}
    \caption{Impact of a global uniform downscaling of $p_T$ \emph{before} the event selection on the significance tests with \ID and $\overline{r}_i$ in the leptonic MSSM case.}
    \label{fig:MSSM_lep_significance_pT_rescaling_cut}
\end{figure}

\begin{figure}[!htb]
\centering
\includegraphics[width=0.49\textwidth]{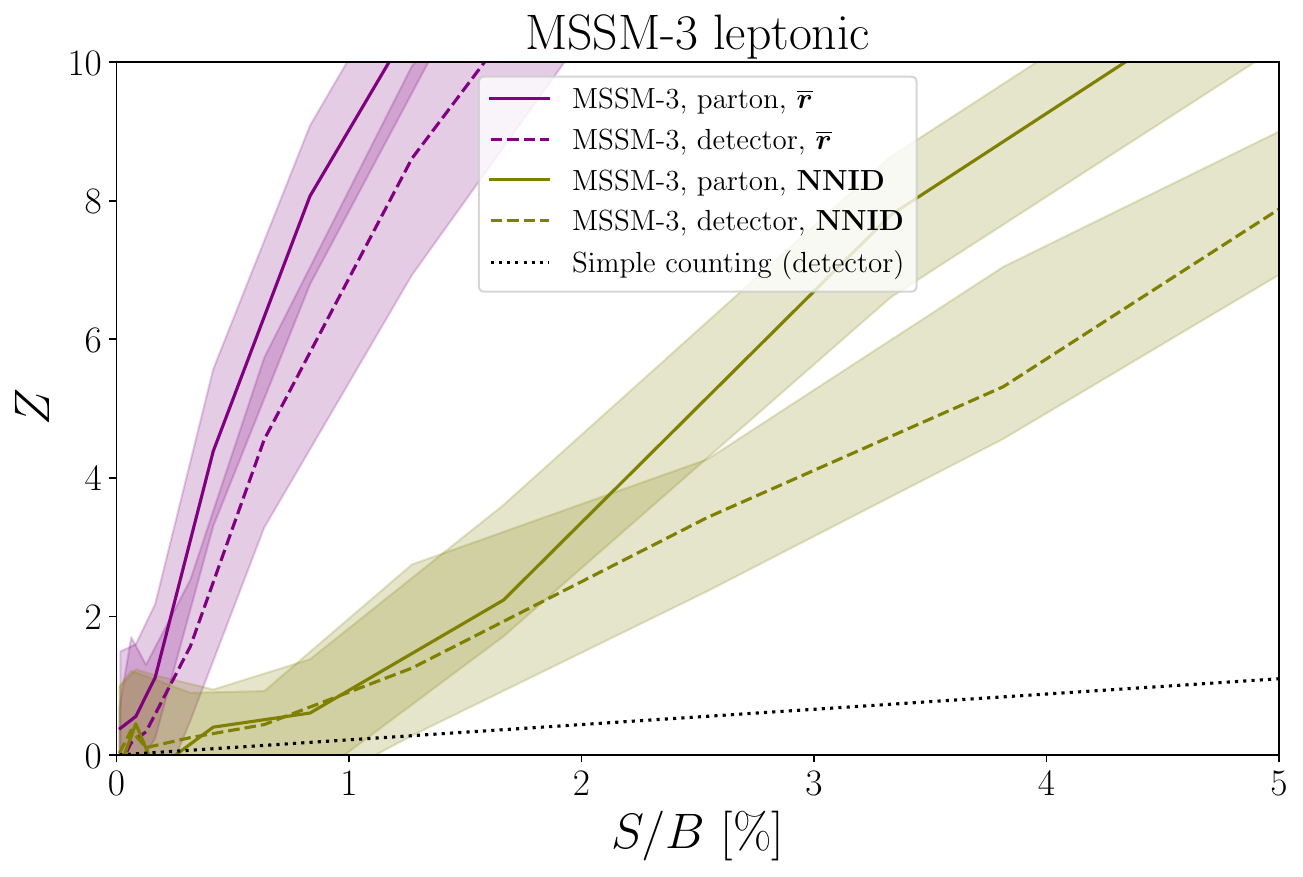}
    \includegraphics[width=0.49\textwidth]{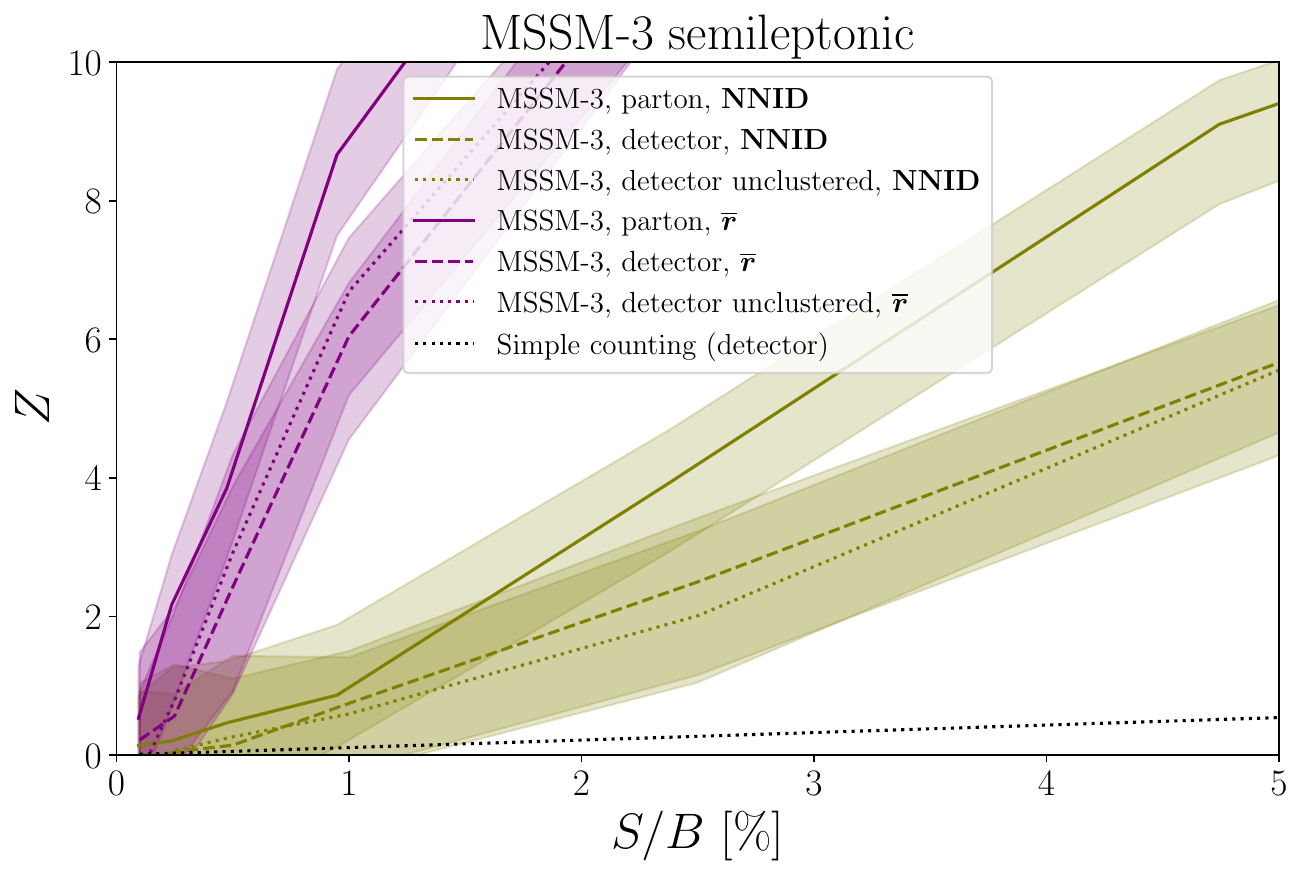}
\caption{Significance bands at 68\% CL of the new physics test in the leptonic (left) and semileptonic (right) MSSM-3 model as a function of the signal over background rate after all cuts.}
\label{fig:MSSM_ZvSB}
\end{figure}

In the leptonic case, the event selection required $p_T > 10\text{ GeV}$ for all leptons, and additionally $p_T < 40 \text{ GeV}$ for all extra jets present in the event. Since the leptons originate from on-shell $Z$ and $W$ bosons, the downscaling has a negligible effect on the lepton $p_T$ cut. The dominant effect instead arises from the jet veto: as $p_T$ is reduced, more events are accepted because they satisfy the veto condition. Quantitatively, the overall cut efficiency increases by about $5\%$ between the no-rescaling and the $20\%$ downscaling cases. Consequently, we expect a small improvement on the significance test, unlike the post-cut rescaling analyzed in \cref{sec:MSSM_lep}, where the test remained exactly invariant. This expectation is confirmed by our results, shown in \cref{fig:MSSM_lep_significance_pT_rescaling_cut}. Interestingly, the effect is more pronounced for the $\overline{r}_i$ test than for the \ID\ one.

To conclude we show the sensitivity curve for the leptonic and semileptonic analysis of the MSSM-3 benchmark as a function of $S/B$ after all cuts in Fig.~\ref{fig:MSSM_ZvSB}.

\bibliographystyle{JHEP}
\bibliography{biblioPaper}
\end{document}